\documentclass{aa}
\usepackage{natbib}
\usepackage{graphicx}
\usepackage{times}
\usepackage{amsmath}
\usepackage{txfonts}
\newcommand{\etal}{et al.,~}
\newcommand{\msun}{{\,\rm M}_{\odot}}
\newcommand{\lsun}{{\,\rm L}_{\odot}}

\newcommand{\kms}{\,{\rm km.s}^{-1}}

\newcommand{\nht}{\ifmmode {{\rm NH}_3} \else {NH{\bas 3}} \fi}
\newcommand{\tco}{\ifmmode {^{13}{\rm CO}} \else {$^{13}{\rm CO}$}\fi}
\newcommand{\dco}{\ifmmode {^{12}{\rm CO}} \else {$^{12}{\rm CO}$}\fi}
\newcommand{\cdo}{\ifmmode {{\rm C}^{18}{\rm O}} \else {${\rm C}^{18}{\rm O}$}\fi}

\newcommand{\htco}{\ifmmode {{\rm H}^{13}{\rm CO}^{+} } \else {${\rm H}^{13}
{\rm CO}^{+}$ }\fi}
\newcommand{\hco}{\ifmmode {{\rm H}^{12}{\rm CO}^{+} } \else {${\rm H}^{12}
{\rm CO}^{+}$ }\fi}
\newcommand{\juz}{\ifmmode {{\rm J}=1\rightarrow 0} \else
{J=1$\rightarrow$0}\fi}
\newcommand{\jdu}{\ifmmode {{\rm J}=2\rightarrow 1} \else
{J=2$\rightarrow$1}\fi}
\newcommand{\jtd}{\ifmmode {{\rm J}=3\rightarrow 2} \else
{J=3$\rightarrow$2} \fi}
\newcommand{\jcq}{\ifmmode {{\rm J}=5\!\rightarrow\!4} \else
{${\rm J}=5\!\rightarrow\!4$} \fi}
\newcommand{\as}{\ifmmode {^{\scriptscriptstyle\prime\prime}}
        \else $^{\scriptscriptstyle\prime\prime}$\fi}
\newcommand{\am}{\ifmmode {^{\scriptscriptstyle\prime}}
        \else $^{\scriptscriptstyle\prime}$\fi}
\renewcommand{\hco}{\ifmmode {{\rm HCO}^+} \else {HCO$^+$} \fi}
\newcommand{\app}{$a_+$}
\newcommand{\amm}{$a_-$}
\newcommand{\vt}{$\Delta V$}
\newcommand{\nhav}{N$_\mathrm{H}$/A$_\mathrm{V}$}

\newcommand{\tabALL}{
\begin{table*}[!ht]
 \caption{Best parameters}\label{tab:all}
\begin{tabular}{l|cc|cccc} 
 \hline \hline
 (1)  & (2) & (3) & (4) & (5) &( 6) &(7) \\
Source   &\multicolumn{2}{c|}{CQ Tau} & \multicolumn{4}{c}{MWC 758} \\
Data      & \dco~\jdu & Dust & \dco~\jdu & \dco~\juz & \dco & Dust \\
 \hline
 $V_\mathrm{LSR}$ (km.s$^{-1}$) & $6.17\pm0.04$ & &  5.79$\pm 0.01$ & $5.90\pm0.02$ & $5.80\pm0.02$  & \\
 Orientation, PA~($^{\circ}$) &  -36.7 $\pm 1.3$ &   $-36 \pm 18 $ & -31 $\pm 1$  & -23 $\pm3$ & -31 $\pm 1$ & $-38 \pm 7 $ \\
 Inclination, $i$~($^{\circ}$) &   29.3 $\pm 1.7 $ &  $29 \pm 9$ & 18 $\pm 6 $ & 16 $\pm 1$ & 16 $\pm 4$  & $40 \pm 20$ \\
 \hline
 \multicolumn{3}{c}{} \\
 \multicolumn{3}{c}{Velocity law:~~~~~~ $V(r) = V_{100} (\frac{r}{100\,\rm{AU}})^{-v}$}\\
 \multicolumn{3}{c}{} \\
 Velocity$(^*)$, (km.s$^{-1}$)    & $4.0 \pm 0.2$  & & $3.6 \pm 1.1$ & [4.00] & $4.0 \pm 0.6$ &   \\
 Velocity exponent,  $v$  &   0.51  $\pm$ 0.02 & &   0.51  $\pm$ 0.03 & 0.47 $\pm 0.07$ & 0.50 $\pm 0.02$& \\
 Stellar mass, M$_*$ ($\msun$) &   1.8 $\pm$ 0.2  & & 1.5 $\pm$ 0.7 [1.80] & [1.80] & $1.80\pm 0.5$ &  \\
 \hline
 $\Sigma$ $(^*)$, (cm$^{-2}$) & $1.7 \pm 0.1\, 10^{16}$ &  $1.7 \pm 0.3\, 10^{22}$ & $3.5 \pm 0.7\, 10^{16}$ &
   $1.6 \pm 2.4\, 10^{16}$ & $4.7\pm 0.9\, 10^{16}$ & $6.0 \pm 2.0\, 10^{22}$\\
$\Sigma_{\textrm{mass}}$ $(^*)$  (g.cm$^{-2}$) &  & 0.075 $\pm$ 0.015 & & & & 0.3 $\pm$ 0.1 \\
 Exponent $p$ & $2.3 \pm 0.2$ &    $1.3 \pm 0.1$ & $2.7 \pm 0.5$ & [3] & $2.9\pm0.4$ &  $1.5 \pm 0.4$ \\
 Outer radius $R_\mathrm{out}$, (AU) & $200 \pm 20$ & $200  \pm 30$ & $300 \pm 20$ & $230\pm30$ & $270 \pm 15$ & $180  \pm 40$  \\
 Temperature$(^*)$,(K) & $150 \pm 50$~~~~[150]  & & $37 \pm 6$ & $24 \pm4$ & $30\pm 1$ \\
 Exponent $q$ &  $0.7 \pm 0.5$~~~~[0.5] &   & $0.05 \pm 0.20$ & $0.6\pm0.3$ & $0.37\pm0.15$ \\
 $\delta V$ $(^*)$, (km.s$^{-1}$) & $0.32 \pm 0.09$ & & $0.50 \pm 0.03$ & $0.28\pm0.10$ & $0.44 \pm 0.02$ \\
 Scale height$(^*)$, (AU) & $22$ &  & $15$ & & 11 \\
 \hline
  $\beta$ & &  $0.70 \pm 0.04$  & & & & 1.0 $\pm$ 0.15 \\
 \hline
\end{tabular}\\
Column (1) contains the parameter name. Columns (2) and (4) indicate the parameters derived from \dco~\jdu,
column (5) parameters derived from \dco~\juz, and
columns (3) and (7) parameters derived from the dust emission, using
the disk temperature from \dco\  and the dust emissivity from Eq.\ref{eq:knu}.
Column (6) indicates the results of a simultaneous fit to both CO lines.
Note that the P.A. is that of the disk axis.
$\delta V$ is the local line width (sum of thermal + turbulent component see \citet{Pietu_etal2007} for a description of the convention). $\Sigma$ is the surface density ((H+2H$_2$)/2) and $\Sigma_{\textrm{mass}}$ the mass surface density assuming a $gas/dust$ ratio of 100.  $(^*)$ values at 100 AU. Square brackets indicate fixed parameters. The error bars correspond to 1 $\sigma$ level of uncertainties.\\
\end{table*}
}
\newcommand{\figMAPS}{
\begin{figure*}[ht]
  \centering
  \resizebox{18.0cm}{!}{
  \textbf{\Large{(a)}}\hspace{-0.5cm}\includegraphics[angle=270.0,width=9cm]{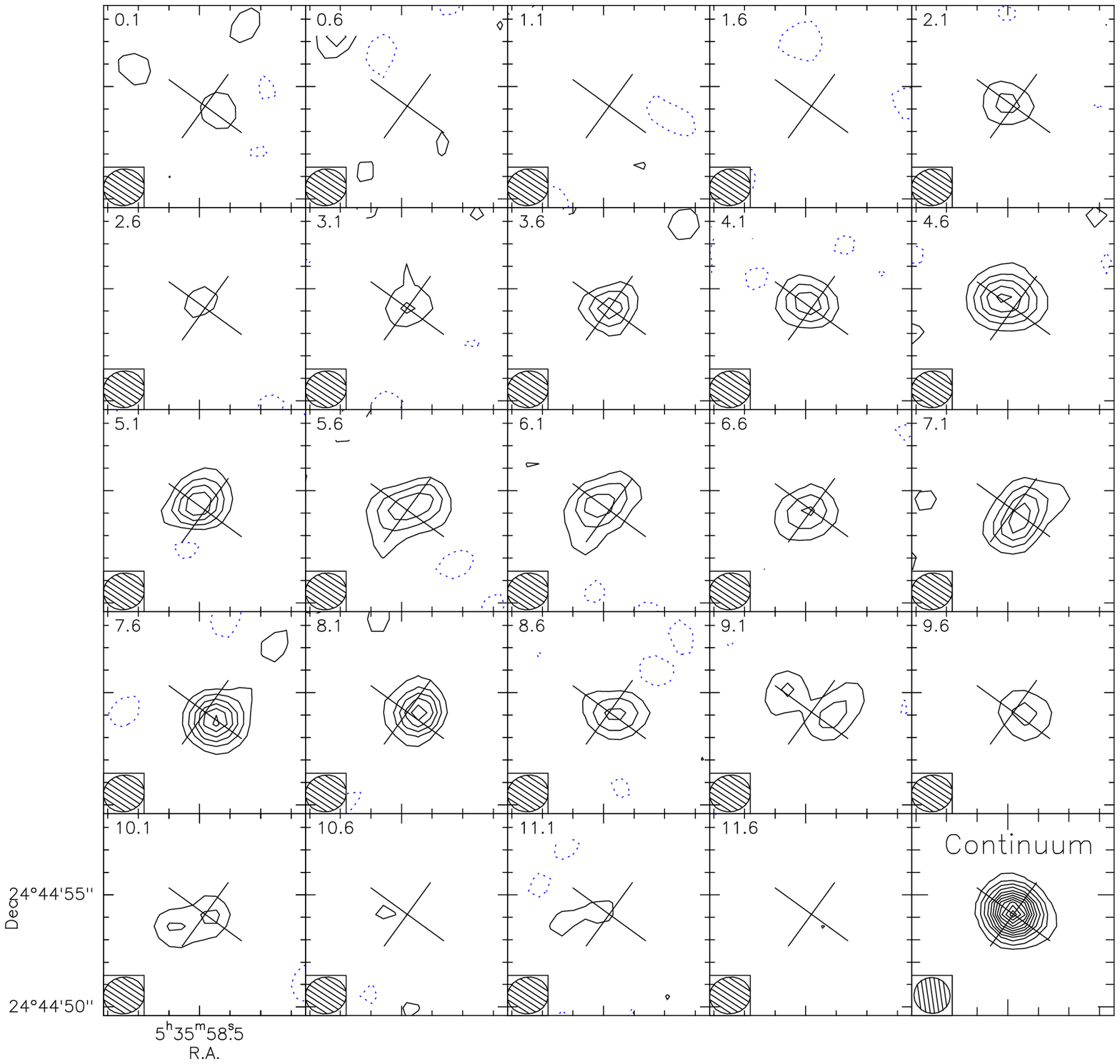}
  \textbf{\Large{(b)}}\hspace{-0.5cm}\includegraphics[angle=270,width=9cm]{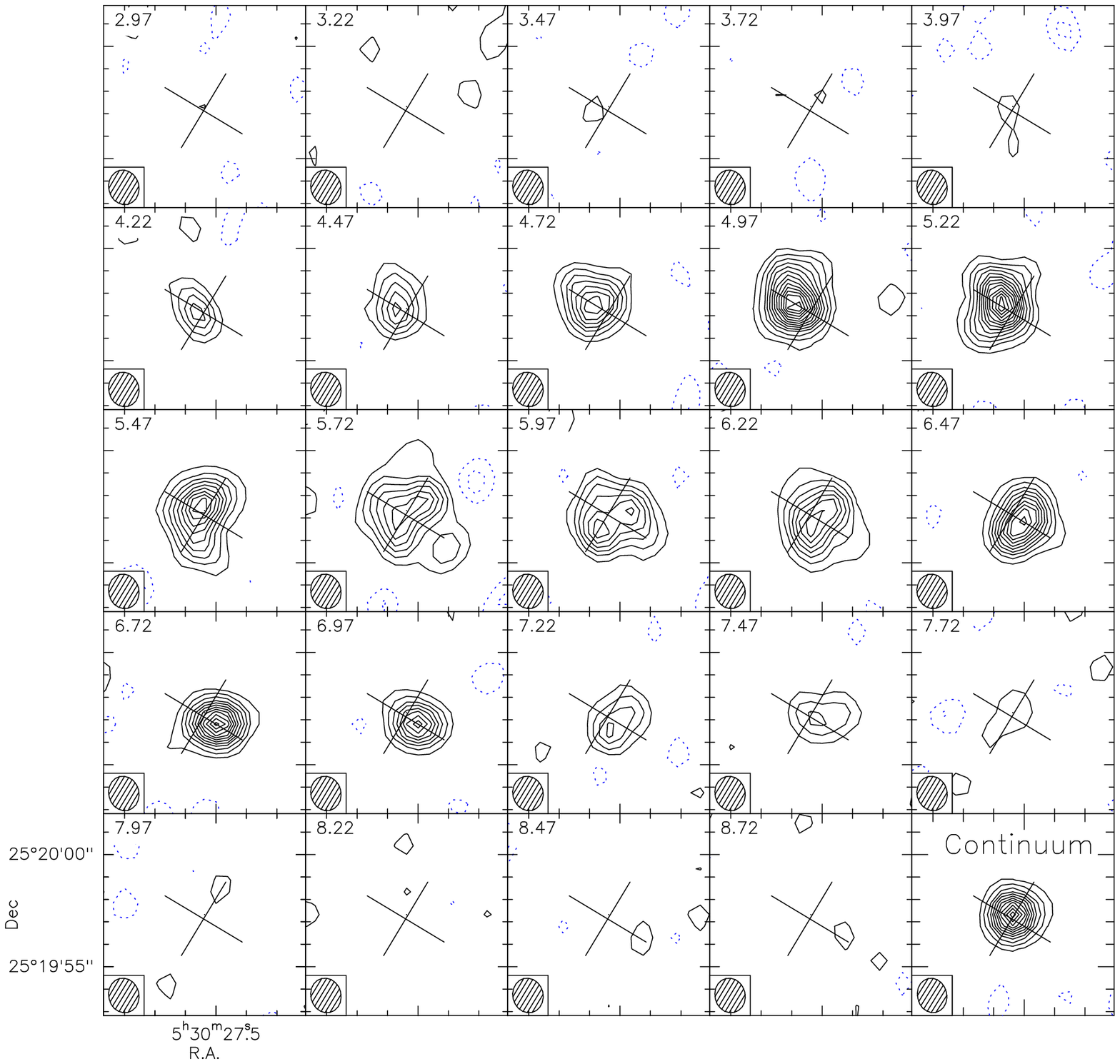}
  }
  \caption{Channel maps of the CO J=2-1 line emission and 1.3 mm continuum emission (bottom left
  panel) towards MWC 758 and CQ Tau. (a) CQ Tau  Contour step is 0.075 Jy/beam or 0.56 K ($2 \sigma$) for
  CO. Contour step is 10 mJy/beam, or 90 mK ($8 \sigma$) for the continuum. (b) MWC 758. Contour step is 0.1 Jy/beam or 1.1 K ($2 \sigma$) for CO.  Contour step is  3 mJy/beam (30 mK, $2 \sigma$) for the 1.3 mm continuum. The crosses
  indicate the position, orientation and aspect ratio of the disk determined from the CO analysis. The velocity (in km.s$^{-1}$) is indicated in the upper left corner.}
\label{fig:maps}
\end{figure*}
}

\newcommand{\figMWC}{
\begin{figure*}[ht]
  \centering
    \resizebox{18.0cm}{!}{
   \includegraphics[angle=270,width=18.5cm]{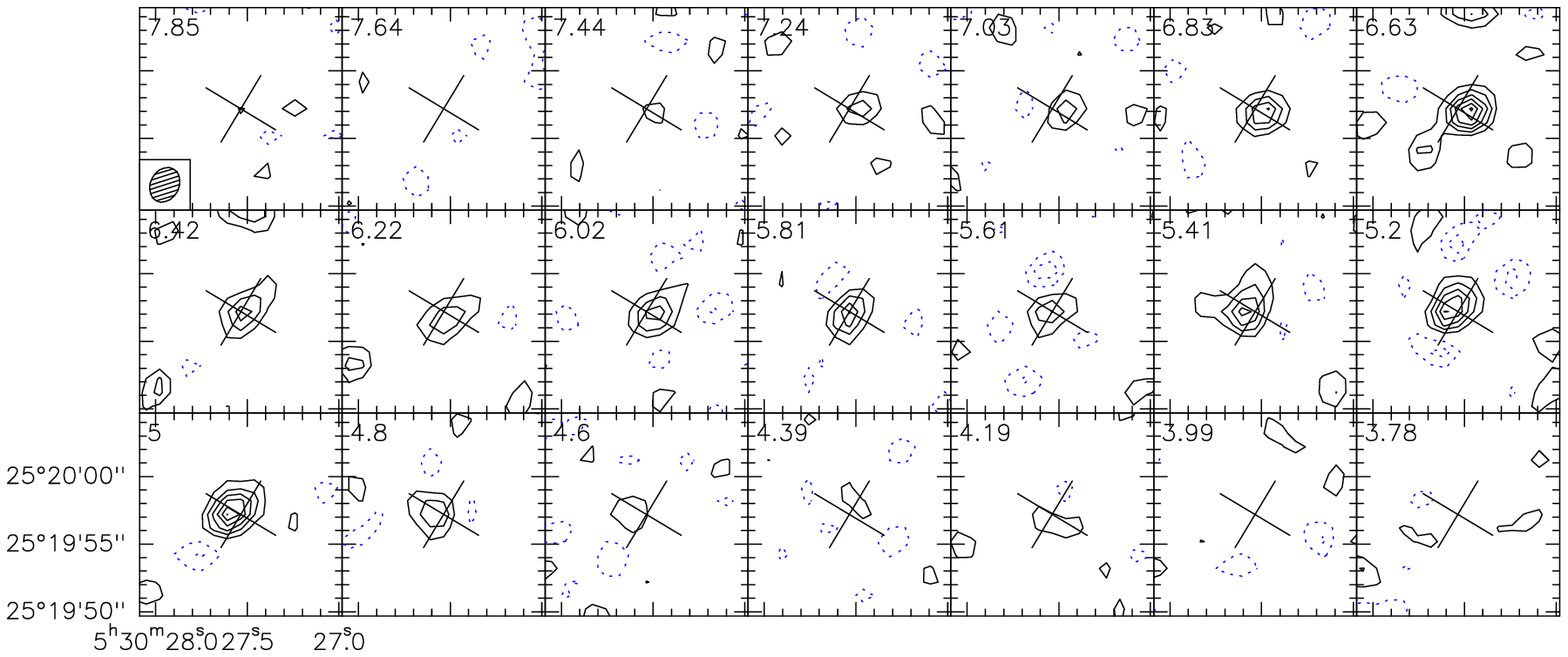}
    }
  \caption{Channel maps of the CO J=1-0 line emission towards MWC 758.
  The angular resolution is $2.7\times 2.1''$.
  Contour step is 0.06 Jy/beam or 1.0 K ($2 \sigma$) for CO. The crosses
  indicate the position, orientation and aspect ratio of the disk determined from the CO analysis. The velocity (in km.s$^{-1}$) is indicated in the upper left corner.}
\label{fig:mwc}
\end{figure*}}

\newcommand{\tableUN}{
\begin{table*}[!ht]
\caption{Properties of HAe stars with CO disks}\label{tab:stars}
\begin{center}
\begin{tabular}{llllllll}
\hline
 (1)  & (2) & (3) & (4) & (5) &( 6) &(7) & (8) \\
Source   & Right ascension      & Declination & Spec.~type & Effective temp.(K) & Stellar lum.($\lsun$) &  Distance & CO paper \\
\hline \hline
MWC\,480  & 04:58:46.264  & 29:50:36.86 & A4   &  8460 & 11.5 & 140 & 1,2\\
AB\,Aur   & 04:55:45.843  & 30:33:04.21  & A0/A1& 10000 & 52.5 & 140 & 3  \\
HD 34282 & 05:16:00.491  & -09:48:35.45 & A1/A0&  9440 & 29   & 400 & 4  \\
HD 163296 & 17:56:21.28 & -21:57:21.9 & A1 & 9300 & 30 & 122 & 5 \\
CQ Tau   & 05:35:58.485  & 24:44:54.19 & A8/F2 & 7200 & 8-16$^*$ & 140$^*$ & 6 \\
MWC 758  & 05:30:27.542  & 25:19:57.32 & A3/A8 & 8200 & 11$^*$ & 140$^*$ & 6\\
 \hline
\end{tabular}
\end{center}
{Columns 2\&3: J2000 coordinates deduced from the fit of the 1.3\,mm continuum map of the PdBI. Errors bars
on the astrometry are of order $\leq 0.07''$. Columns 4, 5, and 6, the spectral type, effective temperature
and the stellar luminosity are those given in \citet{Simon_etal2000} and \citet{Pietu_etal2003} for HD\,34282
\citet{VandenAncker_etal1998} for AB\,Aur and MWC 758, \citet{Natta_etal2001} and
\citet{Mannings_Sargent1997} for CQ Tau. Column 8, CO interferometric papers are: 1~=~\cite{Pietu_etal2007},
2~=~\citet{Simon_etal2000},  3~=~\citet{Pietu_etal2005}, 4~=~\citet{Pietu_etal2003}, 5~=~\cite{Isella_etal2007}.
and 6~=~this work. (*) For MWC\,758 and CQ\,Tau, the luminosity has been
scaled to the assumed distance of 140 pc.}
\end{table*}
}

\begin{document}
\title{Disks around CQ\,Tau and MWC\,758: dense PDR or gas dispersal?\thanks{Based on observations carried out with the IRAM Plateau de Bure Interferometer. IRAM is supported by INSU/CNRS (France), MPG (Germany) and IGN (Spain).}}
\author{Edwige Chapillon\inst{1,2,3}, St\'ephane Guilloteau \inst{1,2}, Anne Dutrey\inst{1,2} and Vincent Pi\'etu\inst{3}}
\offprints{E.Chapillon \email{chapillon@obs.u-bordeaux1.fr}}
\institute{Universit\'e Bordeaux 1; Laboratoire d'Astrophysique de Bordeaux (LAB)
\and{} CNRS/INSU - UMR5804 ; BP 89, F-33270 Floirac, France \\
  \email{chapillon@obs.u-bordeaux1.fr,guilloteau@obs.u-bordeaux1.fr,
    dutrey@obs.u-bordeaux1.fr}
  \and{}
  IRAM, 300 rue de la Piscine, 38400 Saint Martin d'H\`eres, France.\\
  \email{pietu@iram.fr}
  }
\date{Received 6-Feb-2008, Accepted 16-May-2008}
\authorrunning{Chapillon \etal}
\titlerunning{Disks around CQ\,Tau and MWC\,758}

  \abstract
{The overall properties of disks surrounding intermediate PMS stars (HAe) are not
yet well constrained by current observations. The disk inclination, which significantly affect SED modeling, is often unknown.}
{We attempted to resolve the disks around CQ Tau and MWC 758, to provide accurate constraints on the disk parameters,
in particular the temperature and surface density distribution.}
{We report arcsecond resolution observations of dust and CO line emissions with the IRAM array. We also searched for the HCO$^+$~\juz\ transition. The disk properties are derived using a standard disk model. We use the Meudon PDR code to study the chemistry.
}
{The two disks share some common properties. The mean CO abundance is low despite disk temperatures above the CO condensation temperature. Furthermore, the CO surface density and dust opacity have different radial dependence. The CQ\,Tau disk appears warmer, and perhaps less dense than that of MWC\,758.
Modeling the chemistry, we find that photodissociation of CO is a viable mechanism to explain the low abundance. The photospheric flux is not sufficient for this: a strong UV excess is required.  In CQ\,Tau, the high temperature is consistent with expectation for a PDR. The PDR model has difficulty explaining the mild temperatures obtained in MWC\,758, for which a low gas-to-dust ratio is preferred. A yet unexplored alternative could be that, despite currently high gas temperatures, CO remains trapped in grains, as the models suggest that large grains can be cold enough to prevent thermal desorption of CO.
The low inclination of the CQ Tau disk, $\sim 30^\circ$, challenges previous interpretations given for the UX Ori - like luminosity variations of this star.}
{We conclude that CO cannot be used as a simple tracer of gas-to-dust ratio, the CO abundance being affected by photodissociation, and grain growth.
}


\maketitle{}

\section{Introduction}

It is now generally admitted that the disks surrounding intermediate mass ($\simeq 2 \msun$) pre-main-sequence (PMS) HAe stars are warmer and more massive analogs of those surrounding lower mass TTauri stars. \citet{Mannings_Sargent1997,Mannings_Sargent2000} have shown that disks are indeed frequent, if not ubiquitous around HAe stars. Nevertheless, only very few objects have been modeled in detail. Accurate disk orientation, sizes, temperatures and CO abundances are available for a few objects only:  AB Aur \citep{Pietu_etal2005}, MWC 480 \citep{Simon_etal2000,Pietu_etal2007}, and HD 163296 \citep{Isella_etal2007} have been studied in CO isotopologues. Already one disk out of these three stands out as peculiar: the disk around AB Aur exhibits non-Keplerian motions, perhaps due to its youth \citep{Pietu_etal2005,Lin_etal2006}. HD 34382 was observed at high angular resolution in \dco\ by  \citet{Pietu_etal2003}, but due to its larger distance, only limited information could be obtained.

The disk around CQ Tau (HD 36910) has also been imaged at different wavelengths. So far, this is
one of the oldest HAe star ($\sim$ 10 Myrs) surrounded by a resolved dust and gas disk \citep{Mannings_Sargent1997, Testi_etal2001, Doucet_etal2006}. Moreover, CQ Tau appears as a peculiar HAe star exhibiting an UX Ori-like variability \citep{Natta_etal1997}.  MWC 758 (HD 36112) has also been barely resolved in CO by \citet{Mannings_Sargent1997,Mannings_Sargent2000}. Both disks appeared significantly weaker and smaller
in CO lines than the previously studied disks around HAe stars.

\tableUN

In order to improve our knowledge on HAe disks, we have observed CQ Tau and MWC 758 with the IRAM array.
We present here our results. The two stars were imaged in \dco\,\jdu\ and continuum emission at 1.3 and 3.4 mm (angular resolution of 1.5 and 3.4$''$, respectively). 2.5$''$ resolution images of \dco\,\juz\ in MWC 758, and upper limits on the \hco\,\juz\ line are also reported.

Section 2 presents the stellar properties and the observations. Section 3 focusses on the CO and dust modeling. We present in Section 4 the chemical model that we used to interpret the observed CO column densities.
We discuss in Section 5 all the properties of the two sources before conclusion, in Section 6.

\section{Observations}

\subsection{Star properties and the distance problem}
\label{sec:stars}

CQ Tau and MWC 758 are two Herbig Ae stars situated near the edge of the Taurus complex.
The spectral type of MWC\,758 ranges from A3e \citep{The_etal1994} to A8v \citep{Beskrovnaya1999}; that of CQ Tau from
A8 to F2 \citep[e.g.][]{creech-eakman2002}, or even F4/F5 \citep{Mora_etal2001}. MWC 758 has a rotation velocity $v $sin$i = 60$~km\,s$^{-1}$, and CQ Tau $v$ sin$i = 100$~km\,s$^{-1}$ \citep{Bohm_Catala1995,Mora_etal2001}.

Hipparcos measurements place them formally at 100 pc (CQ Tau) and 200 pc (MWC 758) respectively \citep{VandenAncker_etal1998}. However, the formal uncertainties on the parallaxes are such that the two stars could be equally placed (at the $1.5 \sigma$ level) at 140 pc. As the apparent position of the stars place them at the outskirts of the Taurus/Auriga molecular cloud region and their systemic velocities are equal to the Taurus one, we assume here a physical association, and use D=140 pc for both sources. This eases the comparison with other studies of similar objects in the Taurus region.
Furthermore, in Sec.\ref{sec:dist}, we show that the 100 pc distance for CQ\,Tau is not compatible with our new measurements.

Nevertheless, the effects of the distance uncertainties will be considered whenever appropriate in this paper. Most of the results  for the disk parameters can be simply scaled for different distances using the appropriate scaling laws given in \citet{Dutrey_etal2003}.

The stellar properties corresponding to the D=140 pc distance are given in Table \ref{tab:stars}, with those of other HAe stars having detailed disk studies. Note that CQ\,Tau being highly variable \citep[up to 3 magnitudes, see][]{Grinin_etal2008}, the values of its luminosity quoted in the literature vary by almost an order of magnitude.

\subsection{PdBI data}

The \dco\,\jdu\ data for CQ Tau and MWC 758 were obtained in snapshot mode in Nov 1999 and Jan 2001, together with AB Aur \citep{Pietu_etal2005}. The $^{12}$CO J=2-1 data were smoothed to 0.25 $\kms$
spectral resolution for the final analysis.  Baselines up to 170 m provided about 1.5$''$ resolution with robust weighting. The \hco\,\juz\ data were obtained simultaneously: a 20 MHz/512 channels correlator unit provided a spectral resolution of 0.13 $\kms$.

The phase and secondary flux calibrators were 0415+379 and 0528+134. The primary flux calibrator was MWC 349.
The rms phase noise was 8$\degr$ to 25$\degr$ and 15$\degr$ to 50$\degr$ at 3.4 mm and 1.3 mm, respectively, which introduced position errors of $\leq 0.07\arcsec$. The estimated seeing is about 0.2$''$ after calibration.  Since the observations were obtained with those of AB Aur, these data share a common flux scale. Thus the
temperature (and column densities) can be directly compared with those cited by \citet{Pietu_etal2007} for AB Aur,
MWC 480 and DM Tau, as well as the derived values for the dust emissivity spectral index $\beta$.

For MWC\,758, the \dco\,\juz\ line was observed in Dec 1995 and Mar 1996. The longest baseline was 180 m, and provided
an angular resolution of $2.5\times 1.9''$ at PA $149^\circ$.

\figMAPS

Figure \ref{fig:maps} presents the channel maps of \dco\,\jdu\ for CQ Tau and MWC 758. Figure \ref{fig:mwc}
shows the \dco\,\juz\ channel maps of MWC\,758. The characteristic pattern of rotation is clearly apparent. No \hco\ was detected.

\figMWC

We used the GILDAS\footnote{See \texttt{http://www.iram.fr/IRAMFR/GILDAS} for more information about the
GILDAS software.}  software package \citep{pety05} to reduce the data and as a framework to implement our
minimization technique. Table \ref{tab:cont} indicates the result of Gaussian fit to the continuum visibilities. Elliptical Gaussian were used at 230 GHz; at 89 and 113 GHz, the resolution is insufficient to constrain the disk aspect ratio and only a circular Gaussian was fitted.

\begin{table}[ht]
\caption{Continuum results}
\label{tab:cont}
\begin{tabular}{lllll}
Frequency   & Flux & Major & Minor & PA \\
(GHz)  & (mJy) & $''$ & $''$ & $^\circ$ \\
\hline
   \multicolumn{5}{c}{MWC 758} \\
230 &  $56\pm 1$ & $1.00\pm 0.08 $ & $0.81 \pm 0.10$ & $-10\pm 20$\\
113 &  $6.7\pm1.3$ & $1.1\pm0.5$ & -- & -- \\
 89 &  $3.3 \pm 0.4$ &  -- & -- & -- \\
\hline
   \multicolumn{5}{c}{CQ Tau} \\
230 &  $162\pm 2$ & $0.86\pm 0.04 $ & $0.63 \pm 0.04$ & 32 $\pm 7$  \\
 89 &  $13.1 \pm 0.5$ & $0.7 \pm 0.3$  & -- & -- \\
\hline
\end{tabular}\\
The error bars include only thermal noise, but not the flux scale accuracy ($\simeq 10$ \%)
or the effects of seeing. Here, the P.A. is that of the ellipse major axis.
At 89 and 113 GHz, a circular Gaussian was fitted.
\end{table}

\section{CO and Dust Models}

The data were fitted using our standard flaring disk model, with power law distributions for all primary quantities (surface density $\Sigma(r) = \Sigma_0(r/r_0)^{-p}$, temperature  $T(r) = T_0(r/r_0)^{-q}$, velocity and scale height), following the method described in detail by \citet{Pietu_etal2007}. This allows the CO surface density and dust radial emissivity profile to be recovered completely independently, without assumptions on the CO abundances. An essential property of disks is that, because the CO surface density gradient is significantly steeper than the temperature gradient ($p-q > 1$), it is possible, with sufficient angular resolution, to distinguish (from the apparent slope of the surface brightness versus radial distance) the transition between an optically thick core \citep[$T_b \propto r^{-q}$) and an optically thin outer part ($T_b \propto r^{-p} \mathrm{~or~} r^{-p-q}$ for low J CO lines, see][]{Dutrey-PPV2007}. Thus, the fitting procedure can actually reveal whether the CO emission is mostly optically thick or optically thin, even with a single transition.

For the dust emission, we use the standard prescription for the mass absorption coefficient \citep{Beckwith_etal1990}:
\begin{equation}
\kappa_\nu(\nu) = 0.1 (\nu/10^{12}\mathrm{Hz})^\beta \mathrm{cm}^2\mathrm{g}^{-1}
\label{eq:knu}
\end{equation}

For the continuum data, as the angular resolution does not allow to separate the optically thick core, all frequencies observed with the IRAM interferometer were fitted simultaneously, using the temperature (fixed) derived from the CO measurements. 

The results are summarized in Table \ref{tab:all}. The errors bars are derived from the covariance matrix by the minimization routine. We refer to \citet{Pietu_etal2007} for notations (see also Table \ref{tab:all}) and detailed description.
These new observations directly prove the Keplerian nature of the rotation, with a best fit exponent $0.51 \pm 0.02$ for both sources.

\tabALL

\subsection{MWC 758}

The disk outer radius is $\simeq 250$ AU in CO and somewhat smaller in dust. The line and continuum emissions provide very consistent results for the disk orientation.
The \dco\,\jdu\ line appears partially optically thick, thereby allowing the derivation of the disk temperature in the inner regions ($r < 150$ AU), and of the CO column density in the outer regions ($r > 150$ AU) under the assumption of a single temperature power law.
The parameters derived from both transitions are in good agreement.
The exponent $p$ is not well constrained for the \dco\,\juz\ line; we thus fix it to a value compatible with the \dco\,\jdu\ results.  We also present results which both lines fitted together (col 6, Table \ref{tab:all}). The better sensitivity allows to derive more parameters such as stellar mass, temperature and  exponent $q$. 

The difference between the temperatures determined from the \juz\ and \jdu\ lines is in line with the vertical temperature gradients found in some other stars \citep{Pietu_etal2007}, but not determined with sufficient precision to conclude. We thus prefer to use the temperature derived from a simultaneous fit of both transitions. This temperature is moderate, 30 K, with little dependance upon radius (exponent $q \simeq 0.3$). Because the \jdu\ transition is barely optically thick, the temperature is representative of the bulk of the disk, and not only of the surface layers as in the optically thick CO disks studied by \citet{Pietu_etal2007}.

Because the emission is spatially resolved at 1.3\,mm, the dust emissivity index is well constrained, with little
bias due to a possible contribution from an optically thick core: $\beta = 1.0 \pm 0.15$. If we assume that the dust and CO are at the same temperature, the total disk mass is $2.7 \,10^{-3} \msun$ (with $\kappa(230) = 0.023$ cm$^2$g$^{-1}$).

Assuming the same distribution and excitation conditions than for CO, the \hco\,\juz\ line provides an upper limit of
$4\,10^{12}$ cm$^{-2}$  for the column density of \hco\ at 100 AU, i.e.  [\dco]/[\hco]$ > 10^4$

The inclination is poorly constrained ($i \simeq 16 ^\circ$), but, except for the rotation velocity at 100 AU (as the data essentially constrain $v \sin{i}$), all other parameters remain unaffected by this uncertainty. Besides the solution presented here, another solution with $i \simeq 27^\circ$ can be found. However, only the low inclination solution is compatible with the stellar mass derived from the location of the star on a distance independent HR diagram, $log(L/M^2) \simeq 0.5$ and $log(T_\mathrm{eff})= 3.92$. For $i=27^\circ$, the stellar mass is $0.7\pm0.3 \msun$. Reconciling this with the spectral type of the star would imply placing it at a distance of $> 280$  pc. The disk outer radius would be $> 550$ AU, the disk mass $0.08 \msun$, and the temperature at 100 AU
40 K. MWC 758 would have among the most massive disks around HAe stars \citep[with HD 34282][]{Pietu_etal2003}. It would however give $log(L/M^2) \simeq 1.3$, which from the \citet{Siess_etal2000} tracks, require $M_* > 5\msun$.

\subsection{CQ Tau}

For CQ Tau, we obtain for the first time accurate measurements of the inclination $i=29\pm 2^\circ$ of the disk and of the orientation of the disk axis
($PA = -36\pm1^\circ$).
The disk inclination is well constrained from both \dco\ ($29 \pm 2^\circ$) and continuum emission ($i=29\pm 9^\circ$), although with larger
error bars for the continuum. Note that both determinations are affected by different instrumental effects. The dust emission is biased by the phase noise (seeing), while the CO-derived inclination is affected by bandpass calibration errors, which are negligible in our case. The overall agreement indicates that the seing effects remain small compared to the thermal noise.
However, our results on the disk  differ from several previous determinations: for example, \citet{Testi_etal2003} cite a major axis disk orientation $PA \sim 20^\circ$ (or $-70^\circ$ for the disk axis) and inclination ($i\simeq 66^\circ$) very different from ours. Most of the previous measurements were based on deconvolved sizes in the image plane, either on integrated intensity maps of CO \citep{Mannings_Sargent1997} or from relatively low S/N continuum data \citep[e.g. the 7 mm VLA data of][]{Testi_etal2001}. Our results being based on the velocity gradient and a $UV$ plane analysis are much more reliable: in particular, they predict a stellar mass in agreement with the spectral type of CQ\,Tau. It is worth noting that, using the mid-IR camera CAMIRAS installed on the CFHT, \citet{Doucet_etal2006} have imaged the thermal emission of the dust disk at 20.5 $\mu$m and found an inclination angle of $i=33\pm 5^\circ$ in  agreement with our measurements.

The disk of CQ Tau is somewhat smaller, 200 AU in radius, than that of MWC 758. Both continuum and CO measurements give the same outer radius. \citet{Doucet_etal2006} estimated the disk radius to be of order of $\sim 400$~AU at 20.5 $\mu$m (when corrected for the distance of 140 pc). The 20 $\mu$m emission comes from small, warm, grains at the surface layer of a flared disk. On the opposite, the mm data trace large grains in the disk mid-plane. A difference in radial extent between these two populations of grains is a standard feature of protoplanetary disks, as exemplified for example by HH\,30 \citep{Guilloteau_etal2008}.

The \dco\,\jdu\ emission appears essentially optically thin. Since the contribution of the optically thick core is small $\tau(CO) \leq 1$ for R $\geq$ 60\,AU, the disk temperature exponent $q$ is poorly constrained. Best fit values are around 1, but with large uncertainty. We thereby adopted $q = 0.5$. This value is a compromise between the flat temperature distribution ($q \simeq 0$) obtained in the disk mid-planes, as for MWC 758, or DM Tau and MWC 480 \citep{Pietu_etal2007}, and  the steeper laws occurring at the disk surfaces, $q \simeq 0.65$ \citep{Guilloteau_Dutrey1998}.

Even with this reduced degree of freedom, the temperature remains difficult to constrain. A lower limit of about 60 K
can be obtained (in essence, because of the apparent \dco\,\jdu\ surface brightness, taking into account that the Keplerian shear results in significant beam dilution),  but higher values cannot be excluded. We adopted $T_{100}= 150$ K; this choice has no effect on the disk parameters, except for the derived CO and dust surface densities and thus disk mass and CO abundance. The brightness temperature being fixed, for a constant line-width and in the high temperature approximation, the CO surface density is proportional to the assumed temperature (because $T_b(\mathrm{CO}) \propto  \Sigma(\mathrm{CO}) / T_k$) and the dust surface density is proportional to the inverse of the temperature (as $T_b(\mathrm{dust}) \propto \kappa(\nu) \Sigma(\mathrm{dust}) T(\mathrm{dust})$, see \citet{Dartois_etal2003} and their Fig.4 for details).

If we assume that the dust and CO are at the same temperature, and the prescription for the dust emissivity in Eq.\ref{eq:knu}, which gives $\kappa(230) = 0.035$ cm$^2$g$^{-1}$, the total disk mass is small, $1.2 \,10^{-3} \msun$. Note however that this is probably a lower limit, because of the existence of vertical temperature gradients, as directly demonstrated by \cite{Pietu_etal2006}: if the dust is at 50 K, the disk mass increases to $3.6\,10^{-3} \msun$.We stress that our choice of temperature \emph{maximizes} the apparent CO abundance, as this scales as $T^2$.

Our result on $\beta= 0.70 \pm 0.04$ agrees very well with the $\beta= 0.6 \pm 0.1$ determined by \citet{Testi_etal2003} using a broader frequency range including VLA observations at 7 mm.

Assuming the same distribution and excitation conditions than for CO, the \hco\,\juz\ line provides an upper limit of
$4\,10^{12}$ cm$^{-2}$ for the column density of \hco\ at 100 AU, i.e.  [\dco]/[\hco]$ > 4\,10^3$.

\subsection{Prediction for the CO J=3-2 Line}
\label{sec:co32}

The \dco\,\jtd in CQ Tau and MWC\,758 was detected by \citet{Dent_etal2005} using the JCMT. Our best fit model can be
used to predict the intensity of this line. This is given in Table \ref{tab:jcmt}, also for MWC\,480 using the model parameters from \citet{Pietu_etal2007}. The agreement is excellent for MWC\,758 and MWC\,480, but our prediction is
a factor 2 too high for CQ Tau. The discrepancy can be somewhat reduced (by about 30 \%) by using a temperature of 70 K for CQ Tau.  The remaining difference may be due to noise, as the CQ Tau CO J=3-2 line is detected with a S/N of only 6 by \citet{Dent_etal2005}. The spectral baseline uncertainty on the JCMT data can also play a significant role at this sensitivity level.
Nevertheless, this result suggests that the temperature in CQ Tau may not be as high as we estimated.
It is also possible that the \dco\,\jtd\ line is somewhat sub-thermally excited, as its higher opacity makes it more sensitive to the upper layers than the \jdu\ transition.

\begin{table}
\caption{\dco~\jtd\ comparison}\label{tab:jcmt}
\begin{tabular}{ccc}
\hline
Source  & \multicolumn{2}{c}{\dco \jtd\ integrated line flux (Jy\,km\,s$^{-1}$)} \\
name  & observed & predicted  \\
CQ Tau  &  6 $\pm$1 &  14(\#) -- 11(*) \\
MWC758 & 15 $\pm$ 2    & 15 \\
MWC480 & 52 $\pm$ 1.1 & 47 \\
\hline
\end{tabular}\\
\dco~\jtd\ observation from \citet{Dent_etal2005}. The conversion factor from the main beam brightness temperature T$_\mathrm{mb}$(K) to flux density is 19 Jy/K for the 22$''$ beam at 345 GHz. (\#)~for~T=150~K,(*)~for~T=70~K.
\end{table}

\subsection{Distance effects}
\label{sec:dist}

As mentioned in Sec.\ref{sec:stars}, CQ Tau could be at 100 pc and MWC 758 at 200 pc, instead of our assumed D=140 pc. The surface density scales as D$^p$, the temperature as D$^q$, the disk mass as D$^2$,
and the stellar mass, when derived from the Keplerian rotation, as D \citep[see][]{Dutrey_etal2003}.  With the values of $p$ and $q$ from Table \ref{tab:all}, the impact of the distance ambiguity remains limited to a factor 2 on the disk masses. However, the difference in disk mass between CQ Tau and MWC 758 becomes affected by a factor 4.

The CO abundance (at 100 AU) will scale as D$^{p(\mathrm{CO})-p(\mathrm{dust})}$. It is thus little affected by the distance uncertainty (less than 50\%).

\begin{figure}
\includegraphics[angle=270,width=8.5cm]{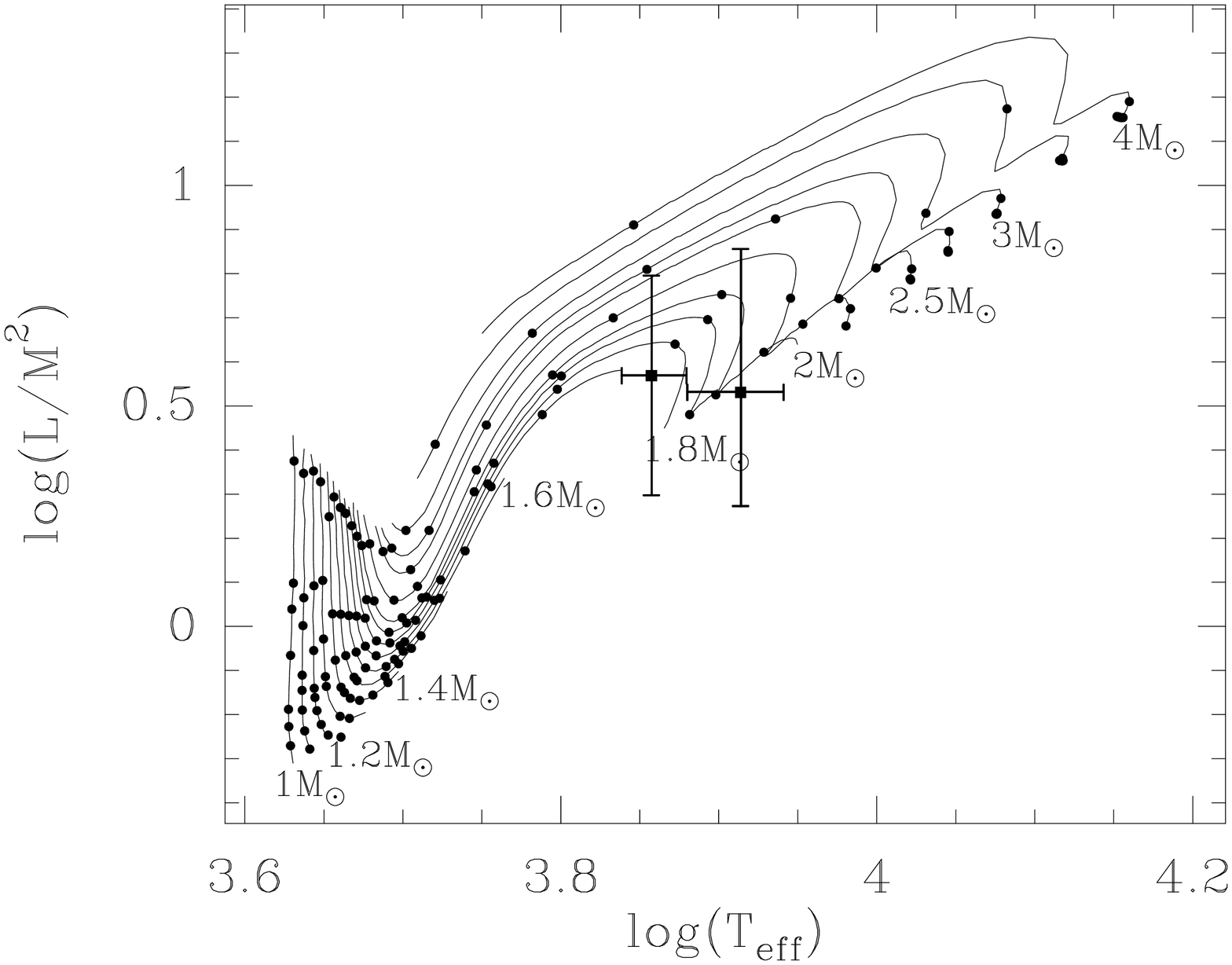}
\caption{Distance-independent HR diagram (L/M$^2$ versus $T_\mathrm{eff}$). Theoretical evolutionary tracks from  \citet{Siess_etal2000} and $Z = 0.02$ metallicity for stellar mass between 1 and $4 \msun$ (from left to right: 1 to 2 by $0.1 \msun$, and then 2.2, 2.5, 2.7, 3.0, 3.5, and $4.0 \msun$). Knots on the curve begin at and are spaced by 1 Myr. CQ Tau is the leftmost cross, MWC 758 is the rightmost one. }
\label{fig:lm2}
\end{figure}

Figure \ref{fig:lm2} shows the location of CQ\,Tau and MWC\,758 on a distance independent HR diagram, as used by \citet{Simon_etal2000}, using the evolutionary tracks of \cite{Siess_etal2000}. The location of CQ\,Tau in this diagram is a further argument for the distance of 140 pc for this object: at 100 pc, the kinematic data would require a mass of $1.3 \msun$, which is clearly incompatible with the effective temperature and L/M$^2$, which indicate a
$\sim 1.8 \msun$ star. We point out that, with these new results, CQ\,Tau could actually be younger than initially thought (5 Myr), and even younger than MWC\,758.

\subsection{Comparison between sources}

Both sources have many characteristics in common:
\begin{enumerate}\itemsep 0pt

 \item Partially optically thin CO line emission, in contrast to most previously studied disks.

 \item Different surface density profiles for CO and dust, with the CO surface density falling faster ($p \simeq 2.5 - 3$) than the dust opacity ($p\simeq 1.5$). This property is apparently shared with all other observed sources so far, with the exception of HD\,163296 which displays a relatively shallow CO surface density profile ($p \simeq 1$, according to \citet{Isella_etal2007}).

 \item Similar values for the dust emissivity exponent $\beta = 0.7 - 1 $. In both sources, we have resolved the continuum emission, so this value is free of the bias due to a possible optically thick core. For CQ Tau, \citet{Testi_etal2003} also found that the free-free contribution is negligible. These values are smaller than those found for MWC\,480 and AB\,Aur, $\simeq 1.4$ \citep{Pietu_etal2006,Pietu_etal2007}.

 \item Similar lower limit for the $[\dco]/[\hco]$ ratio, $\simeq 4000$, corresponding to $[\tco]/[\hco] > 100$
 when taking into account a reasonable (average) CO isotopologue ratio. This limit is consistent with what has been measured in other sources by \citet{Pietu_etal2007}.

\item An essential result, which remains whatever the distance, is that the apparent CO abundance is very low in both sources, less than 10$^{-6}$, corresponding to a CO depletion $> 100$ compared to typical molecular clouds \citep{Ohishi_etal1992}.

 \item Both sources have also relatively small disk masses, less than or about $5\,10^{-3} \msun$. Because of the distance problem, and  dust temperature uncertainty for CQ\,Tau, disk masses remain the least accurate parameters. In the most extreme case, the MWC\,758 disk could be up to 8 times more massive than the CQ\,Tau one.
\end{enumerate}

Although the two sources are very similar, CQ Tau stands out among the sources studied so far because it accumulates a number of peculiarities. It is at least as warm as AB\,Aur, the warmest source detected in CO J=2-1 so far, and only the second source in which \dco\,\jdu\ is optically thin, after BP Tau \citep{Dutrey_etal2003}. Like BP Tau, the CO and dust outer radii are identical within the current error bars.  This is an important difference with most other sources, either TTauri stars or HAe stars, for which the outer radius of the bulk of the dust emission is in general smaller than the CO outer radius (see the T Tauri stars DM\,Tau, LkCa\,15 and the HAe stars AB\,Aur, MWC\,480 \citep{Pietu_etal2006,Pietu_etal2007}, and HD\,163296 \citep{Isella_etal2007}).

\subsection{CO abundance: an anomalous gas to dust ratio ?}

For both sources, one of the surprising findings is the very low apparent CO abundance, $\simeq 10^{-6}$ (within a factor 3, as this value decreases with radius), if a standard gas to dust ratio of 100 is assumed, i.e. an apparent depletion of $\simeq 100$. Note that this result is completely distance independent.
Since the temperature of these two disks is large, depletion of CO due to sticking on grains cannot be invoked in a simple way. The situation here is similar to that found for the T Tauri star BP Tau by \citet{Dutrey_etal2003}.

This suggests  either \textit{i)} that the gas (H and H$_2$) to dust ratio is actually much lower than our assumed value of 100, because we are observing disks in the process of dissipating their gaseous content, or that the dust distribution has evolved (grain growth and settlement along the mid plane) either \textit{ii)} by simple coagulation and size redistribution, and/or \textit{iii)} by mantle accretion, taking off elements from the gas phase. In all cases, the CO/dust ratio is no longer a direct reflection of the gas to dust ratio, because of changes in the H$_2$ shielding of the CO photo-dissociation lines.

\citet{Jonkheid_etal2007} have studied the chemistry of HAe disks, and find that for very low masses of
{\em small grains}, $\simeq 10^{-6} \msun$, the CO/H$_2$ ratio can be of order of a few $10^{-6}$ (see their models B4/BL4). Thus, our low observed CO column densities might be the result of {\em grain growth} and/or {\em efficient dust settling}. However, only in the lowest mass disks are the predicted column densities as low as we observe, a few $10^{16}$ (their Fig.8). As \citet{Jonkheid_etal2007} do not present calculations with grain growth for disk masses in the range of what we observe, we have performed new chemical models which are described in the next section.

\section{The Chemical Models}

We use the PDR code from the Meudon Group \citep{LeBourlot_etal1993, LePetit_etal2006}, with the modifications from  \citet{HilyBlant_etal2007} for the grain size distribution.
The model is a one-dimensional stationary plane-parallel slab of gas and dust illuminated by an ultraviolet (UV) radiation field. The radiative transfer in the UV (which takes into account the self-shielding of H, H$_2$ and CO lines and absorption in the continuum by dust grains), the molecular abundances and optionally the thermal balance are calculated iteratively at each point in the ``cloud''. The chemical network is similar to that of \citet{Goicoechea_etal2006}. No freeze-out onto grains is considered: this assumption will be discussed in Sec.\ref{sec:largegrains}.

In order to study the chemical and physical effects of grain growth on disks, in particular the UV penetration, the PDR code was modified by \citet{HilyBlant_etal2007} to introduce a standard power law grain size distribution $n(a) \propto a^{-\gamma}$ with \app~and \amm~ being the maximum and minimum cutoff radii, respectively. This modification affects the UV extinction curve, the chemistry (H$_2$ formation) and the thermal balance and is self-consistently introduced. The resulting extinction curve is calculated using the Mie theory for homogeneous isotropic spherical particles and presented in Appendix \ref{a:extinction}.

To have a two-dimensional molecular distribution we compute the model at different radii. The output of this 1+1D model is the vertical distribution of molecular abundance calculated at different radii.

\subsection{Parameters of the Chemical Model}

As an input we can impose physical conditions relevant from the disk structure (i.e. temperature and vertical density laws as derived from Table \ref{tab:all}) at each radius. Since the disk masses are the most poorly constrained parameters, the models are not intended to fit any particular object, but are to be used to understand the major effects of the model parameters on the CO abundance, and more specifically, to seek what conditions and physical processes prevent CO to reach high surface densities. Two different disk density profiles were used. The ``low mass'' case correspond to a dust mass (in the inner 200 AU) of $11\,10^{-6} \msun$, the ``high mass'' case of $33\,10^{-6} \msun$, both with $p=1.3$ and a scale height of 22 AU at 100 AU.

We investigate several gas to dust ratio ($g/d$) under various UV field conditions, and different maximum grain size \app. A wide range of parameters has been explored (see Table~\ref{tab:chem}). The whole ensemble of results
is presented in Appendix \ref{a:extinction}, \ref{a:uv} and \ref{turbulence}.

\textbf{}

\begin{table}.
\caption{Modeling parameter values}\label{tab:chem}
\begin{tabular}{ll} 
\hline
Parameters & values in calculations \\
\hline
 $g/d$  & 1, 10, and 100 \\
 \app & 1\,mm, 10, 1, and 0.1 $\mu$m \\
 $\chi$ (at 100 AU) & $10^2$,$10^3$, $5\,10^3$, $10^4$, plus a modified shape\\
  \vt (km.s$^{-1}$)  & 0.5, 1.0, and 2.0 \\
\hline
\end{tabular}\\
$g/d$ is the Gas to Dust ratio, \app\ the maximum grain radius, $\chi$  the UV scale factor over the Draine interstellar field, and \vt\ the local line width for H$_2$. See Section 4.1 for details.
\end{table}

{\bf UV field}
To estimate the UV field, we assumed UV is scattered by the dust: half towards the mid-plane and half towards the exterior. We have checked that the opacity toward the star is strong enough to justify this hypothesis, as the amount of small grains remains sufficient even when increasing \app. The shape of the UV field is assumed to follow
a Draine field (as provided by \citet{Sternberg_etal1995}). With the above scattering assumption, it is described by a scale factor $\chi$ (\footnote{ $\chi$ is the ratio of the local UV field over the Draine interstellar UV field}) at 100 AU, and decreases as $1/R^2$. As the actual UV fields from the stars are unknown, several simulations have been done with different $\chi$ factors ($10^3$ and $10^4$). As expected, a decrease of the incident UV field implies a decrease of the photodissociation rates, and makes CO more abundant (Figure \ref{fig:uv} and \ref{fig:1e-1_eqth_coldens_rad} in Appendix B).

{\bf Thermal Balance}
Models can be calculated with and without thermal balance.
The dust temperature is derived following \citet{Burton_etal1990}. With thermal balance, 
this dust temperature is used to compute the energy transfer between the gas and the grains as described in \citet{Burke_Hollenbach1983}. The gas temperature is then calculated, taking into account heating and cooling processes (see \citet{LePetit_etal2006}). Without thermal balance, the gas temperature is taken from the CO observations.
The thermal balance has little effects on the CO abundances, except at very low densities. Provided that the assumed gas and dust temperatures are close enough to those derived in the disk mid-plane from the thermal balance, the results on the disk structure and CO column densities are rather similar. This is clearly seen by comparing Fig.\ref{fig:eqth_coldens_rad} and Fig.\ref{fig:grain_coldens_rad}.

{\bf Grain Distribution}
The composition of the dust is 50\% of silicates and 50\% of graphite with $\gamma=-3.5$ and $a_-=$3nm in all calculations. The minimum grain radius $a_-$ is small enough to properly represent the photoelectric process, the UV extinction curve and the formation of H$_2$ on grain surface. Keeping the dust mass constant, we have simulated grain growth by varying the maximum radius \app. The amount of small grains is therefore reduced to the benefit of large ones and the extinction curve is clearly modified. With $\gamma=-3.5$, the UV opacity scales as $1/\sqrt{a_+}$ for $a_+ > 10 \mu$m.

We present the various models in Appendix A, with \app varying from 0.1 $\mu$m to 1 mm.
The main result is a significant decrease of the opacity towards the mid-plane and therefore an extension of the photo-dissociation layer when \app increases. As a consequence, the transition between the region where Carbon is predominantly atomic or ionized, and  the region where CO dominates appear closer to the disk mid-plane (figure \ref{fig:grain}) and the CO column density is reduced (\ref{fig:grain_coldens_rad}).

\begin{figure}[ht]
  \centering
  \includegraphics[angle=270.0,width=8.5cm]{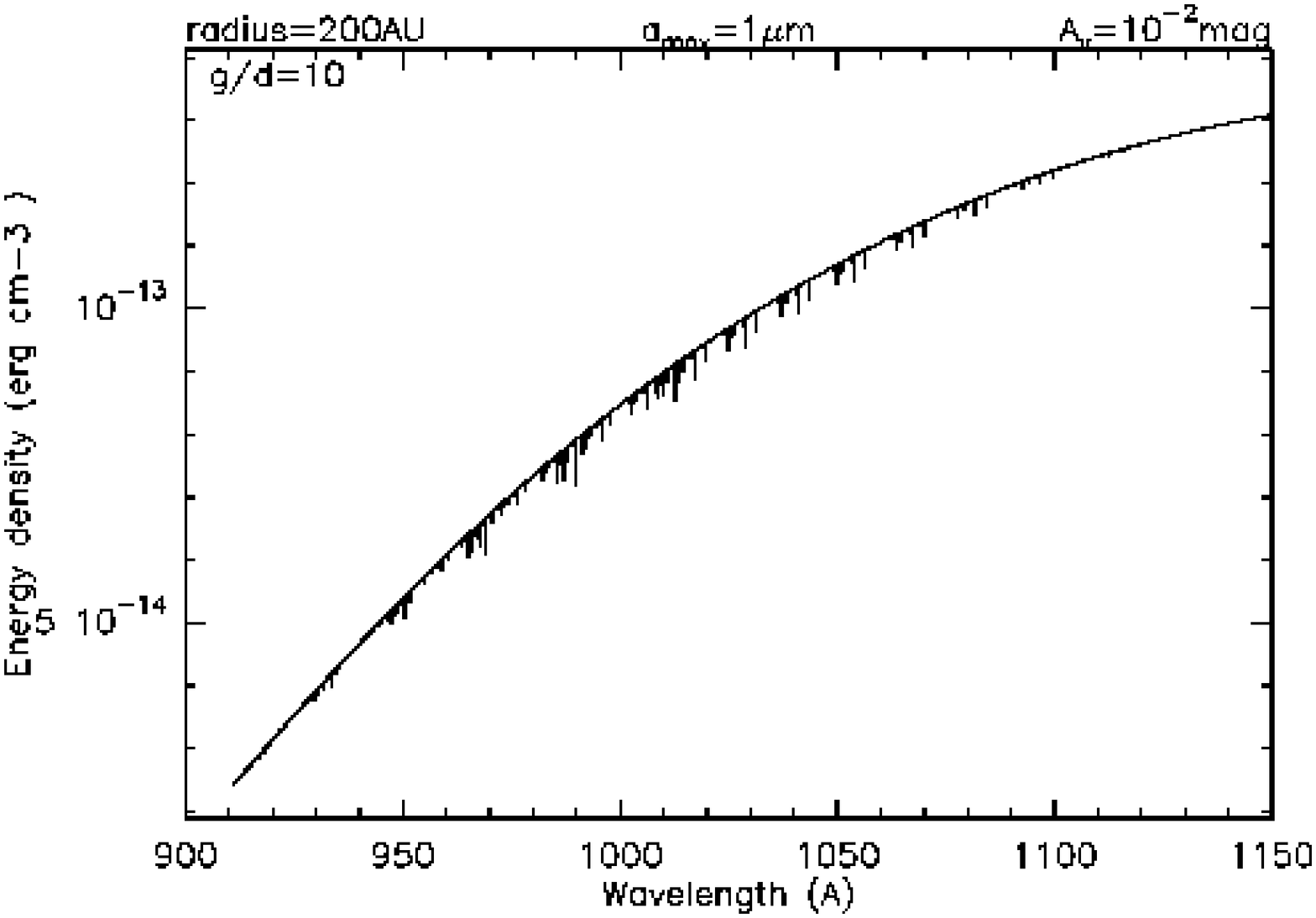}
  \includegraphics[angle=270.0,width=8.5cm]{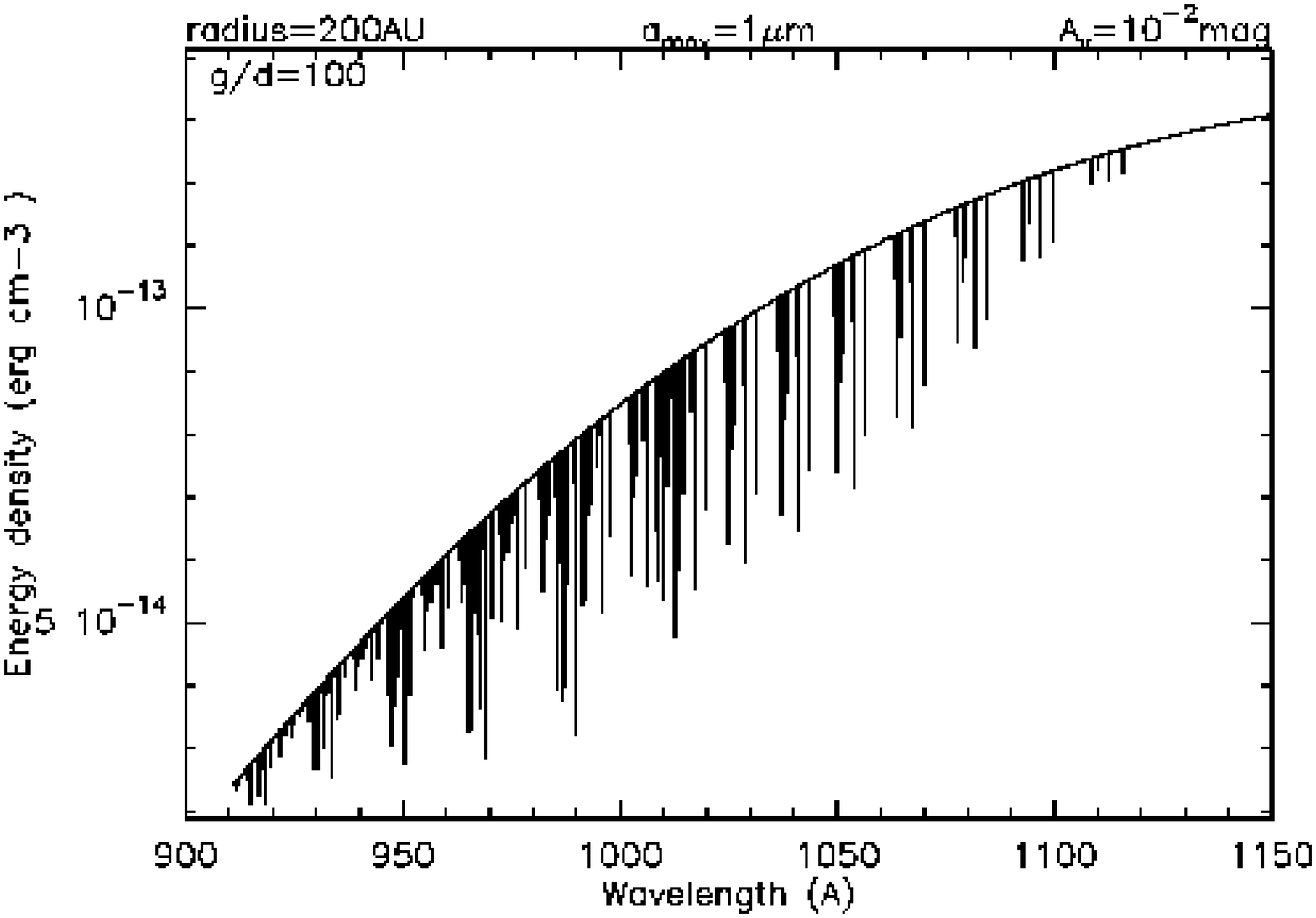}
  \caption {Energy density between 900 and 1200 \AA\ for gas/dust = 10 (top) and 100 at R=200 AU and A$_\mathrm{v} = 0.01$ with $a_+ = 1 \mu$m, $\chi=10^4$, $\Delta$V=1 km\,s$^{-1}$}
  \label{fig:rfield}
\end{figure}

{\bf $g/d$ ratio} We modify the ratio by \emph{decreasing the gas mass} (H) while \emph{keeping the dust mass} constant to simulate the dissipation of the gas. We performed calculations for $g/d=100$ and $g/d=10$ with several grain size distributions. The main effect of decreasing the gas to dust ratio is a modification of the spectral distribution of the radiative energy density in the disk (see
figure \ref{fig:rfield}). The gas being less abundant, the absorption in the lines of CO and H$_2$ is less efficient and the transition H - H$_2$ occurs closer to the mid-plane, and the disk is more ionized (e.g. for $a_+ = 1 \mu$m,  C$^+$ is the dominant form of carbon in the mid-plane for radius greater than 200 AU).

{\bf Turbulence}
 We also check the effect of the intrinsic line width (turbulence) on the photo-dissociation line shielding. We have made three runs with three different values of the Doppler width of the H$_2$ lines (0.5 - 1.0 - 2.0 km\,s$^{-1}$) for  $a_+ = 1 \mu$m and 1~mm with $\chi = 10^4$. The stronger effects (obtained for $a_+ = 1$ mm) are presented in Appendix \ref{turbulence}, but the effect on CO is negligible, essentially because the CO pre-dissociation intrinsic line widths are much larger.

\subsection{Chemical Models and comparison with the CQ Tau Disk}

We only discuss in this section the results relevant for the
CQ Tau case. The MWC\,758 case
is treated in the discussion by comparison to this example. More models are given in Appendix
\ref{a:extinction}, \ref{a:uv} and \ref{turbulence}. We assume the physical and chemical conditions
derived from the CO analysis (see Table~\ref{tab:all}). Our goal is to reproduce the low CO column density which is
observed, and corresponds (for a normal gas-to-dust ratio) to a CO depletion factor of $\sim 100$ or an abundance CO/H$_2 \simeq 10^{-6}$.

\begin{figure*}[!ht]
  \centering
  \hspace{0.4cm}\textbf{\Large{(a)}}\hspace{-0.5cm}
  \includegraphics[angle=270.0,width=15.3cm]{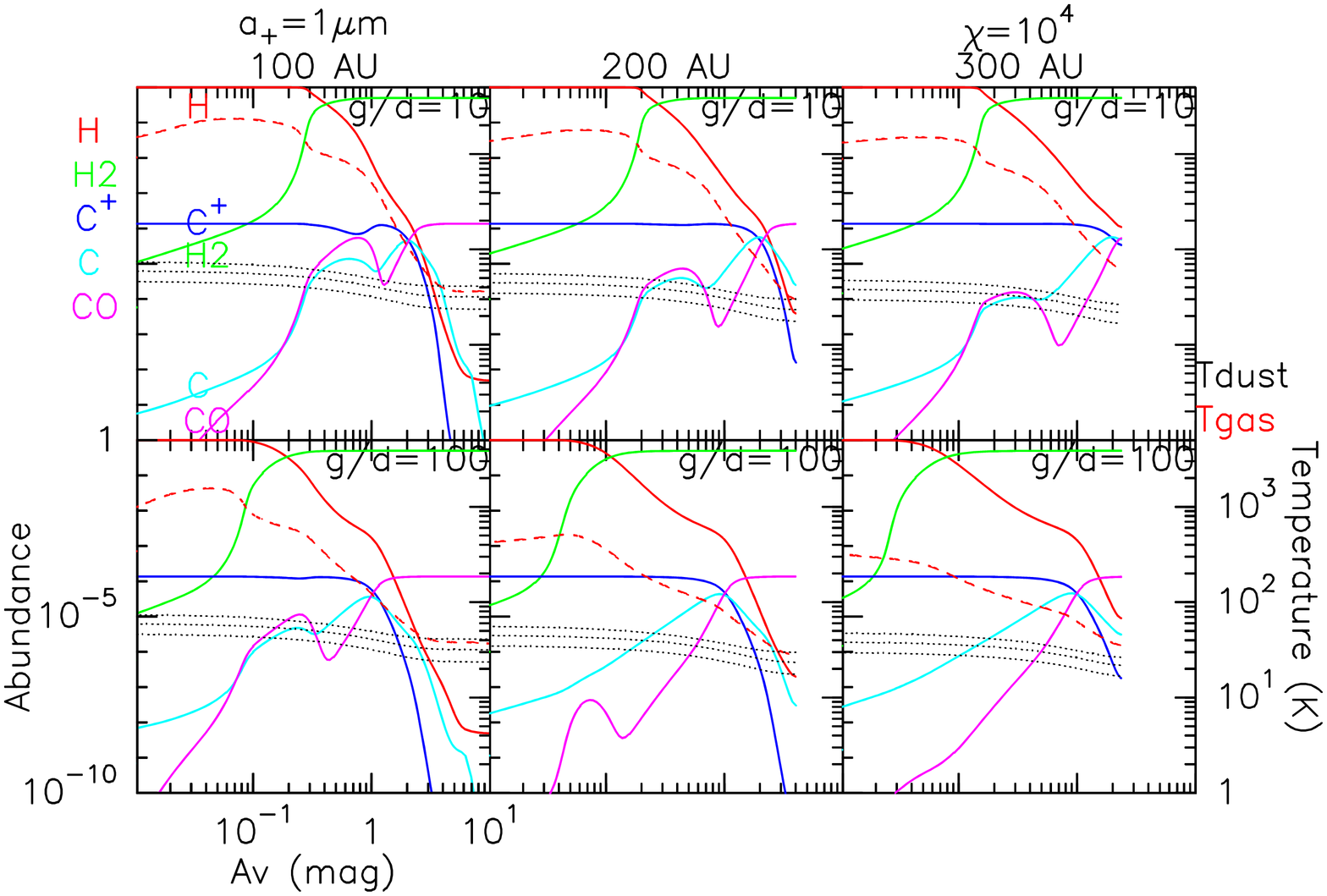}
\\
  \hspace{0.4cm}\textbf{\Large{(b)}}\hspace{-0.5cm}
  \includegraphics[angle=270.0,width=15.3cm]{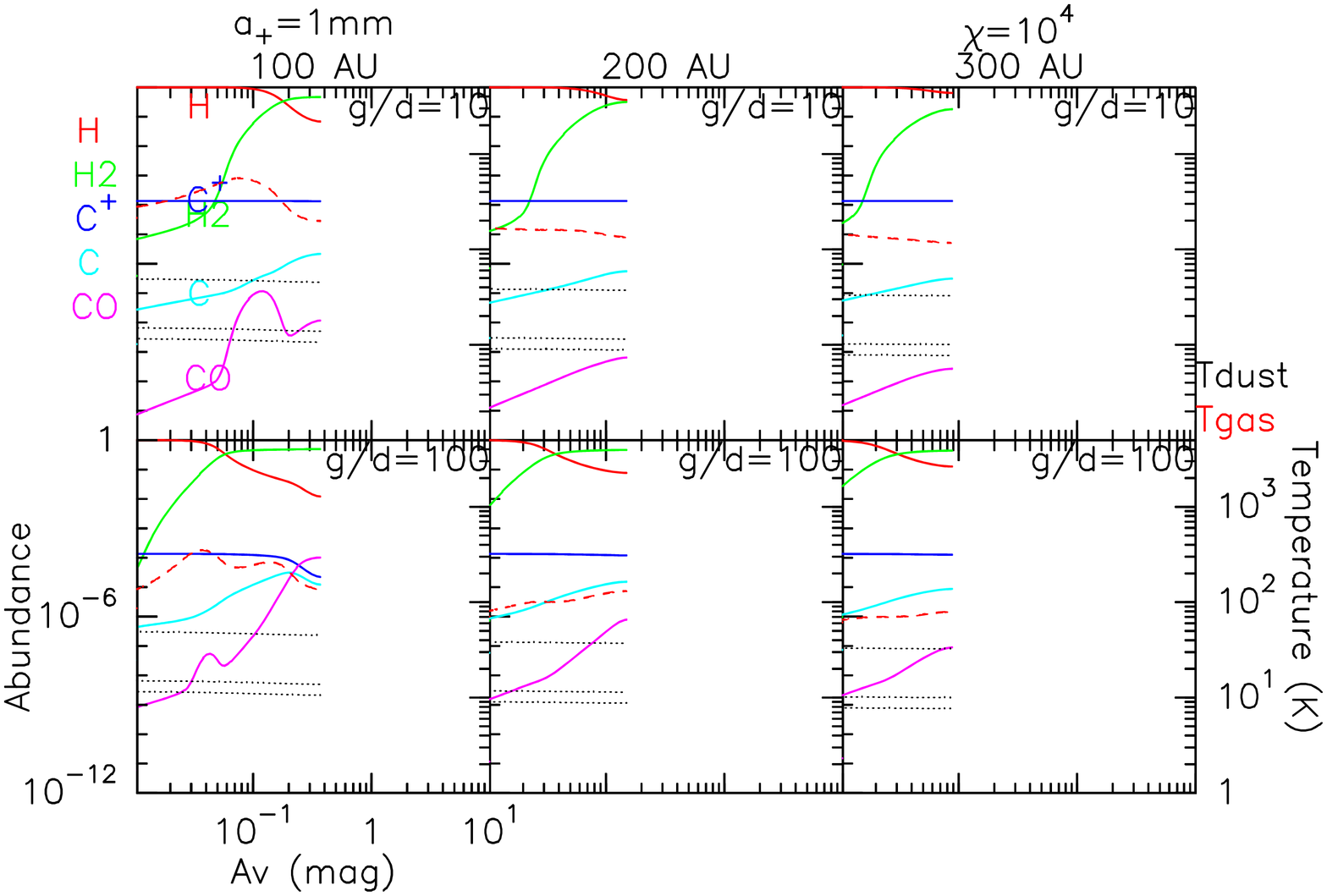}
  \caption { Vertical distribution through the disk of the abundance of H, H$_2$, C$^+$, C and CO and gas (dashed line) and dust (dotted line) temperature at the radii 100, 200 and 300 AU for the models with standard UV field ($\chi = 10^4$ at 100 AU), $a_+ = 1\,\mu$m (top), $a_+ = 1$\,mm (bottom) and with the thermal balance calculated. Dust temperatures are plotted for the extreme grain sizes (\app and \amm) and an intermediate value ($\frac{a_++a_-}{2}$). In the big grains case (bottom), the dust disk is so optically thin  that we never reach A$_\mathrm{v} = 1$.  Note that the vertical and radial scales are the same in all figures.}
  \label{fig:eqth}
\end{figure*}

\begin{figure*}[ht]
  \centering
      \includegraphics[angle=270.0,width=8.2cm]{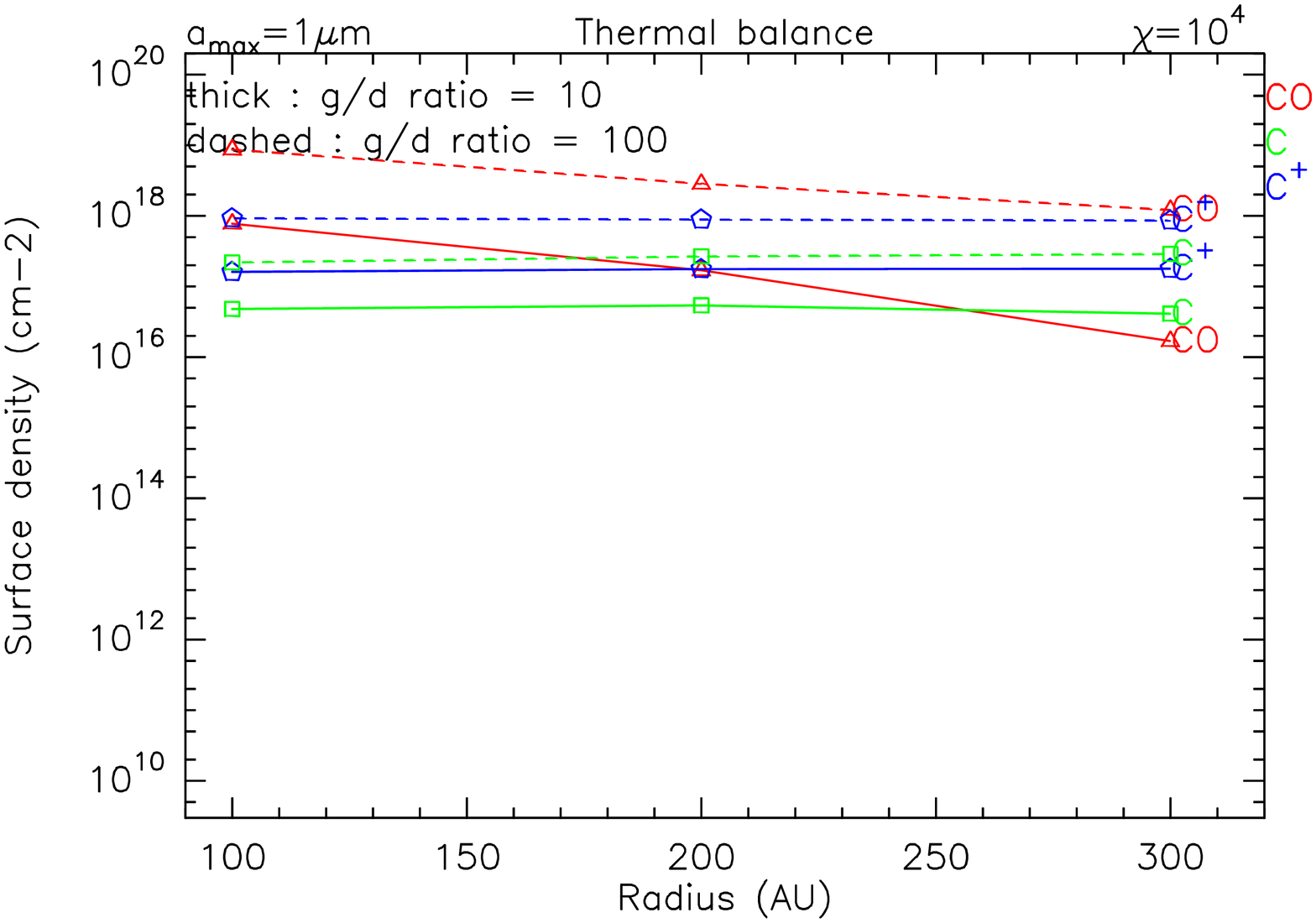}
      \includegraphics[angle=270.0,width=8.2cm]{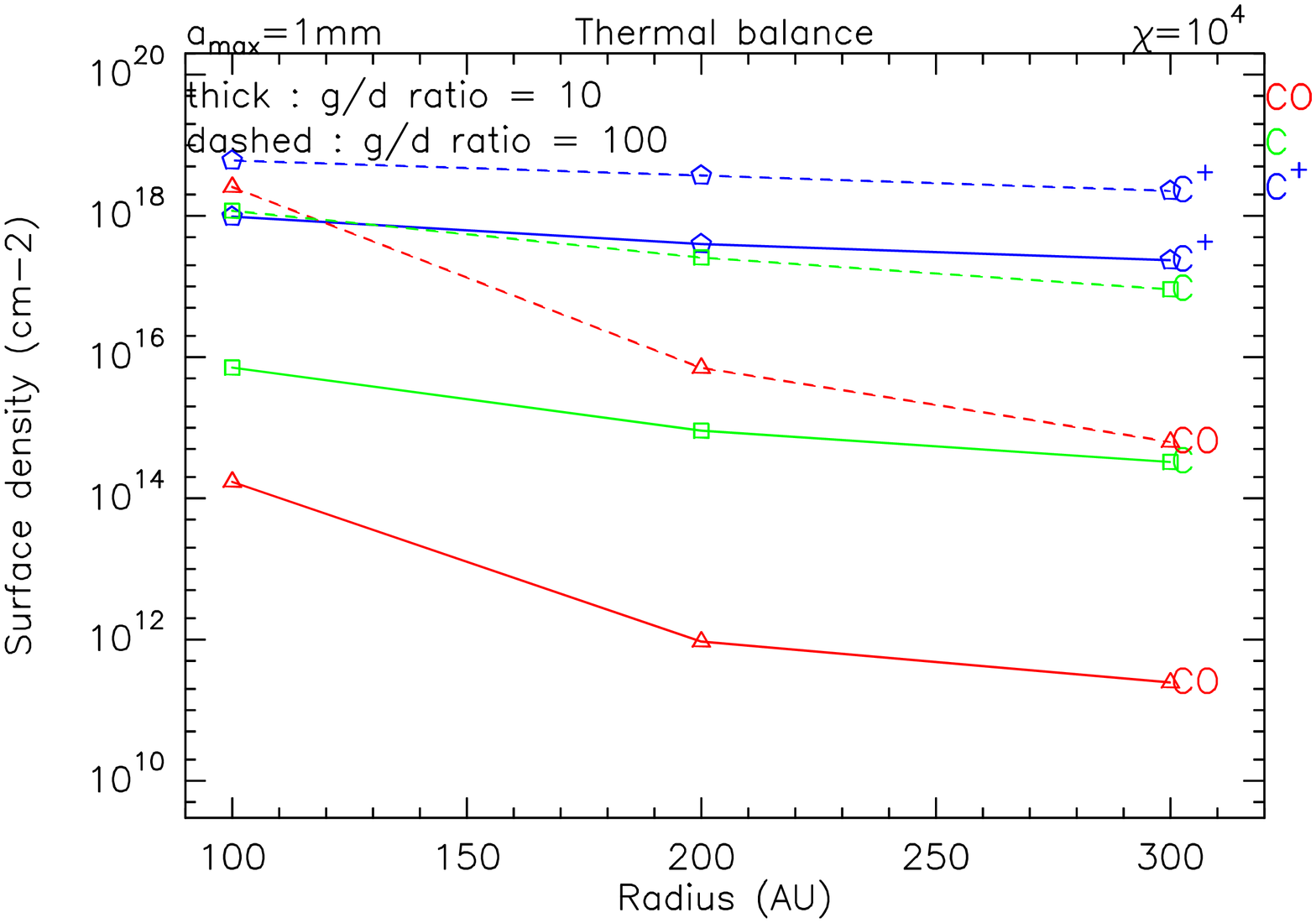}
  \caption{Radial distribution of the column density of C$^+$, C and CO  for the model with standard UV field, the thermal balance calculated, $a_+=1\,\mu$m  (left) and  $a_+ = 1$\,mm (right).}
  \label{fig:eqth_coldens_rad}
\end{figure*}

Figures \ref{fig:eqth} and \ref{fig:eqth_coldens_rad} shows the abundances with respect to H+2H$_2$ and the column densities obtained at radii 100, 200 and 300~AU under a UV field of $\chi = 10^4$.
We assume two different values for \app (1$\mu$m and 1 mm) and for $g/d$ (10 or 100) and the thermal balance is calculated.
The gas temperature goes up to several thousand Kelvin in the disk atmosphere. In the mid-plane in the small grain case, the gas temperature is of the same order of magnitude as the dust temperature because the density is high enough to have thermal coupling by collisions. The shape of the CO abundance distribution is clearly affected by reactions at high temperature with OH leading to a secondary peak not located in the disk mid-plane. Although this peak is quite high in abundance, it occurs when the density is low and the column density is very weakly affected (Figure \ref{fig:eqth_coldens_rad}).

\subsection{Chemical uncertainties and comparison with other results}
\citet{Jonkheid_etal2007} have also calculated the chemistry and gas temperature of evolving Herbig Ae disks. Their model differs in several ways from the above approach. Contrary to us (see above), they have modified the gas to dust ratio by decreasing the mass of the small grains (i.e. interstellar grains) in order to mimic the dust growth and/or settling. The UV transfer is performed with a 2-D code \citep{vanZadelhoff_etal2003} whereas in our case this is a 1-D approach. But the self-shielding of H$_2$ and CO are calculated assuming a constant abundance whereas we compute it explicitly  at each point of the ``cloud''. Their chemical network is described in \citet{Jonkheid_etal2004} and \citet{Jansen_etal95}. It also incorporates chemical reactions with PAHs which are not considered in our model. 

Indeed, we also find that CO becomes underabundant if and only if photodissociation is important. This requires a reduction in the UV opacity which is obtained by grain growth to 1\,mm, or larger. Note that dust sedimentation towards the disk mid-plane should not play a significant role, as the small dust grains which are responsible for the UV opacity remain coupled to the gas. However, contrary to \citet{Jonkheid_etal2007}, we find that Carbon is predominantly in its ionized form C$^+$ when CO is photo-dissociated.

\section{Discussion}

\subsection{CO Abundance: Photodissociation with grain growth}

Figure \ref{fig:eqth_coldens_rad} (right) suggests that a case with \app $\geq$ 1\,mm and $g/d \simeq$ 100 can explain CO column densities of order $10^{16}$ cm$^{-2}$ around 200 AU.
Figure \ref{fig:eqth_coldens_rad} (left) allows us to conclude that the case with $a_+ = 1\,\mu$m  cannot explain the observed CO column densities, even with a low $g/d$ ratio.  For CQ\,Tau  this is in agreement with the spectral index we measure ($\beta = 0.7$) which indicates that significant grain growth has occurred as also found by \citet{Testi_etal2003}. Grain growth is also supported by the $\beta =1$ value found for MWC\,758.

This result suggests that grain growth, or more precisely the enhancement of the UV penetration resulting from grain growth is the dominant process explaining the measured CO column density. Note that we have assumed that the dust size distribution follows $d n(a) \propto a^\gamma$ with $\gamma=-3.5$. The conclusion will not significantly change for realistic slopes $\gamma$ (\citet{Draine_2006} indicates $\gamma = -3.2$ for $\beta = 0.7$).
As the extinction curve is governed by the amount of small grains, it is reasonable to conclude that any solution which guarantees the same amount of small grains, leading to a similar extinction curve, would be sufficient to explain the CO column density without changing $g/d$. Note that for $\gamma = -3.5$, the mass within grains smaller than $a$ is about $\sqrt{a/a_+}$ times the total dust mass, i.e. 3 \% of the dust is in grains smaller than $1 \mu$m for $a_+ = 1$ mm.

It is worth pointing that the photodissociation offers a natural explanation for the steeper surface density law of CO compared to dust. As UV penetration is more efficient in the outer parts of the disk, CO is more heavily dissociated, and the CO abundance decreases with radius (see Fig.\ref{fig:eqth_coldens_rad}), resulting in a larger slope $p$ for CO than for dust and Hydrogen, in qualitative agreement with the observations. However, the $p$ values predicted by the models (4-6) seem larger than observed (2.5-3).

The $[\dco]/[\hco]$ ratio can also be used as a secondary diagnostic of the physical conditions. For large grains, a low $g/d$ ratio predicts a value around 1000 for this ratio, and much higher values for $g/d=100$. Although this is relatively inconclusive compared to our current limit (4000), longer integration times on HCO$^+$ could bring a useful diagnostic.

Note however that the CO column density is very sensitive to the UV penetration (and thus to the disk mass). We computed models twice less massive than those presented in Figs.\ref{fig:eqth}-\ref{fig:eqth_coldens_rad}. In such cases, C$^+$ is the dominant form of Carbon throughout the disk, and the CO column densities become much smaller than observed. Similar results can be obtained using a larger maximum grain size.

\subsection{The UV problem}
\label{sec:UVproblem}
A significant uncertainty is linked to the knowledge of the UV field and to its vertical diffusion in the disk. Comparing Figure \ref{fig:coldens-uv1e3} with Figure \ref{fig:grain_coldens_rad} in the \app = 1mm  case shows that decreasing the UV field to $\chi = 10^3$ essentially brings back the Carbon into CO and increases the CO column densities by a factor 100 in both cases ($g/d = 10$ and 100).
It is important to stress that what matters here is the number of available photons between 900 and 1200 $\AA$ where all the important photodissociation processes occur. Such large values of the UV field cannot be provided by the relatively cool stellar photospheres, and must come from the UV excess due to accretion, see Appendix for details.
On the other hand, the heating will be significantly affected by the mean intensity up to 2000 \AA, which is dominated by the stellar photosphere. In this respect, note that many authors specify the strength of the radiation field in reference to \citet{Habing1968}, i.e. by the integral of the UV flux between 912 and 2400 $\AA$ (=\,5.166\,eV) or between 912 and 2050 $\AA$ (=\,6e\,V) \citep{vanDishoeck_etal2006}. As the shape of the UV spectra from T Tauri or Herbig Ae stars differs quite significantly from the interstellar UV field between 900 and 2400 $\AA$, this can lead to some inappropriate results.

\begin{figure}
  \includegraphics[angle=270,width=8.5cm]{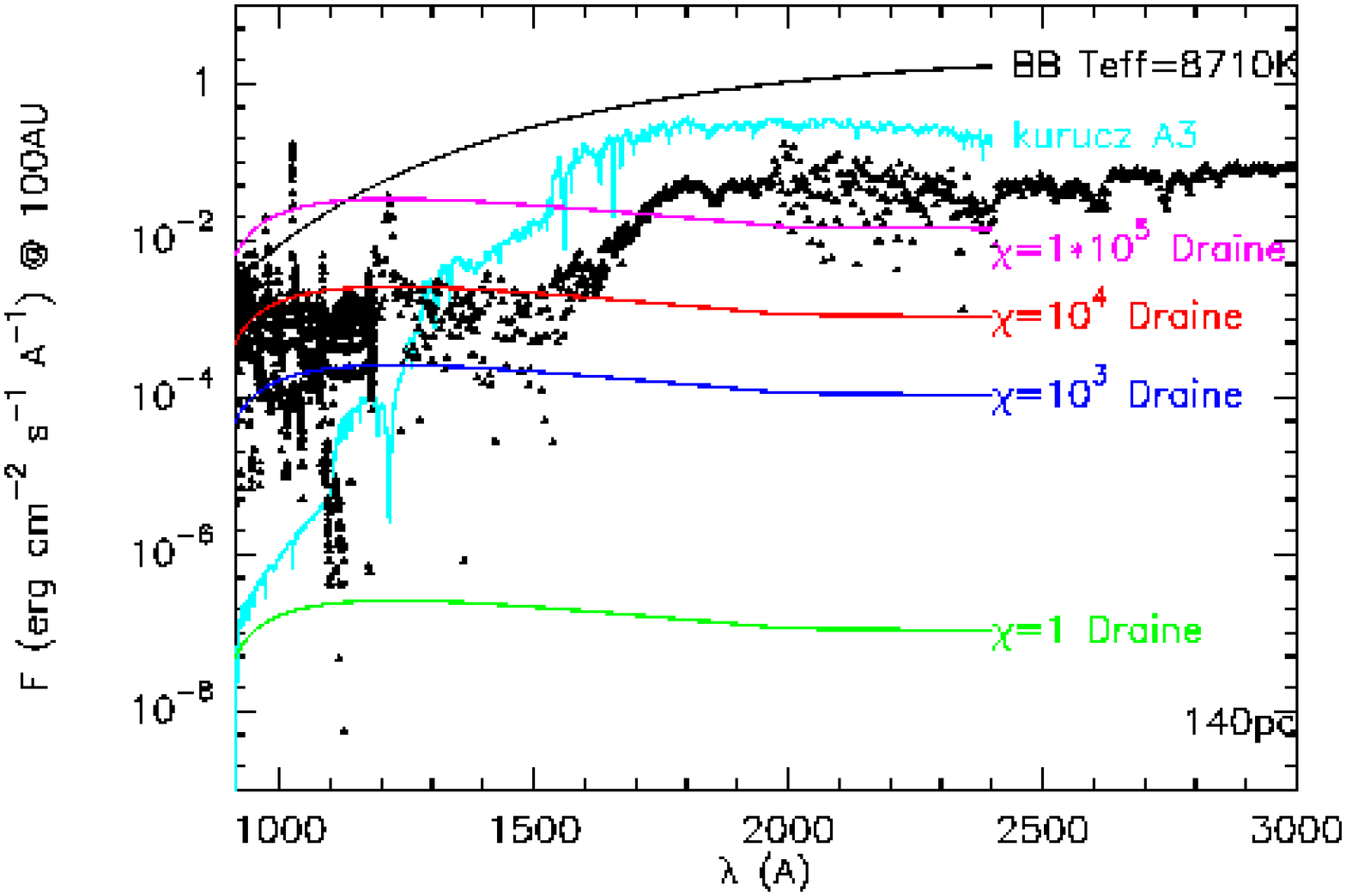}
  \caption {Flux at 100 AU of FUSE and IUE observations MWC\,758 (points), an A3V star according from the Kurucz atlas, a black body with the same temperature as an A3 star and several scaled Draine fields. The $10^{5}$ one is calculated to have the same integrated intensity between 912 and 2400 \AA\ as the observations. The star is assumed to be at 140~pc.}
  \label{fig:flux}
\end{figure}

For example, the FUSE and IUE spectra for MWC\,758 at 100 AU from the star (Figure \ref{fig:flux}) are relatively well fit, between 900 and 1200 $\AA$, by a Draine field scaled up by $\chi \simeq 10^4$
\citep[the apparent best fit is for a few $10^3$, but correction for the star extinction, E(B-V)=0.07,][must be added to the curve presented in Fig.\ref{fig:flux}]{phd_claire}. On the other end, a scale factor of $\chi = 10^5$ would be needed to match the integrated intensity between 912 and 2400 \AA . This scale factor would overestimate by a factor 10 the available UV flux in the 912 -- 1200 $\AA$ domain relevant for photodissociation.

\begin{figure}
  \includegraphics[angle=270,width=8.5cm]{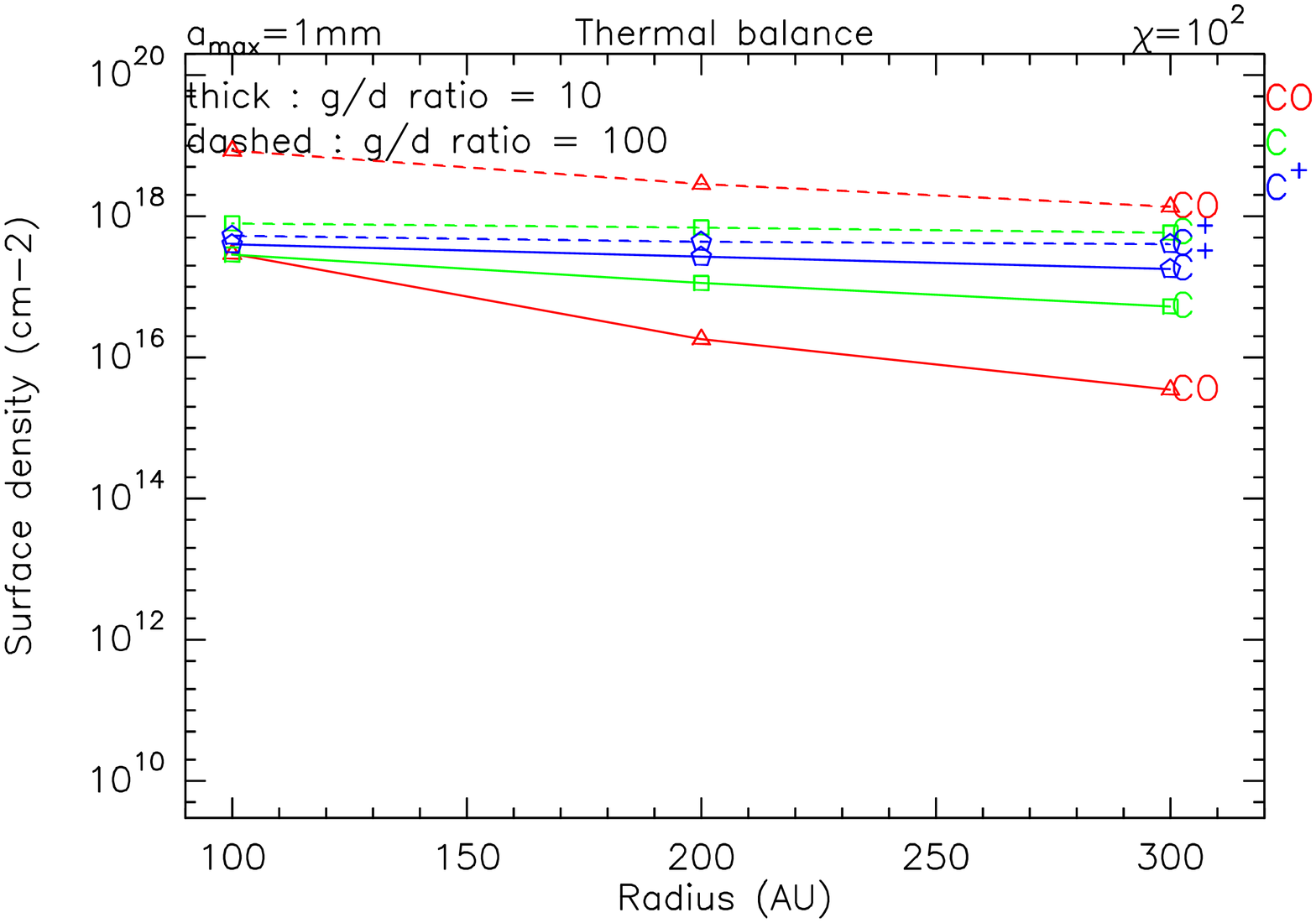}
  \caption{ Radial distribution of the column density of C$^+$, C and CO for the model with \app = 1mm and $\chi = 10^2$.}
  \label{fig:uv1e2}
\end{figure}

For CQ\,Tau, the situation is less clear because of its variability. From the IUE spectrum, the CQ\,Tau UV flux is 10-30 times lower than that of MWC\,758 \citep[see][their Figure 5]{Grady_etal2005}. However, the star may have been observed during a minimum by IUE. We performed another calculation with big grains and a lower UV field ($\chi =10^2$). The results for $g/d=10$ are in good agreement with the observations: the CO column density is about $10^{16}$cm$^{-2}$ at 200 AU (Fig.\ref{fig:uv1e2}) and the gas temperature determined from the thermal balance is about 60~K. The $[\dco]/[\hco]$ ratio is consistent with our limit. Thus a solution with gas dispersal cannot be excluded if the actual UV field is indeed relatively low in CQ Tau.

\subsection{MWC\,758 and the Temperature problem}

The observed CO temperature and that calculated from the (approximate) thermal balance are in reasonable agreement for CQ Tau, around 60 -- 100 K. MWC 758 poses a different challenge, as the observed gas temperature is much lower, about 30 K, while the estimated CO column density is similar.

Reducing the UV flux inside the disk is the simplest way to obtain a lower temperature. For a similar incident UV flux, increasing the attenuation through the disk, i.e. having smaller grains, will lower the temperatures. However, this will also reduce photodissociation and produce more CO. Indeed, Fig.\ref{fig:eqth} shows that low gas temperatures (in the mid-plane) can be obtained only with small (maximum size 1$\mu$m) grains, in which case the CO abundance is high.  For all cases compatible with low CO column densities, the gas temperature is high ($> 50$~K). In fact, in the PDR interpretation, to have a similar CO surface density, the UV fields in the disk must be similar in both sources.

What can explain the temperature difference between the two objects? The thermal balance of the gas is controlled by several mechanisms. In the upper layers, the energy release due to H$_2$ formation dominates the heating, and the gas cools by radiation in the fine structure lines of C and O. Deeper in the disk, photoelectric effect on dust grains become important (and sometimes dominant), and then the energy deposition from cosmic rays (which is largely transported via the chemical reactions). CO line cooling becomes important. At still higher densities, the gas-grain coupling become dominant: this can be either a heating or a cooling mechanism, depending on whether dust is warmer than the gas or vice versa.

The similarities and differences between CQ Tau and MWC\,758 are
  \begin{itemize}
  \item CQ Tau apparently has a lower UV field than MWC\,758, by a factor 10 to 30 (see Sec.\ref{sec:UVproblem})
  \item The UV attenuation is most likely larger in MWC\,758 than in CQ\,Tau, because of the larger density and the (slightly) smaller grains (as its $\beta$ value is larger).
  \item The smaller grains in MWC\,758 lead to a higher dust temperature, although the effect is expected to be small \citep[e.g.][]{krugel_siebenmorgen1994}.
  \item The larger densities in MWC\,758 result in a more efficient gas-grain coupling, as the coupling constant for thermal exchange
    \begin{equation}
      \Lambda \propto \frac{n_\mathrm{dust} n_\mathrm{gas}}{\sqrt{a_- a_+}} \propto  \frac{g/d \, n_\mathrm{dust}^2}{\sqrt{a_- a_+}}
    \end{equation}
    \citep[see][]{Burke_Hollenbach1983}, where we assume the grain size distribution has an exponent $\gamma = -3.5$.    \item The above equation also indicates that smaller grains in MWC\,758 also result in a more efficient gas-grain coupling.
  \end{itemize}
Can these differences explain the lower gas temperature in MWC~758? In addition to the above effects, dust sedimentation may play a role. If dust grains are more sedimented in CQ\,Tau, the gas-grain coupling will be less efficient, and this may preserve a higher temperature of the gas above the disk mid-plane.

\begin{figure*}
  \centering
\begin{tabular}{cc}
\includegraphics[angle=270,width=8.2cm]{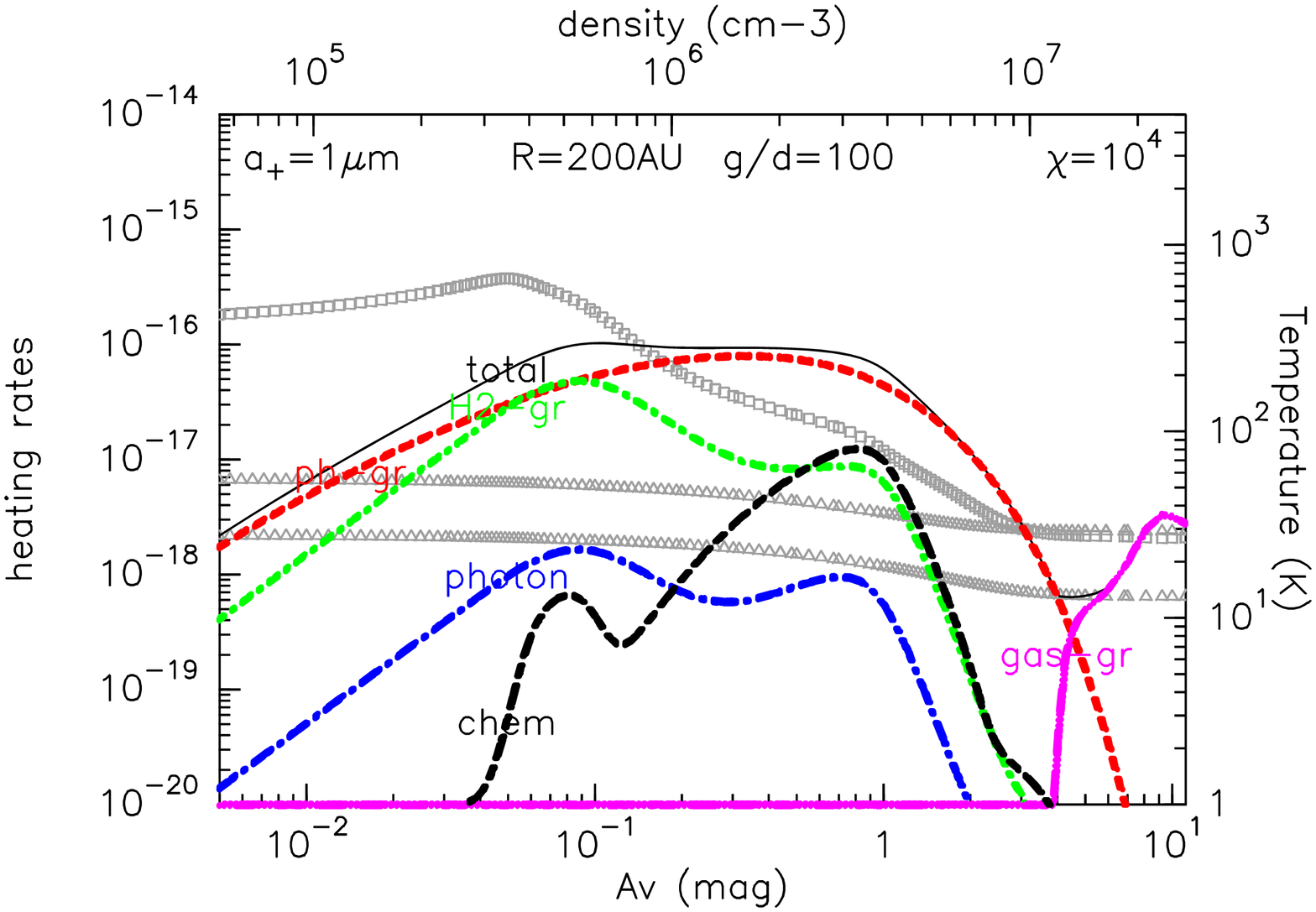}
&
\includegraphics[angle=270,width=8.2cm]{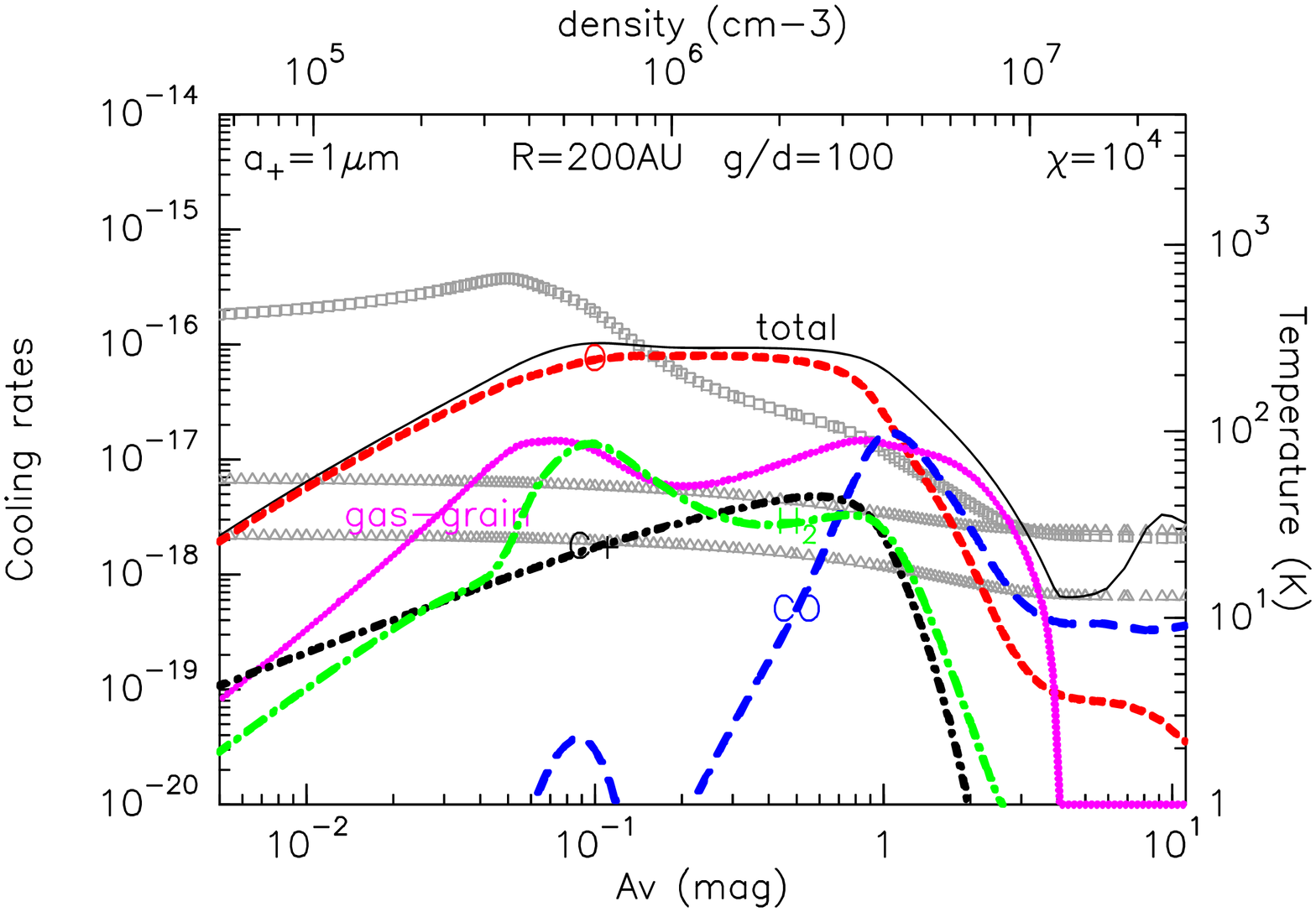}
\\
\includegraphics[angle=270,width=8.2cm]{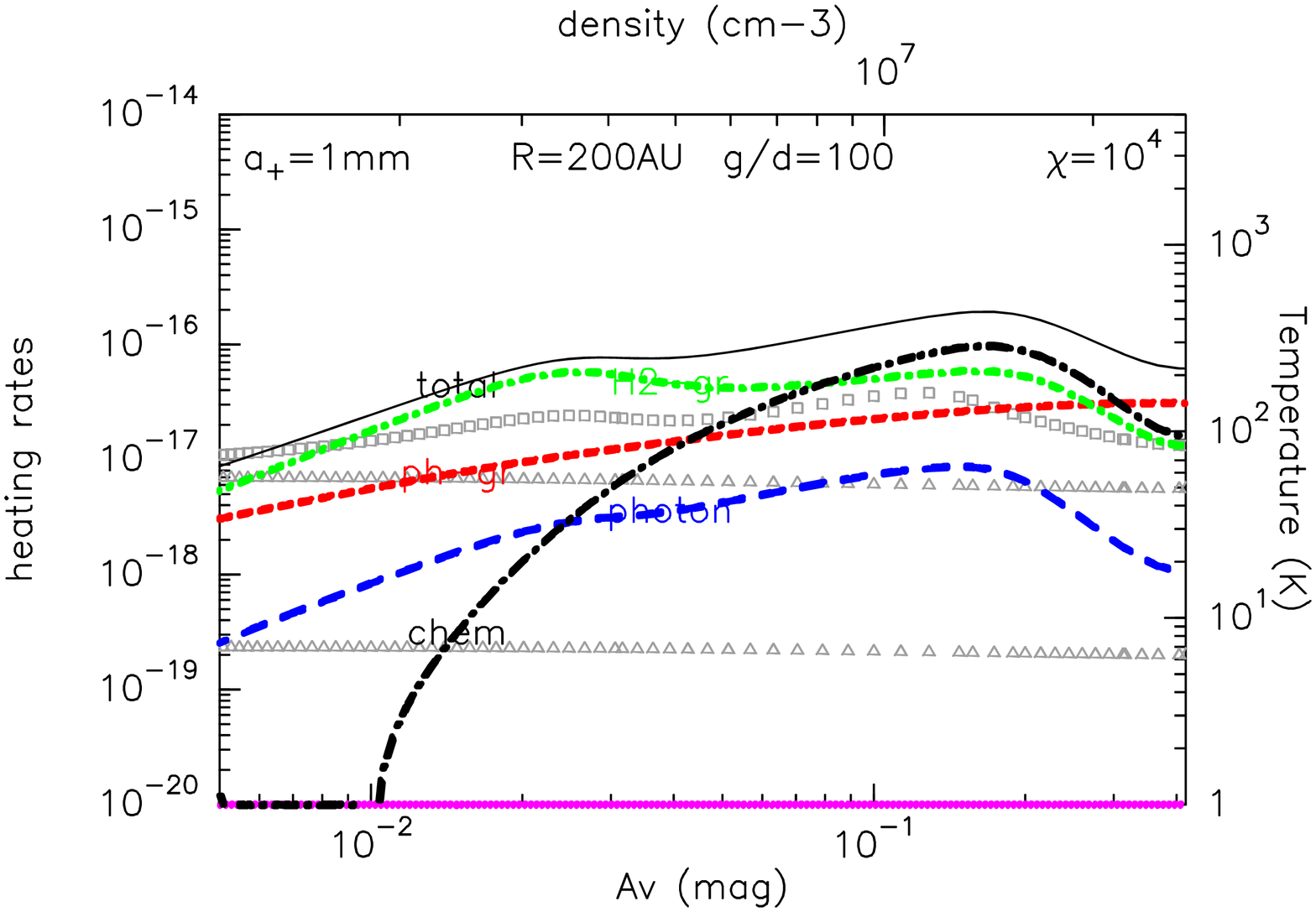}
&
\includegraphics[angle=270,width=8.2cm]{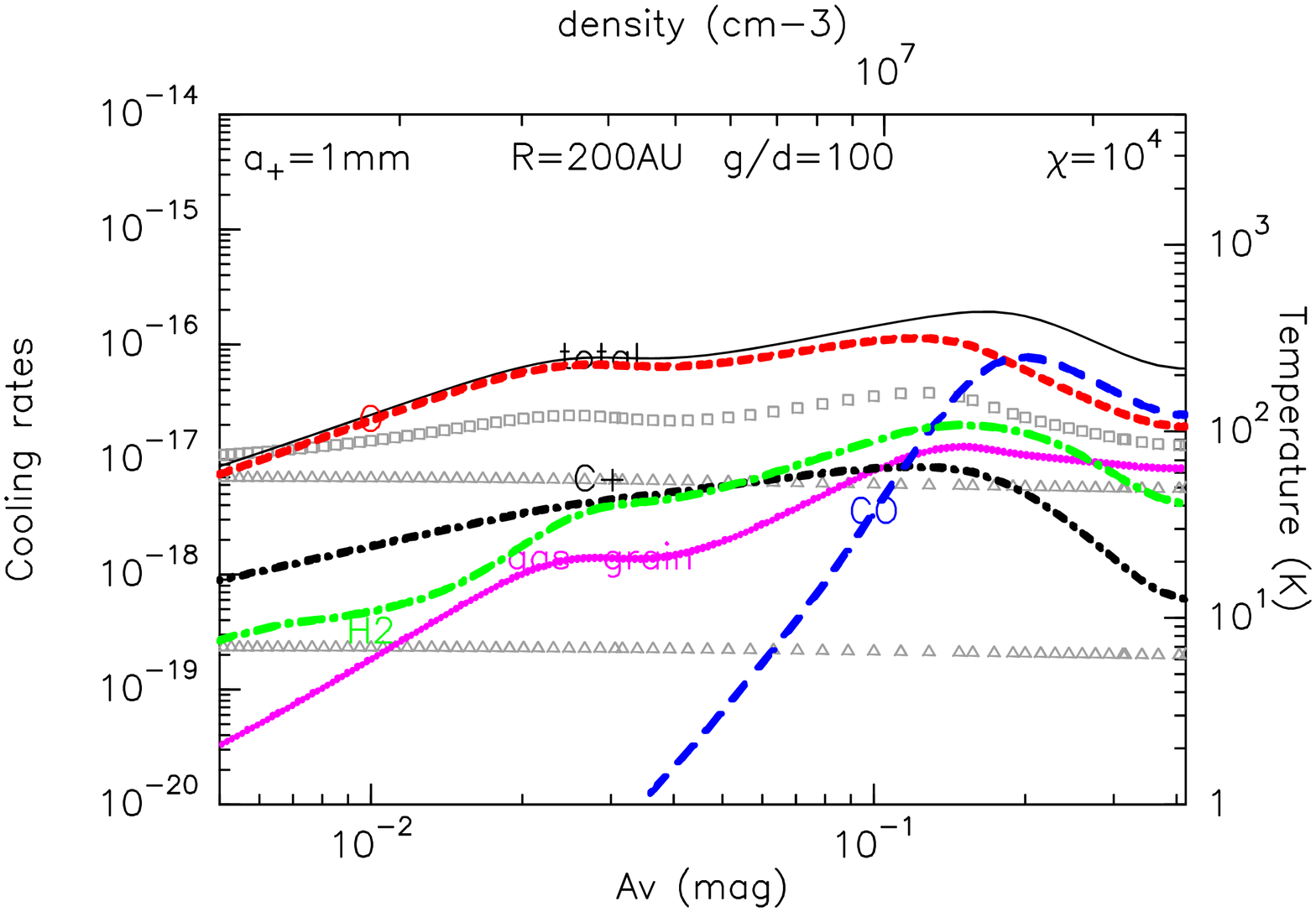}
\end{tabular}
\caption {Heating (left) and cooling (right) process for the models with small grains (top) and big grains (bottom) and a $g/d$ ratio of 100 in erg\,cm$^{-3}$\,s$^{-1}$. \emph{H$_2$-gr} is the heating due to H$_2$ formation on grains. \emph{ph-gr} is the photoelectric effect on grain. \emph{photon} is the direct heating of the gas by the photons and \emph{chem} is the heating due to chemistry reactions (see text for details). Squares: gas temperature, triangles: upper and lower limit of dust temperature.}
\label{fig:heat}
\end{figure*}

The magnitude of these effects is difficult to evaluate without detailed modeling. Figure \ref{fig:heat} indicates the heating/cooling rates for a disk with a (dust) surface density compatible with that of MWC\,758 as computed in our models. In this figure, the ``chemical'' heating rate is the total energy released by chemical reactions: one of the prime energy source for these is actually cosmic-rays through dissociative recombination ion-neutral reactions, the primary ions being produced by cosmic-rays. Note that it is probably an upper-limit \citep[see][section 6.1.5]{LePetit_etal2006}. The direct heating by cosmic-rays is too low to appear here.

Taken at face value, the results of the thermal balance study would favor the low $g/d$ solution (i.e. $g/d=10$), which results in a somewhat more efficient cooling. Reducing even further the $g/d$ ratio ($g/d=1$) results in \emph{higher} gas temperature because the gas is then mostly heated by H$_2$ formation  and photoelectric effect on grains.
However, there are several significant uncertainties in our modeling procedure. For example, the amount of small grains which controls the photo-electric heating \citep{bakes_Tielens1994} is poorly constrained, so the efficiency of this process may be overestimated in our model. Similarly, the gas-grain coupling is dominated by the small grains, because of their larger cross section and higher temperature, and remains also rather uncertain. Furthermore, we use a very crude approximation for the diffusion of UV photons towards the disk plane, and even the  unattenuated UV flux is uncertain (see Sec.\ref{sec:UVproblem}). Note that the photodissociation is totally dominated by the UV excess, while the stellar UV flux will play a role in the heating processes.

In fact, the situation of CQ\,Tau and MWC\,758 is comparable to that of BP\,Tau, where \citet{Dutrey_etal2003} have found warm, optically thin CO, with a very low ``apparent'' abundance (of order 10$^{-6}$). It is thus tempting to interpret BP\,Tau in the same framework of CO photodissociation by the UV. This possibility is further supported by the fact that BP\,Tau is a very active T Tauri star, with a large accretion rate \citep[$2\,10^{-7} \msun$/yr, see e.g.][]{Bertout_etal1988,Gullbring_etal1996}.

\subsection{An alternative: Grain growth by ice mantle accretion}
\label{sec:largegrains}

We have not explored so far the possibility that Carbon (and Oxygen too) nuclei are locked into grains.
The gas temperature determines whether gaseous components can stick onto grains, while the dust temperature determines whether ices can evaporate. In both sources, the gas is warm enough to avoid new CO molecules to stick on grains. However, a fraction of C and O nuclei can nevertheless remain trapped in grains, as discussed below.

This can happen if Carbon has been removed from the gas phase through the building of ice mantles onto grains. Such a process does not change the dust mass significantly (at most a factor 2 with standard elemental abundances),  and changes even less the grain sizes. However, it removes selectively from the gas heavy nuclei with respect to Hydrogen. It is then conceivable to have a low CO / dust ratio. The difficulty of such a possibility is the thermal history of the dust and gas during the disk evolution. The dust must have been cold enough to accrete CO, which is the main reservoir of Carbon, and not released later, either thermally or by photo-desorption due to the enhanced UV field.

This can happen if the CO is converted onto grains into other species (such as H$_2$CO or CH$_3$OH) with larger desorption temperature.

In this respect, it is worth noting that in our simulations, the temperatures of the big grains are low enough (around 10~K) to prevent CO from being thermally evaporated. In a dynamical process where the small grains build up ice mantles, and then grow by coagulation to form larger and larger grains, the bulk of the CO trapped on grains will end up in the large grains, which contains most of the mass. Hence, despite the apparent ''large'' gas temperature traced by CO, a substantial fraction of C and O may be actually depleted onto grains.  In this explanation, the apparently low CO to dust ratio would not imply a low $g/d$ ratio, but strong C and O depletion like in disks around T Tau stars.
The achieved depletion depend on the past thermal history but also critically on the grain growth processes. The maximal depletion depends on the dust mass fraction in grains which are large enough to remain below the CO desorption temperature. \\

\begin{figure*}
\centering
\includegraphics[angle=270,width=15cm]{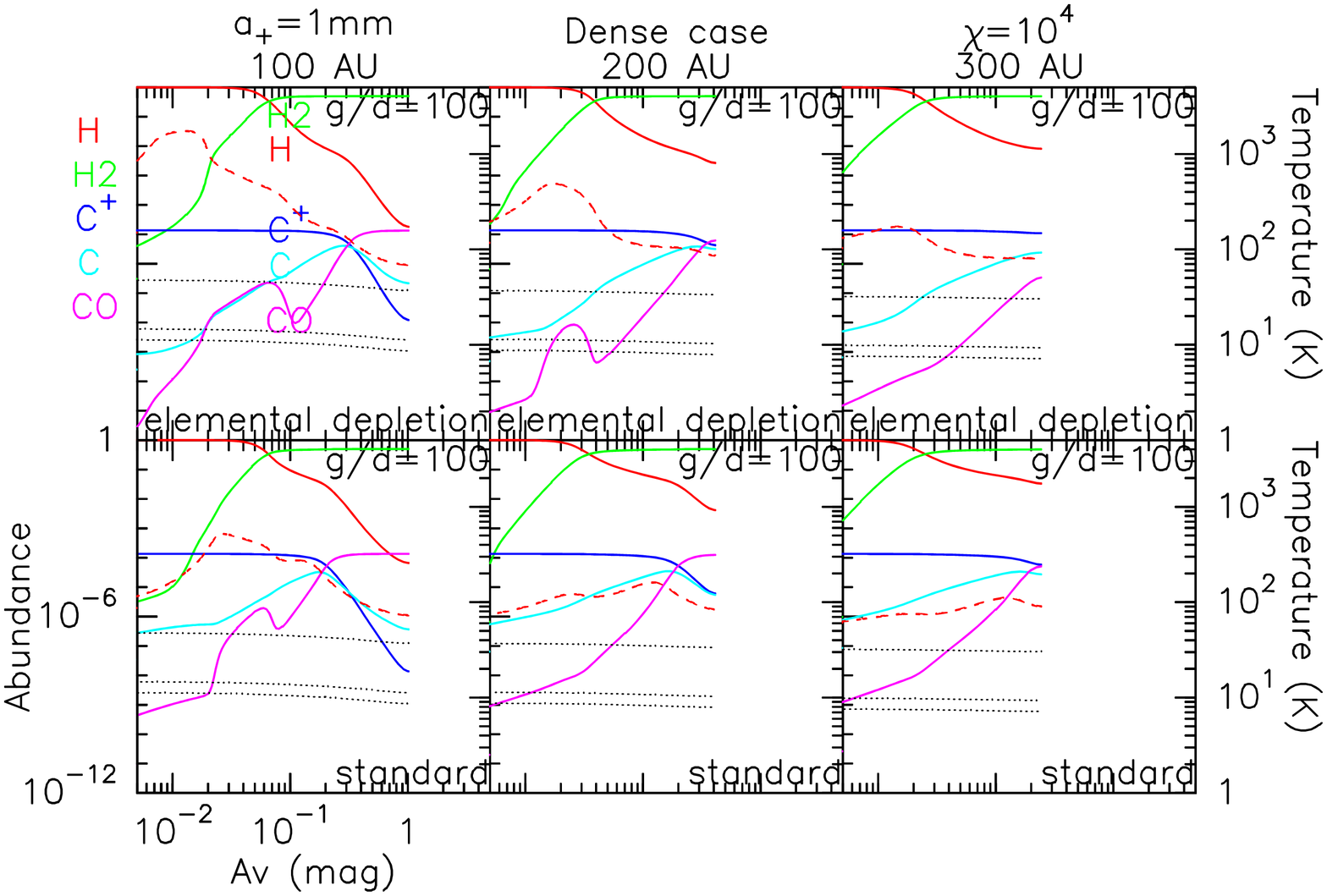}
\caption{Vertical distribution through the disk of the abundance of H, H$_2$, C$^+$, C and CO and gas (dashed line) and dust (dotted line) temperature at the radii 100, 200 and 300 AU for the models with standard UV field ($\chi$=10$^4$), \app = 1mm, $g/d$=100 in the dense case (MWC758 like). Top: all the elemental abundances (except H and He) are depleted by a factor 10. Bottom: standard elemental abundances}
\label{fig:depletion}
\end{figure*}

We have computed some models with such depleted elemental abundances. As expected, the CO abundance reflects the elemental depletion (see Figure \ref{fig:depletion}). Note that the gas remains warm even in the ``dense'' case (factor \textbf{3} in density compared to the previous models).

A possible way to check this hypothesis would be a direct measurement of the \emph{mm dust} temperature. The only such available measurement is for MWC\,480 by \citet{Pietu_etal2006}, who indeed found a very low dust temperature in this source, much lower than found from the SED fitting, and perhaps as low as 10 K near 100 AU. Although this can be attributed to vertical temperature gradients, the temperature dependence on the grain size may also play a role in explaining such a large difference.

\subsection{The pending UX Ori Problem for CQ Tau}

The large photometric variations observed in CQ Tau classify it as an UXOr variable, and have been interpreted as due to variable extinction. The accurate determination of the inclination of CQ Tau challenges this interpretation since
UX Ori phenomenon can only be observed if the disk inclination is larger than $\sim 45^o$ \citep{Natta_etal1997, Natta_Whitney2000}.

\citet{Eisner_etal2004} have measured the distribution of the NIR emission from the disks of CQ Tau and MWC\,758 using the PTI interferometer. For CQ Tau, they find a position angle for the \emph{disk axis} \footnote{ \citet{Eisner_etal2004} give the PA of the emission major axis whereas we indicate the PA of the disk axis} of $14\pm 6^\circ$ and $i=48\pm5^\circ$  (for MWC\,758, they find PA$ = 40\pm6^\circ$ and $i=33\pm4^\circ$).
 Taken at face value, this indicates that the orientation and inclination of the two inner disks {\em differ markedly}  from that of the outer disks, sampled by CO, the axes being separated by $30 \pm 7^\circ$.
One possibility would be that the inner disk is warped due to dynamical interactions with (yet undetected) inner bodies, as in the case of the more evolved disk of $\beta$ Pictoris \citep{Mouillet_etal1997}. Such large warps have never been observed so far. In the case of the spectroscopic binary UZ Tau-E, for which \citet{Simon_etal2000} suggested a mis-alignment between the orbit and the surrounding disk by a similar amount, a recent re-evaluation of the binary orbit \citep{Prato_etal2002} concludes that the orbit and disk planes are aligned to within 4 degrees.
In all confirmed cases, disk warps are small: only a few degrees in $\beta$ Pic for example.

Moreover, the inner disk results are based on very few interferometric data points: a confirmation of theses values would be required before one can seriously invoke such large disk warps.

\section{Summary}

Using the IRAM array, we have observed in continuum and CO J=2-1 and J=1-0 the disk surrounding MWC758
and in CO J=2-1 that of CQ Tau. We also got upper limits in HCO$^+$ J=1-0
in both sources. We detected the dust and gas emission and were able to derive
their physical conditions applying a standard disk model. In both cases, we found
two relatively small CO and dust disks compared to those observed and analyzed
so far around other Herbig Ae stars such as AB\,Aur, MWC\,480 or HD\,34282.
We also used the PDR model developed by the Meudon Group in order to qualitatively
interpret the observed CO abundances. Our main results are:

\begin{itemize}

\item The two disks have  partly optically thin CO J=2-1 line
and high CO depletion factor with respect to H$_2$ of the order $\sim 100-200$.
We also found a similar lower limit on the ratio [$\dco$]/[HCO$^+$] $\simeq 4000$.
In both cases, their CO surface densities fall down faster than the dust surface density.
The dust emissivity spectral index is of the same order with $\beta \simeq 0.7 - 1$, implying
grain sizes up to a few mm.

\item The main difference between the two disks is the temperature.
The disk orbiting CQ Tau is the hottest one observed so far with
T$_{100} \simeq 150 \pm 50$~K, while the MWC\,758 disk is not as warm, with
T$_{100} \simeq 30 \pm 1$~K.

\item Although the disk masses are uncertain, due to distance ambiguities and, in CQ Tau, to limited knowledge of the temperature, they appear lower than those of similar objects studied so far such as MWC\,480, HD\,163296, and HD\,34282

\item  We found that change in UV opacity due to grain growth, and thus the resulting enhancement of
the UV penetration, is a major process to explain the low CO column densities.

\item An important uncertainty in the PDR model resides in the estimate of
the UV field intercepted by the disk. The amount of stellar UV photons is too small to
produce the required CO photodissociation. The stars must be accreting to produce a sufficient
UV excess. Since the accretion rates and the UV fields are both poorly known
for most similar sources, this remains a limiting factor for chemical modeling of disks.

\item CQ Tau may be explained by a high UV flux and a normal $g/d$ ratio.
However, obtaining column densities just around $10^{16}$~cm$^{-2}$ requires a rather fine balance between
incident UV flux and UV penetration.

\item The lower gas temperature of MWC\,758 is not well reproduced by the PDR
model. A reduced $g/d$ ratio may be favored for this source. Alternatively, the uncertainties
in the overall thermal balance modeling may be the cause of the discrepancy.

\item Although the observed CO gas is too warm to allow sticking onto grains, we find that the temperature of large grains to be low enough to prevent CO from being released from the grain surfaces. The apparent low CO to dust ratio may thus be the result of the disk thermal history.

\item Finally, the inclination angle we measured in the outer disk of CQ Tau
($29 \pm 2^o$) appears to be incompatible with the fact that this object
is classified as an UX Ori star.
One possibility left is that the inner disk is warped due to
dynamical interactions with large bodies.  Accurate determinations of the inclination of the
inner and outer disks in more UXOr are required to progress in resolving this issue.
\end{itemize}
Our study clearly indicates that a low apparent CO to dust ratio does not necessarily imply a low H$_2$ to dust ratio, and thus that CO is not an unambiguous tracer of gas dispersal.

\begin{acknowledgements}
This research made use of the SIMBAD
database, operated at CDS, Strasbourg, France.
We acknowledge all the Plateau de Bure IRAM staff for their help during the observations.
We thank Pierre Hily-Blant for his modifications of a previous version of the PDR code.
We also acknowledge Franck Le Petit for many fruitful discussions about the PDR code.
Thierry Forveille kindly provided us with the CO J=1-0, and Gail Schaeffer helped with its
initial processing. The FUSE spectrum was provided by Claire Martin-Za\"idi. This research was
supported by the INSU program PCMI.
\end{acknowledgements}

\bibliography{9523}

\bibliographystyle{aa}
\appendix

\section{The Extinction Curve with Grain Growth}
\label{a:extinction}
\begin{figure}[ht]
  \centering
  \includegraphics[angle=270.0,width=8.5cm]{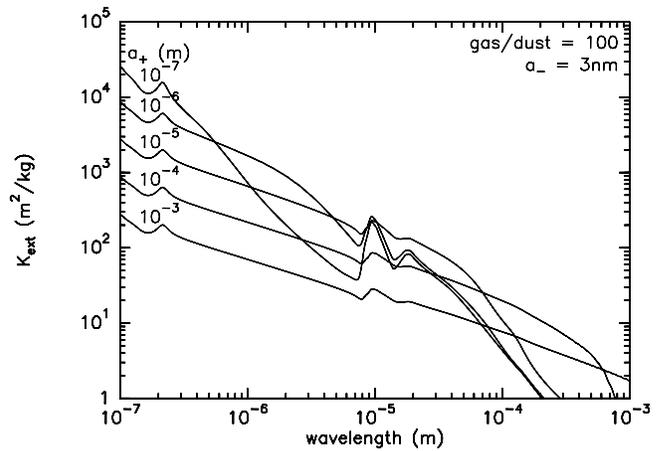}
  \caption {Extinction curves as a function of the maximum radius of the grain size distribution for a gas to dust ratio of 100, \amm = 3nm and $\gamma = -3.5$. }
  \label{fig:nhav}
\end{figure}

The extinction curves for several \app\ can be seen on figure \ref{fig:nhav}. The peak around $\lambda = 2.2\,10^{-7} $m is due to small graphite grains \citep{Mathis_1994}, the feature around $\lambda = 1\,10^{-5}$m is due to silicates. The curve for \app = $10^{-5}$ crosses the one for \app = $10^{-4}$ cm at $5.55\,10^{-7}$m, and the A$_V$ extinction is calculated at $\lambda = 5.5\,10^{-7}$m. This explains why the \nhav\ ratios for these two dust size distributions are so similar (table \ref{tab:nhav}).

\begin{table}[ht]
\caption{Extinction factor $\textrm{N}_H/\textrm{A}_v$} \label{tab:nhav}
\begin{tabular}{cc|cc|cc}
\hline
a$_+$ (m) & N$_H$/A$_V$ & a$_+$ (m) & N$_H$/A$_V$ & a$_+$ (m) & N$_H$/A$_V$ \\
\hline
1$\,10^{-8}$ &  4.02$\,10^{21}$ & 5$\,10^{-6}$ & 3.66$\,10^{21}$ & 1$\,10^{-3}$ & 4.86$\,10^{22}$\\
5$\,10^{-8}$ &  3.08$\,10^{21}$ & 1$\,10^{-5}$ & 5.07$\,10^{21}$ & 5$\,10^{-3}$ & 1.08$\,10^{23}$\\
1$\,10^{-7}$ &  1.80$\,10^{21}$ & 5$\,10^{-5}$ & 1.10$\,10^{22}$ & 1$\,10^{-2}$ & 1.53$\,10^{23}$\\
5$\,10^{-7}$ &  1.43$\,10^{21}$ & 1$\,10^{-4}$ & 1.55$\,10^{22}$ & 5$\,10^{-2}$ & 3.42$\,10^{23}$\\
1$\,10^{-6}$ &  1.82$\,10^{21}$ & 5$\,10^{-4}$ & 3.44$\,10^{22}$ & \\
\hline
\end{tabular}\\
$\textrm{N}_H/\textrm{A}_v$ ratio as a function of the maximum radius of the grain size distribution for a gas to dust ratio of 100, \amm = 3nm and $\gamma = -3.5$.
\end{table}

\begin{figure*}[ht]
  \centering
    \begin{tabular}{cc}
  \includegraphics[angle=0.0,width=7.75cm]{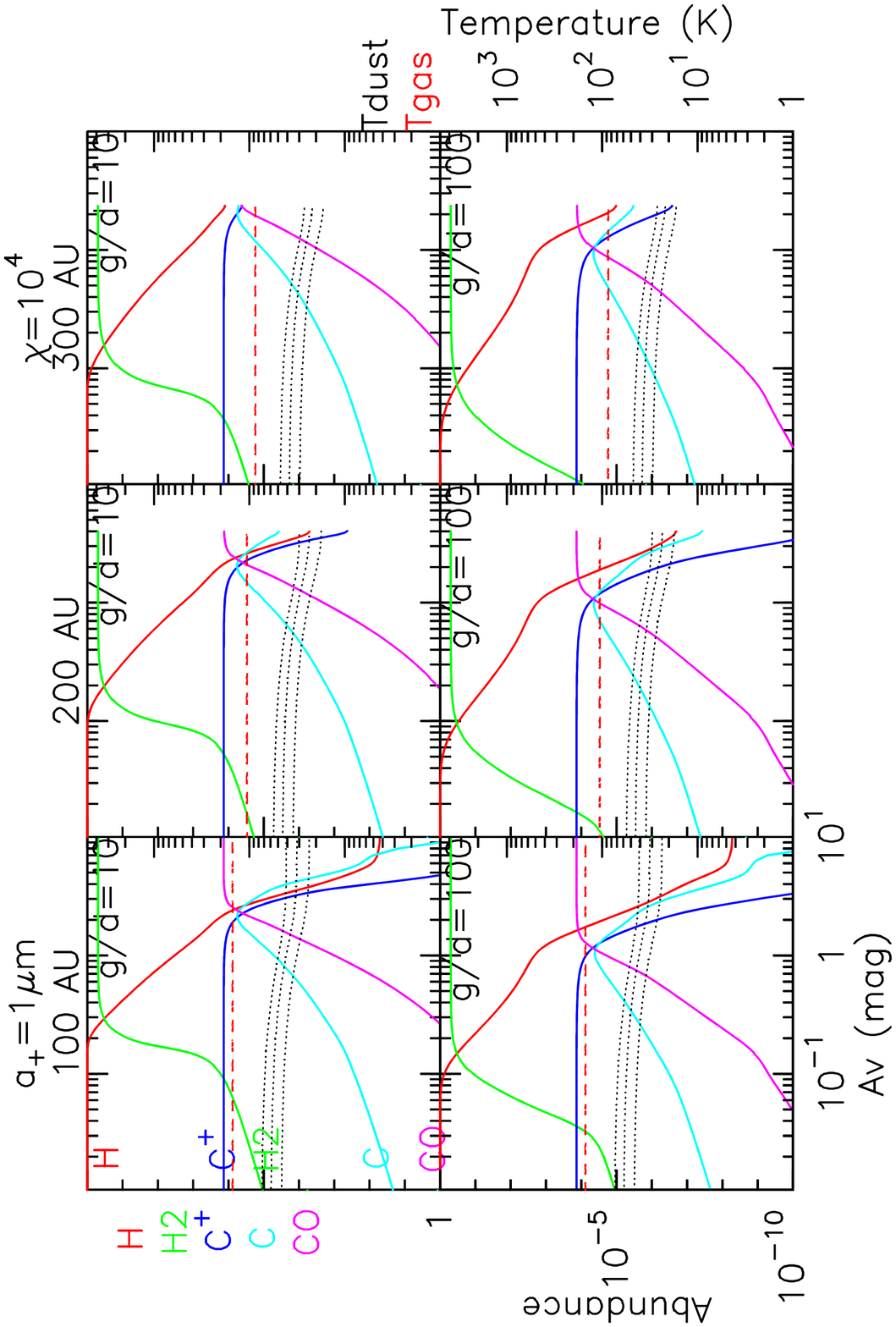}
&
  \includegraphics[angle=0.0,width=7.75cm]{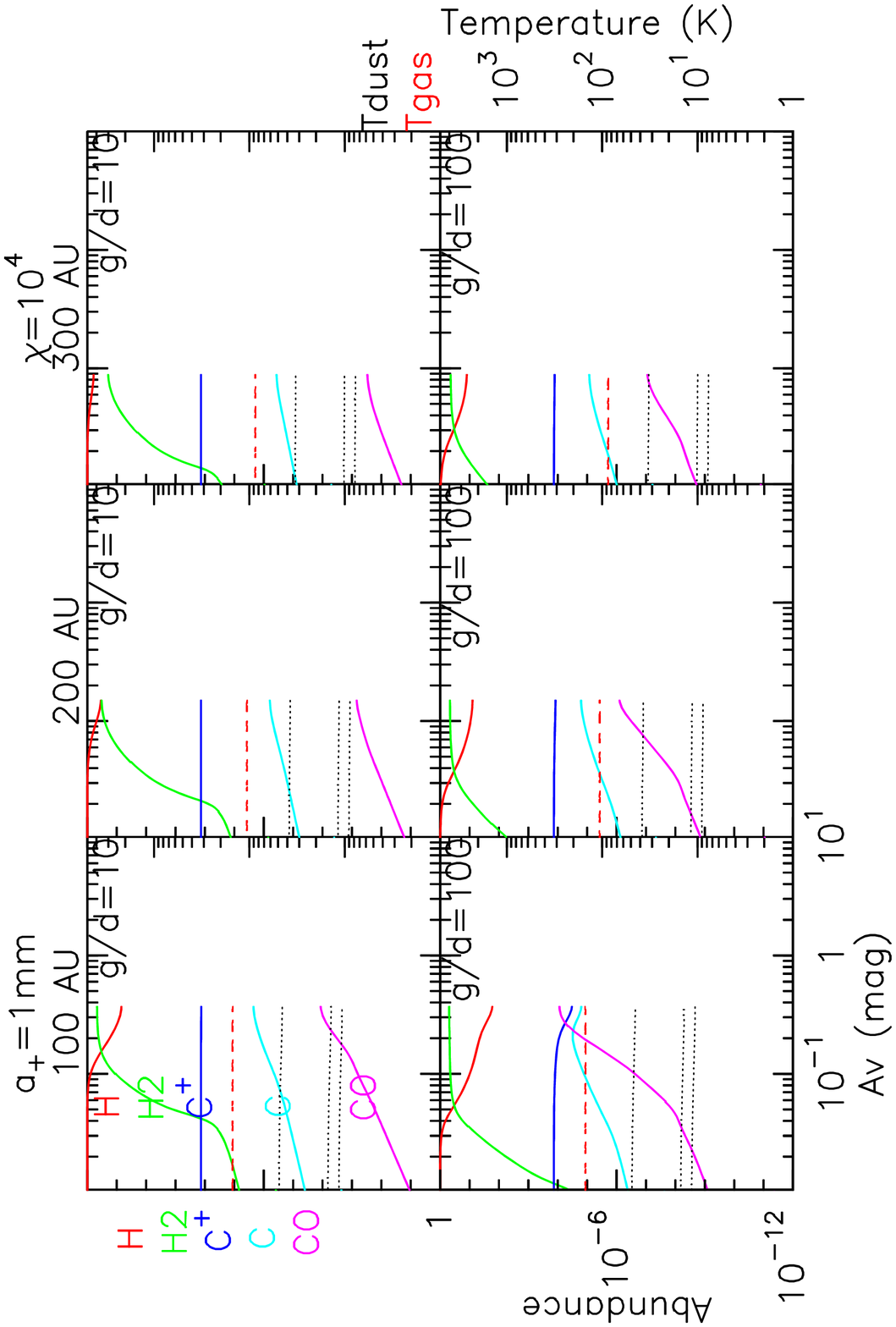}
\\
  \includegraphics[angle=0.0,width=7.75cm]{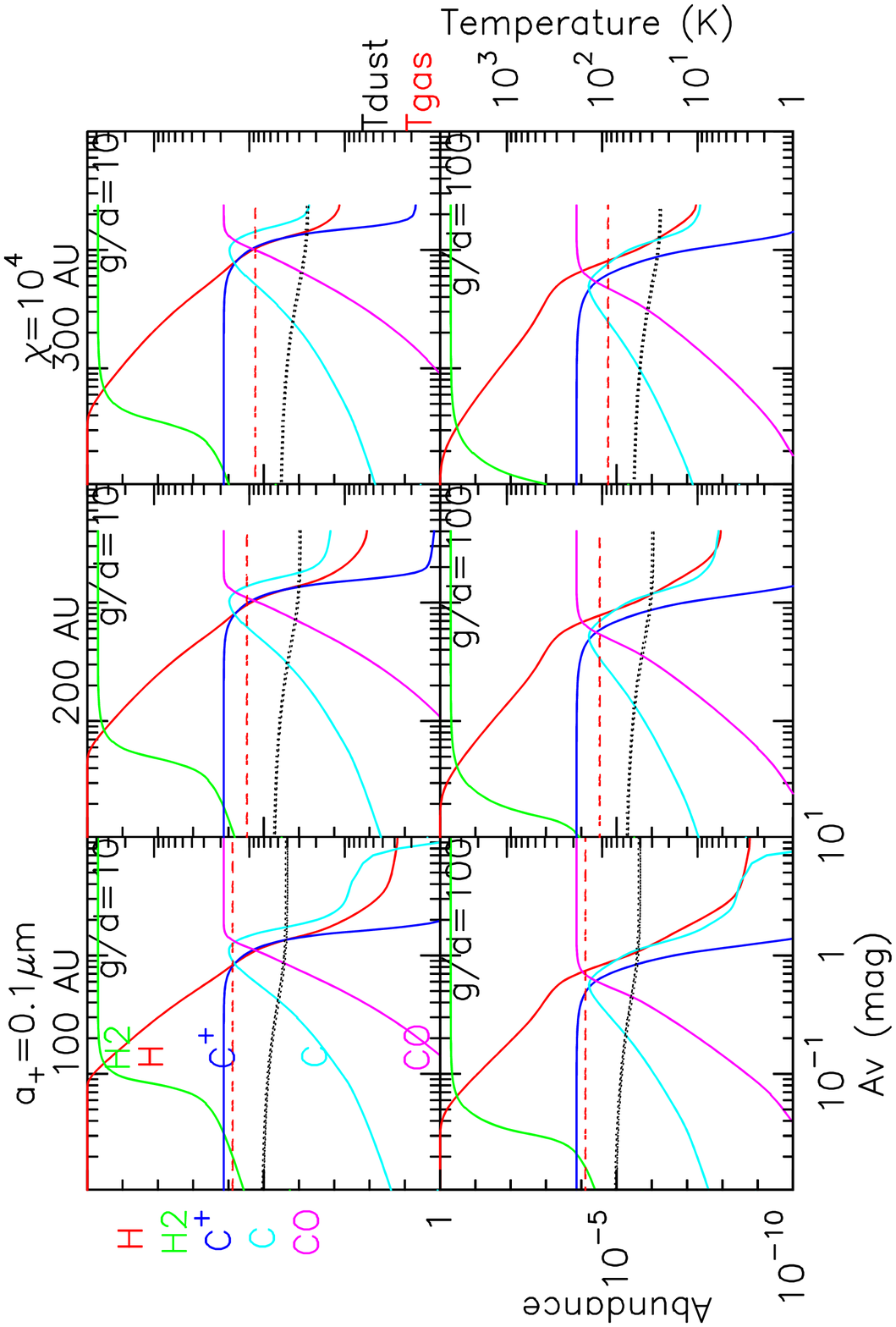}
&
  \includegraphics[angle=0.0,width=7.75cm]{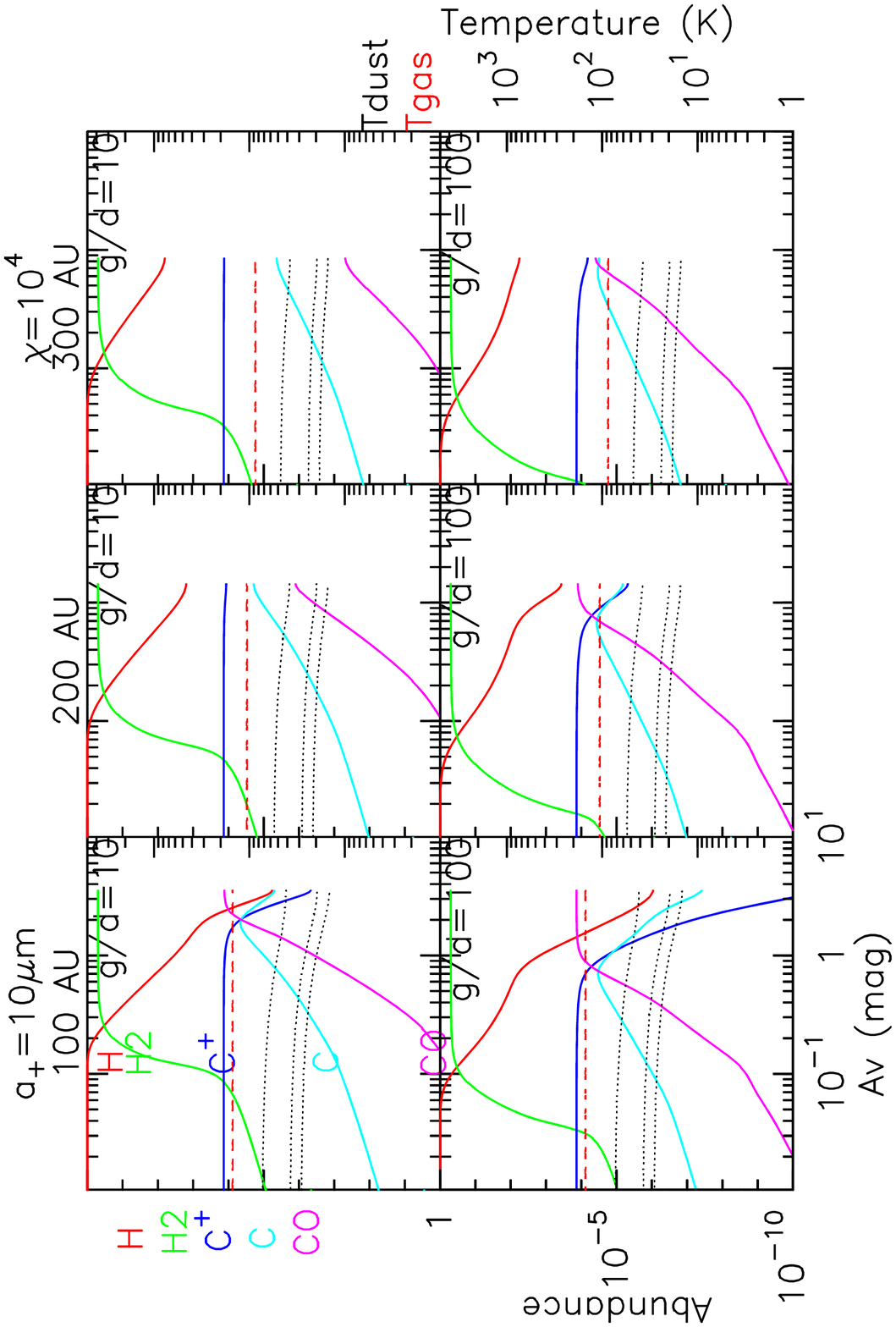}
\\
    \end{tabular}
  \caption {Grain size effect on molecular abundances. Vertical distribution through the disk of the abundance of H, H$_2$, C$^+$, C and CO and of the temperature ( no thermal balance calculated)
 at the radii 100, 200 and 300 AU for the models with \app= 0.1, 1, 10\,$\mu$m and 1\,mm. In the latter case the dust disk is so optically thin  that we never reach A$_v = 1$.}
  \label{fig:grain}
\end{figure*}

\begin{figure*}[ht]
  \centering
    \begin{tabular}{cc}
      \includegraphics[angle=270.0,width=8.0cm]{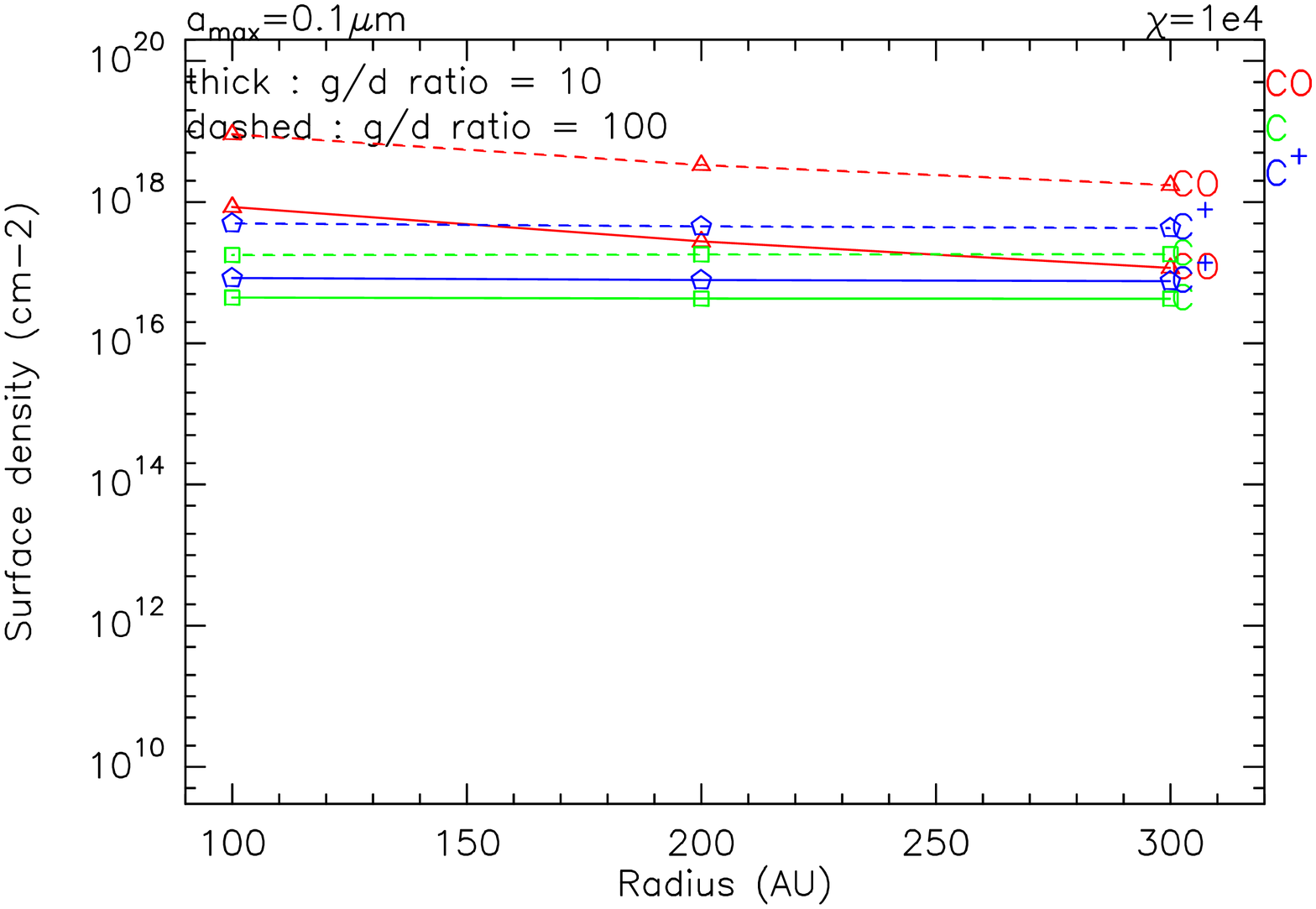}
      &
      \includegraphics[angle=270.0,width=8.0cm]{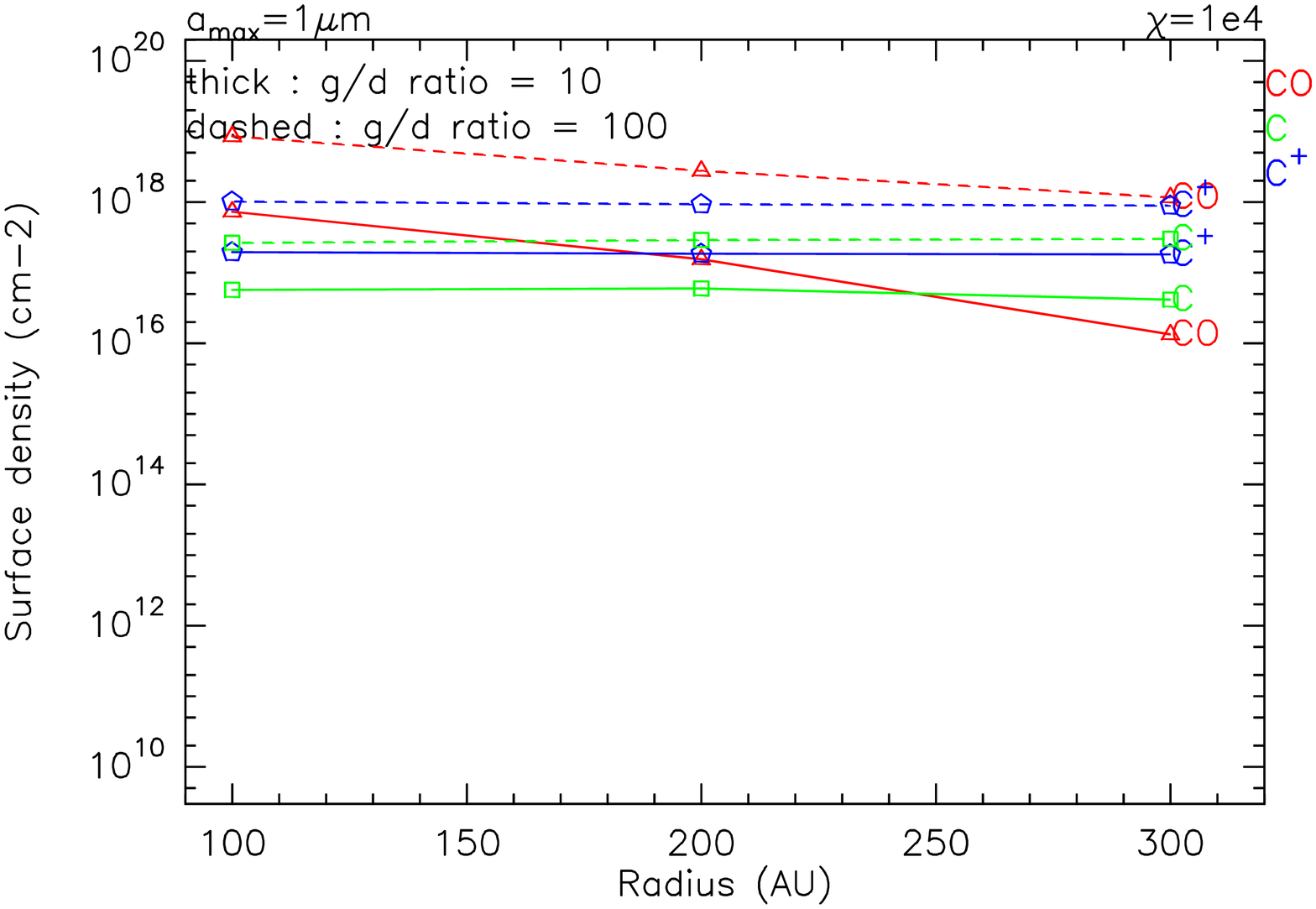}
      \\
      \includegraphics[angle=270.0,width=8.0cm]{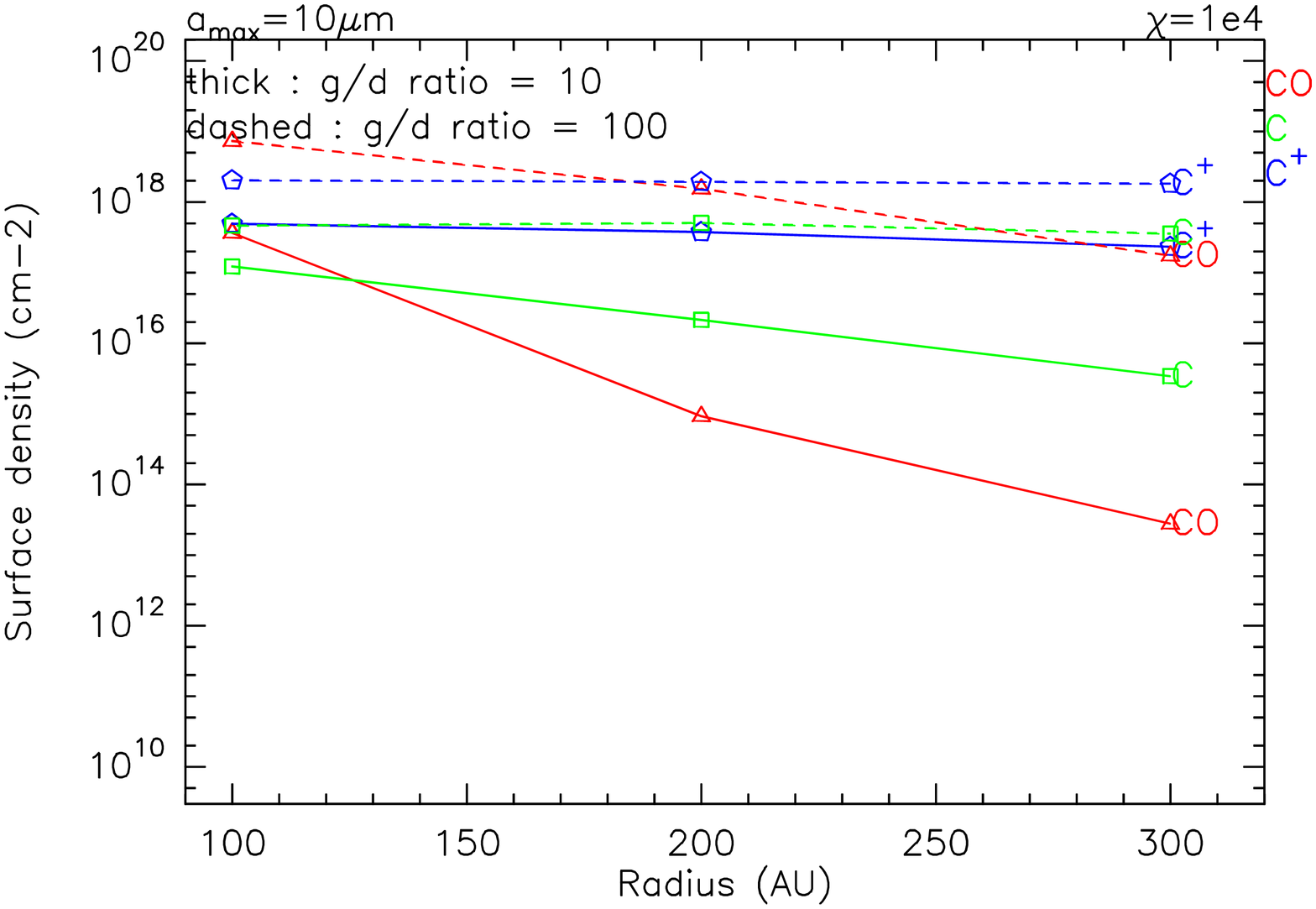}
      &
      \includegraphics[angle=270.0,width=8.0cm]{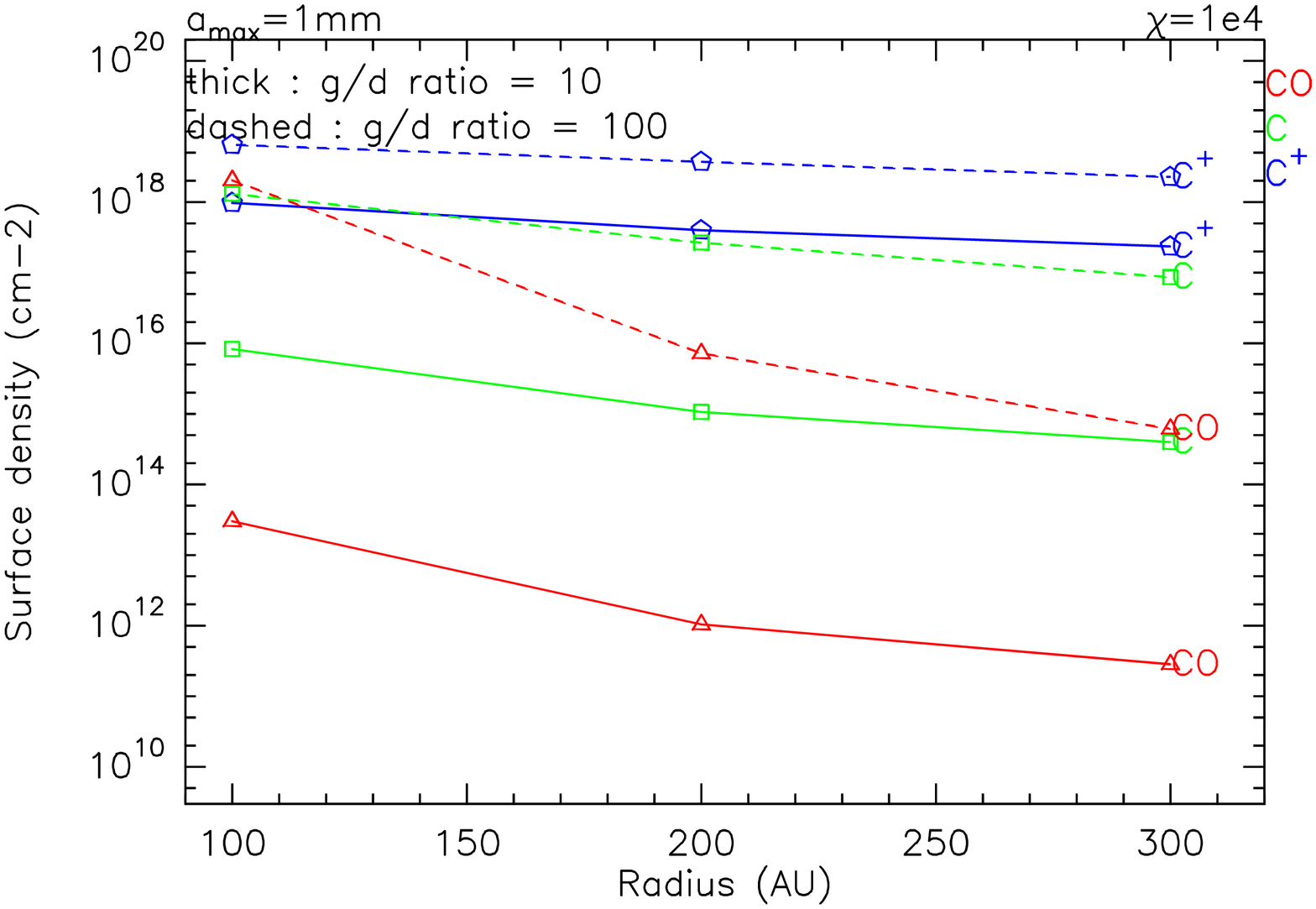}
      \\
    \end{tabular}
  \caption { Grain size effect on column densities. Radial distribution of the column density of C$^+$, C and CO  for the model with standard UV field without the thermal balance calculated. \app = 0.1, 1, 10\,$\mu$m, and 1\,mm from left to right and top to bottom}
  \label{fig:grain_coldens_rad}
\end{figure*}

\label{a:grains}
When the grains are bigger the opacity in the mid-plane is smaller so that the photo-dissociation layer is more extended toward the mid-plane and CO is less abundant (Figure \ref{fig:grain}). In the case of \textit{relatively} big grains (\app = 10 $\mu$m) with a gas to dust ratio of 10 C$^+$ is the dominant form of carbon in the mid-plane for R $\geq$ 200 AU. The radial distribution if the column density are presented on figure \ref{fig:grain_coldens_rad}.

%


\section{Effects of UV field}
\label{a:uv}
\begin{figure*}[t]
  \centering
    \begin{tabular}{cc}
      \includegraphics[angle=0.0,width=7.75cm]{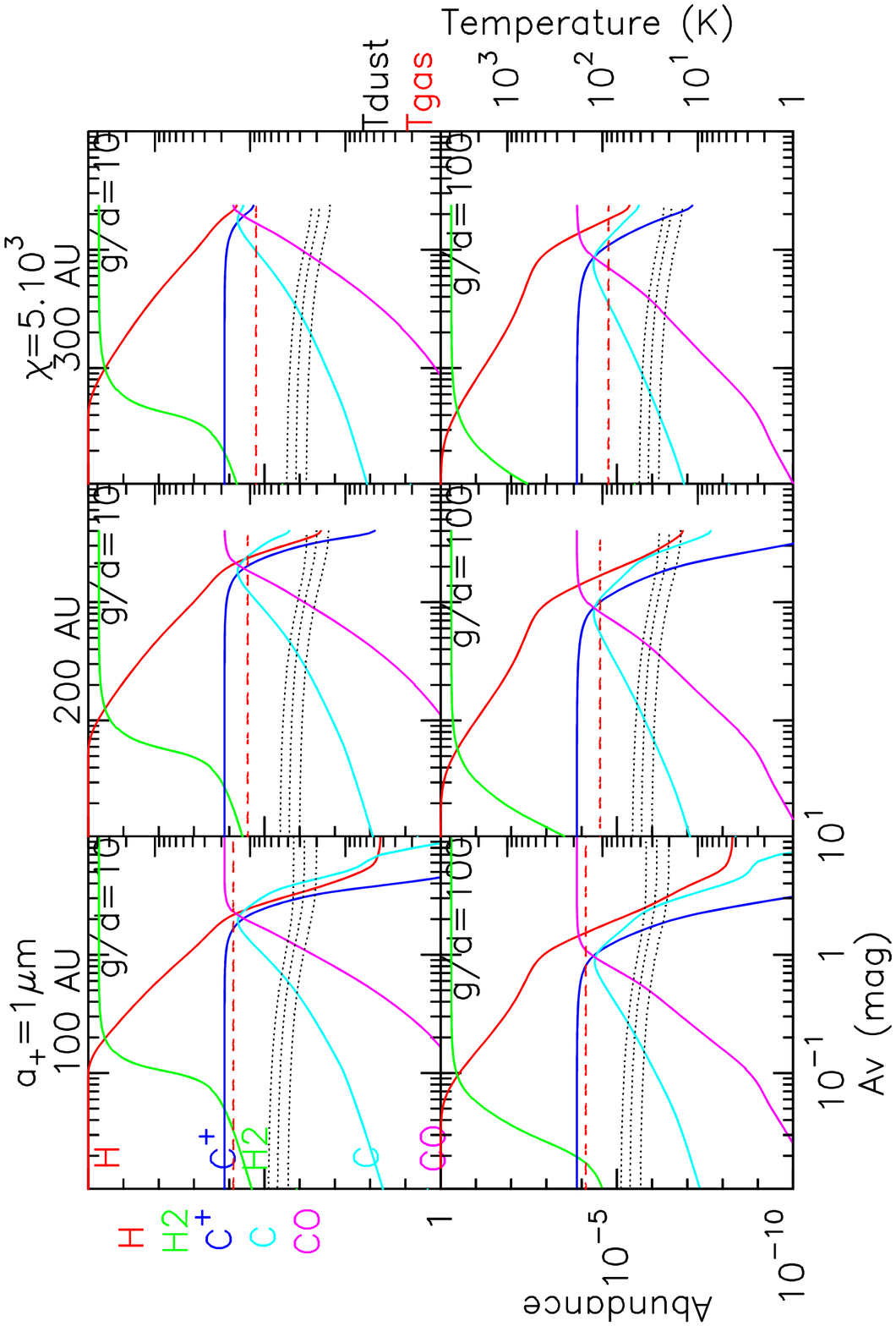}
      &
      \includegraphics[angle=0.0,width=7.75cm]{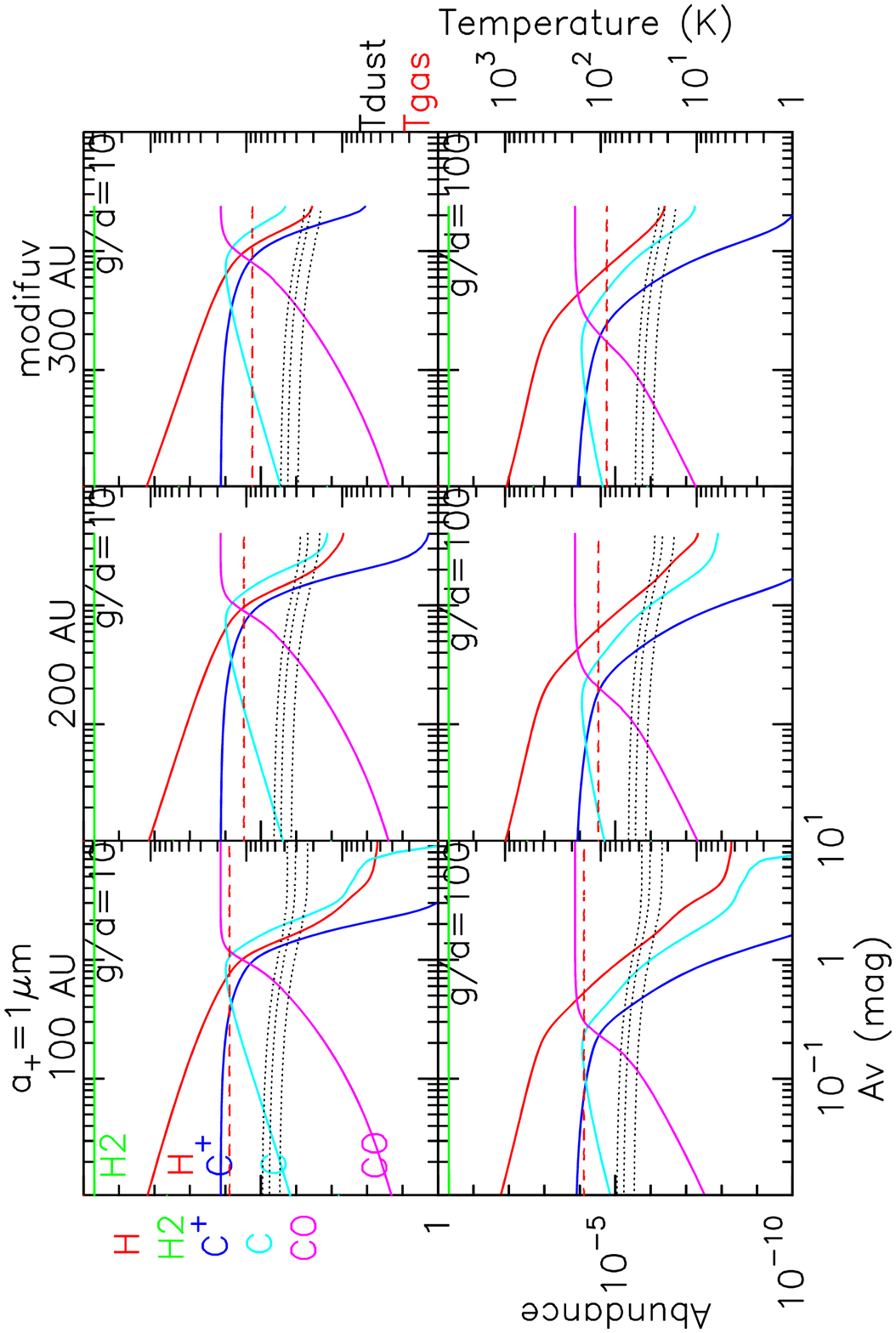}
      \\
      \includegraphics[angle=0.0,width=7.75cm]{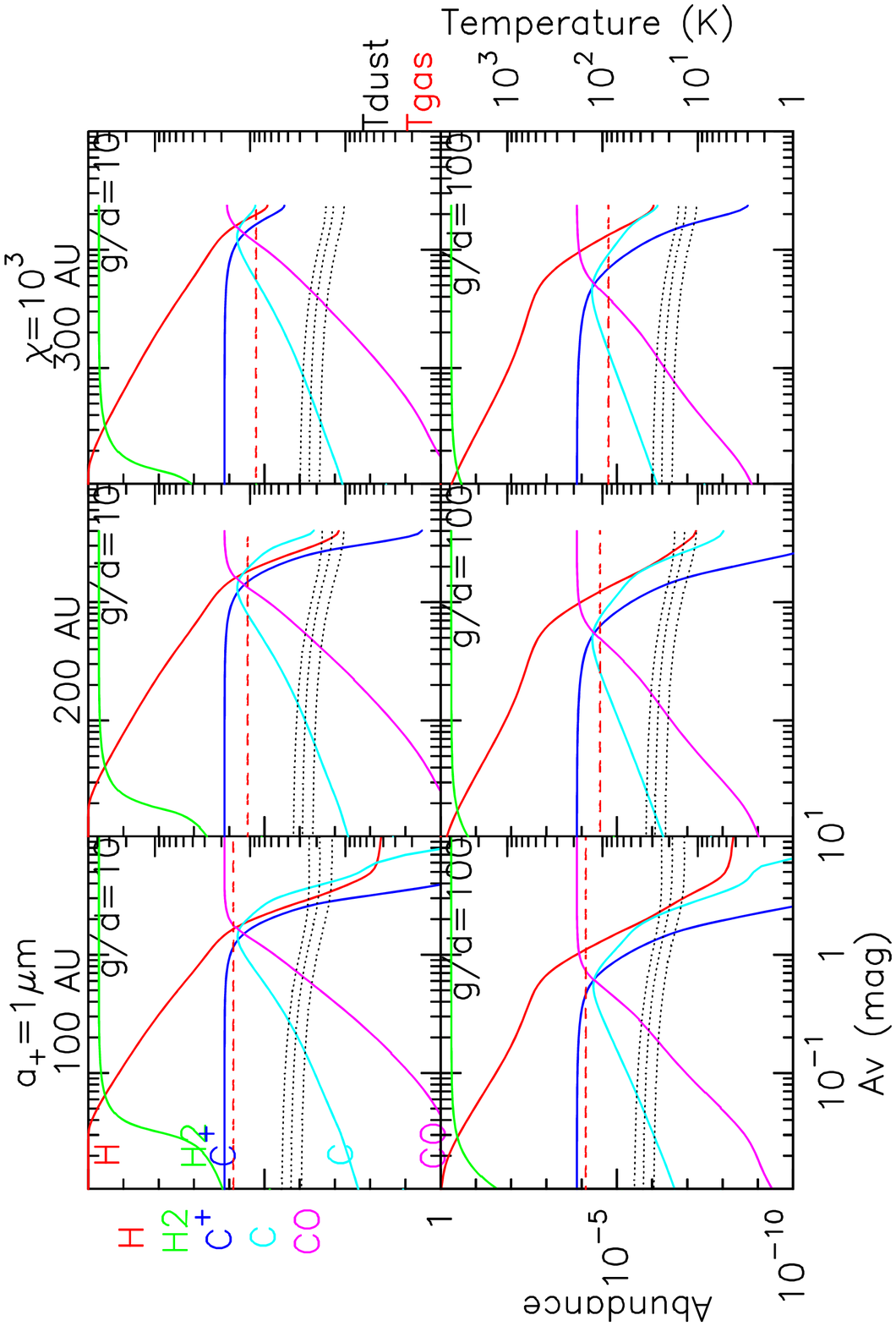}
      &
      \includegraphics[angle=0.0,width=7.75cm]{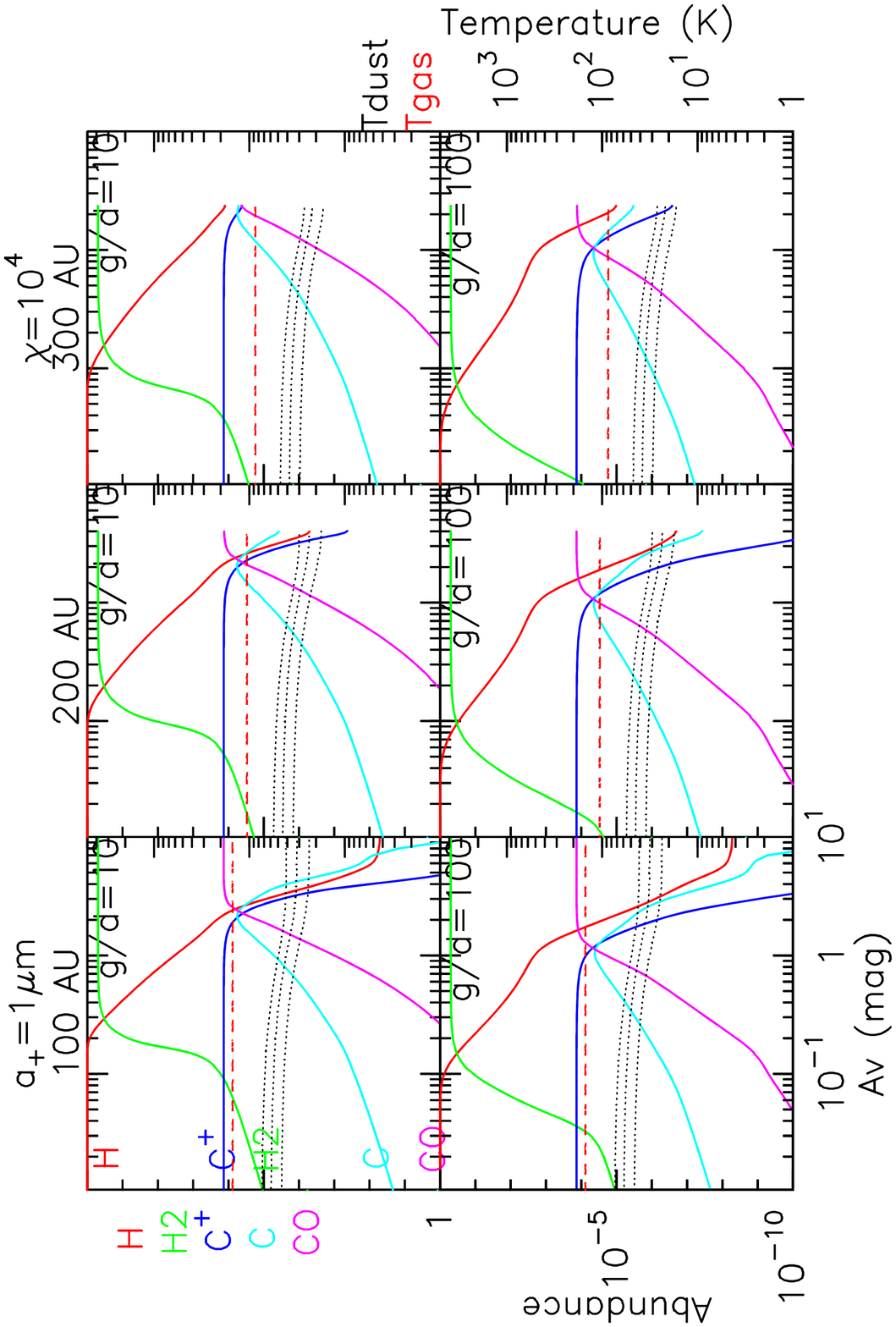}
      \\
    \end{tabular}
    \caption {UV field effect on abundances. Vertical distribution through the disk of the abundance of H, H$_2$, C$^+$, C and CO and of the temperature (no thermal balance calculated)
 at the radius 100, 200 and 300 AU for the models with small grains , a scale factor on the Draine field = 10$^3$, 5.10$^3$ and 10$^4$ at 100 AU and with the modified UV shape (see text for explanation).}
    \label{fig:uv}
\end{figure*}

\begin{figure*}[t]
  \centering
    \begin{tabular}{cc}
      \includegraphics[angle=0.0,width=7.75cm]{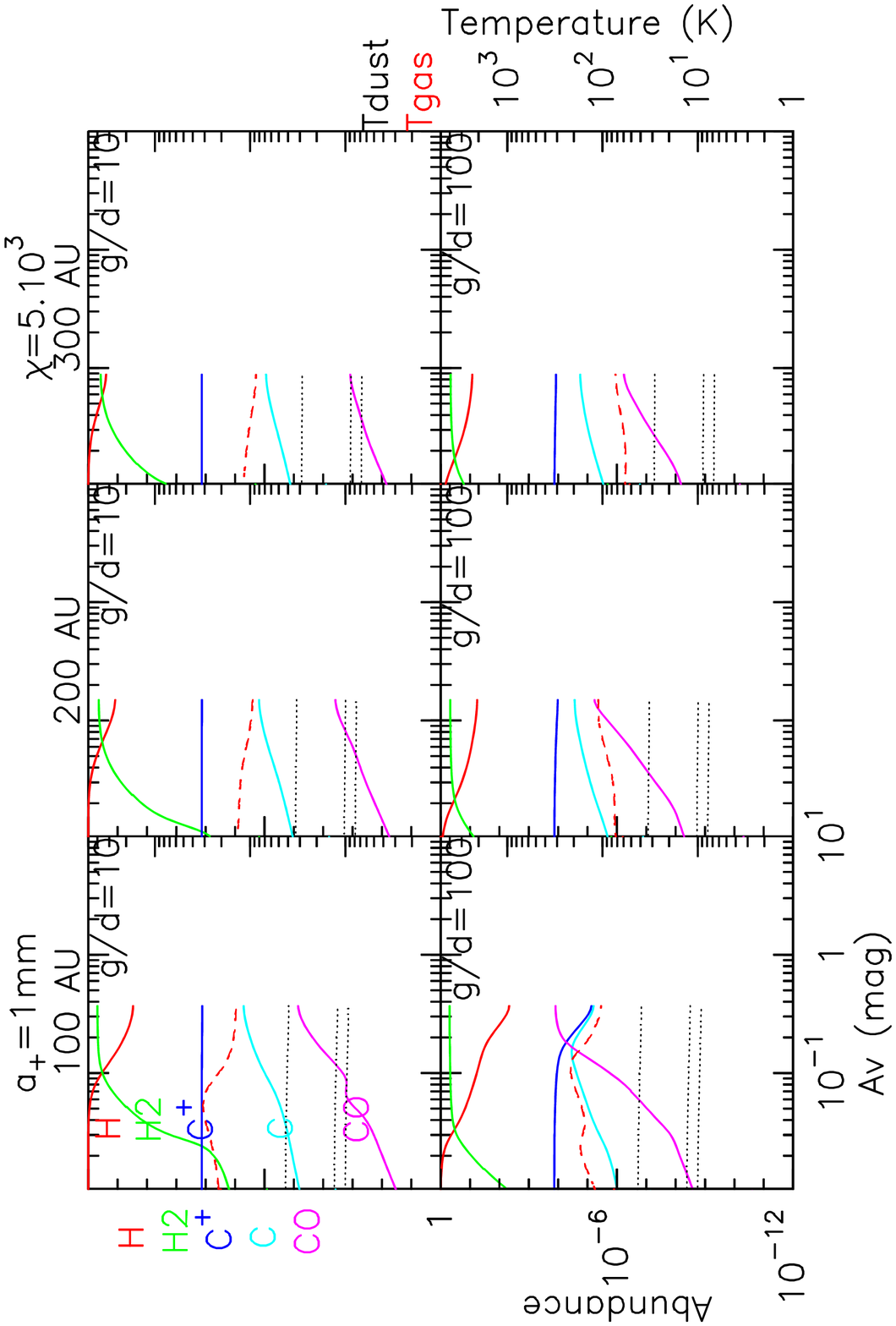}
      &
      \includegraphics[angle=0.0,width=7.75cm]{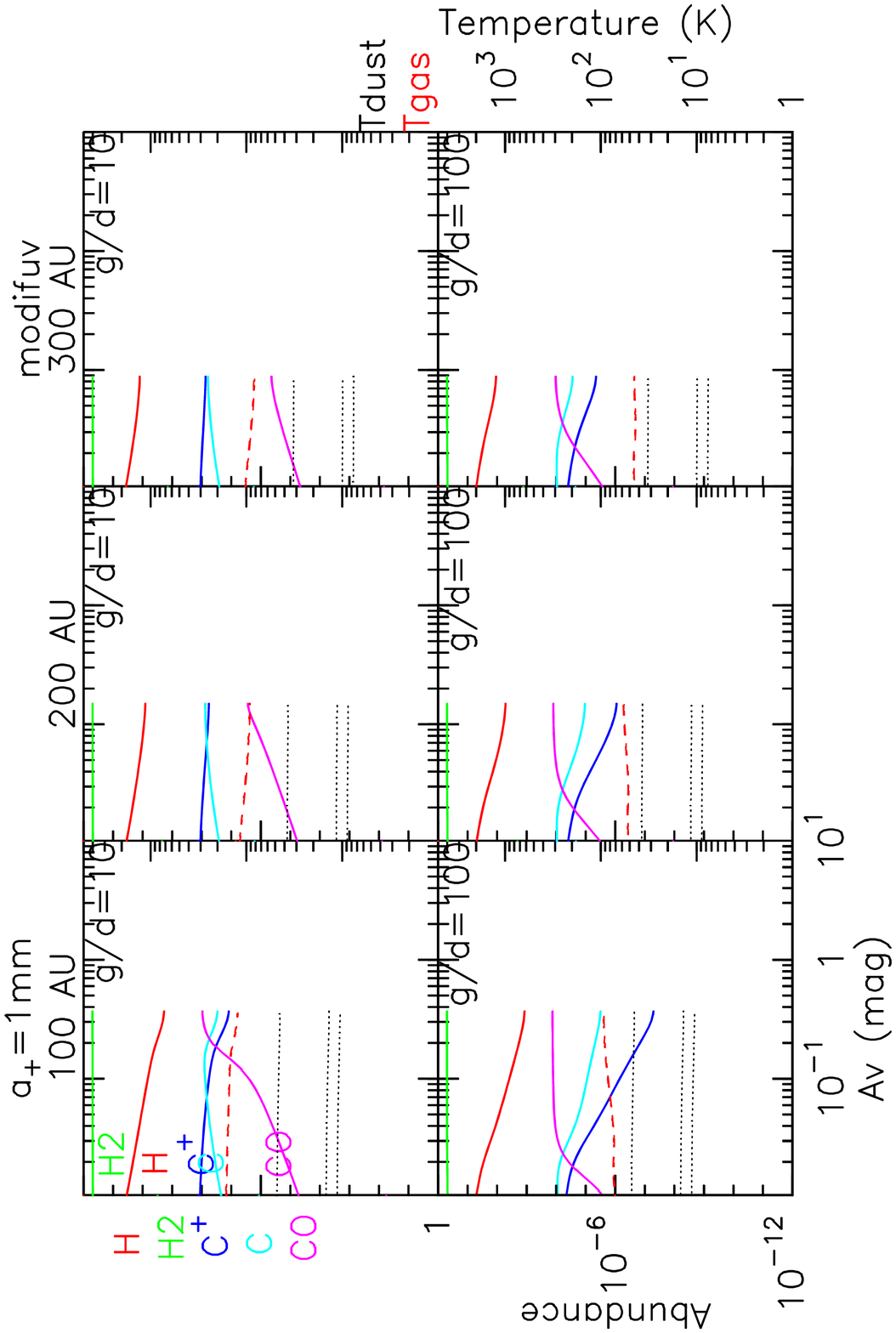}
      \\
      \includegraphics[angle=0.0,width=7.75cm]{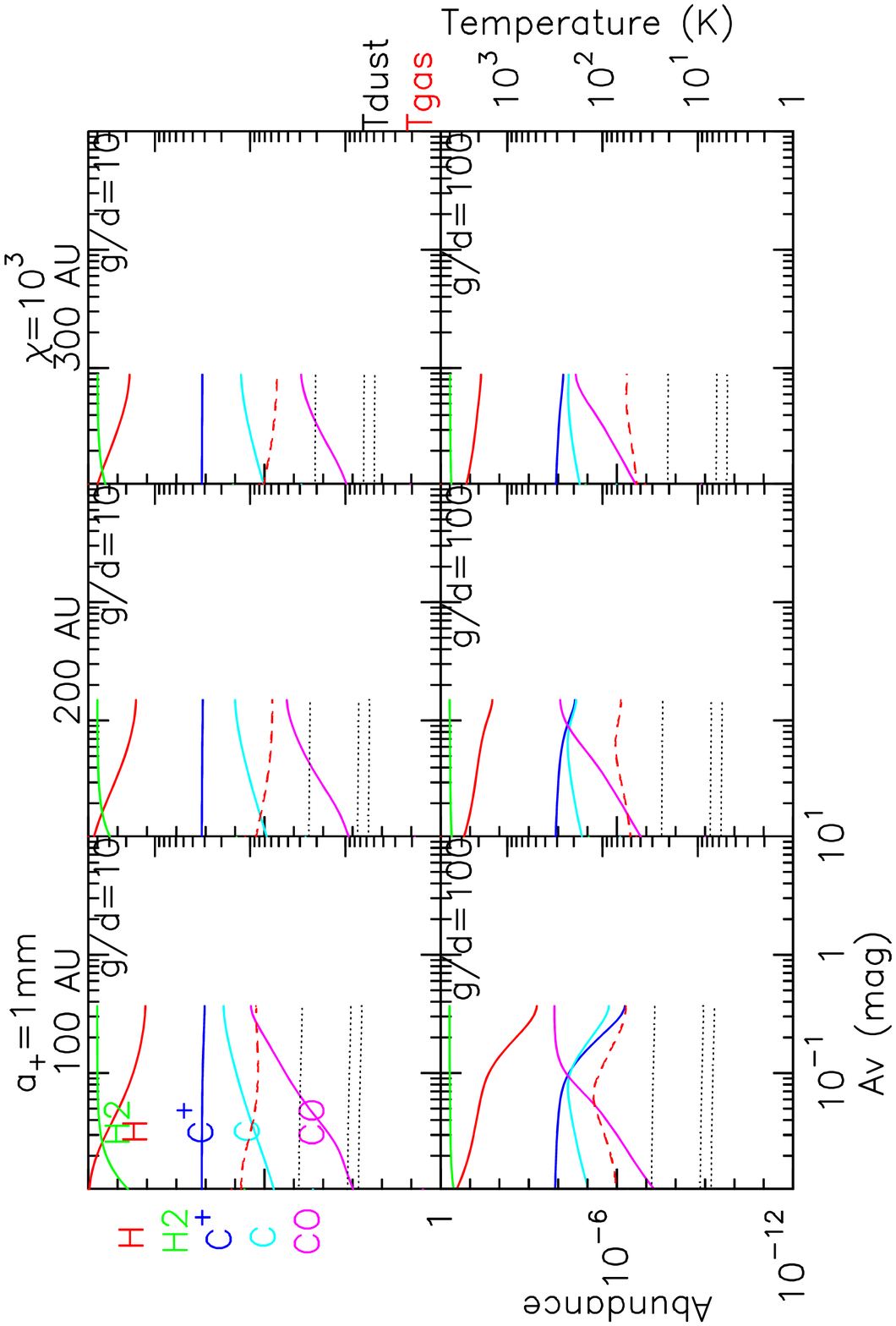}
      &
      \includegraphics[angle=0.0,width=7.75cm]{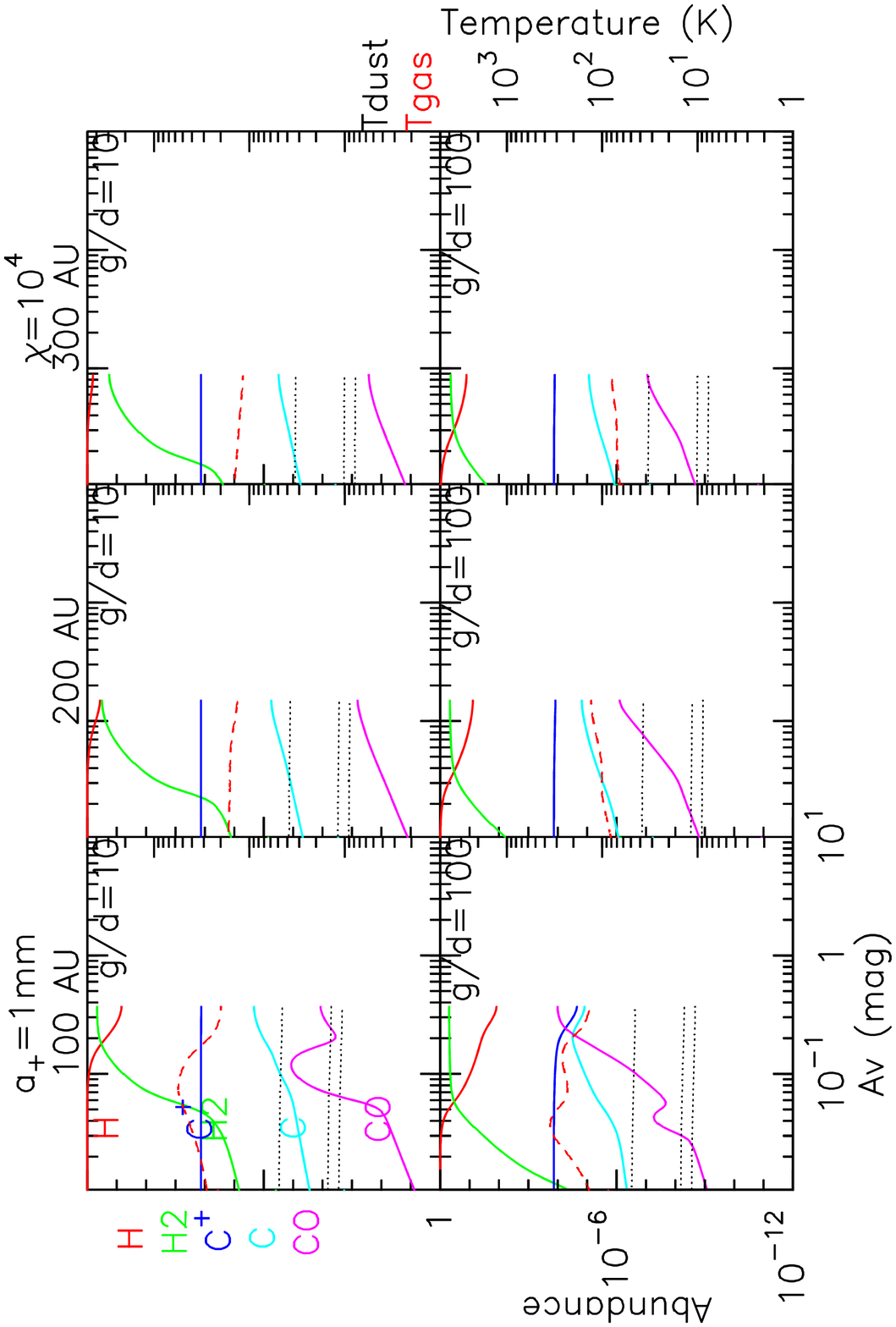}
      \\
    \end{tabular}
    \caption{UV field effect on abundances. Vertical distribution through the disk of the abundance of H, H$_2$, C$^+$, C and CO and of the temperature (\textit{thermal balance calculated}) at the radius 100, 200 and 300 AU for the models with big grains, a scale factor on the Draine field $\chi = 10^3$, $5\,10^3$ and $10^4$ at 100 AU and with the modified UV shape (see text for explanation). Dust temperatures are plotted for the extreme grain sizes (\app and \amm) and an intermediate value ($\frac{a_++a_-}{2}$).}
    \label{fig:uv1e-1}
\end{figure*}

For smaller UV fields, CO is more abundant (figure \ref{fig:uv},\ref{fig:uv1e-1}).
The most important part of the UV spectrum is the one between 900 and 1200 \AA\ where the lines of H$_2$ and CO are, because the self-shielding of CO is important (see the effect for the two gas to dust ratio figure \ref{fig:rfield}).
We perform a calculation with a modified UV field to mimic the shape of the spectrum taken by \citet{Jonkheid_etal2007} (a cut-off below 1200 \AA) with the standard scaling factor $10^4$. The H-H$_2$ transition is dramatically affected: it occurs for a visual opacity less than $10^{-3}$ magnitude instead of about $10^{-1}$ for the $\chi = 10^4$ and $10^{-2}$ magnitude with $\chi =10^3$ for the Draine field shape. CO is less affected but the transition C-CO is also shifted towards the low opacity region. However the UV excess in Herbig Ae star is important and the spectrum shape between 912 and 1600 \AA\ (where photodissociation takes place) is quite well represented by a scaled Draine field (see Figure \ref{fig:flux}).


\begin{figure*}[ht]
  \centering
      \includegraphics[angle=270.0,width=8.0cm]{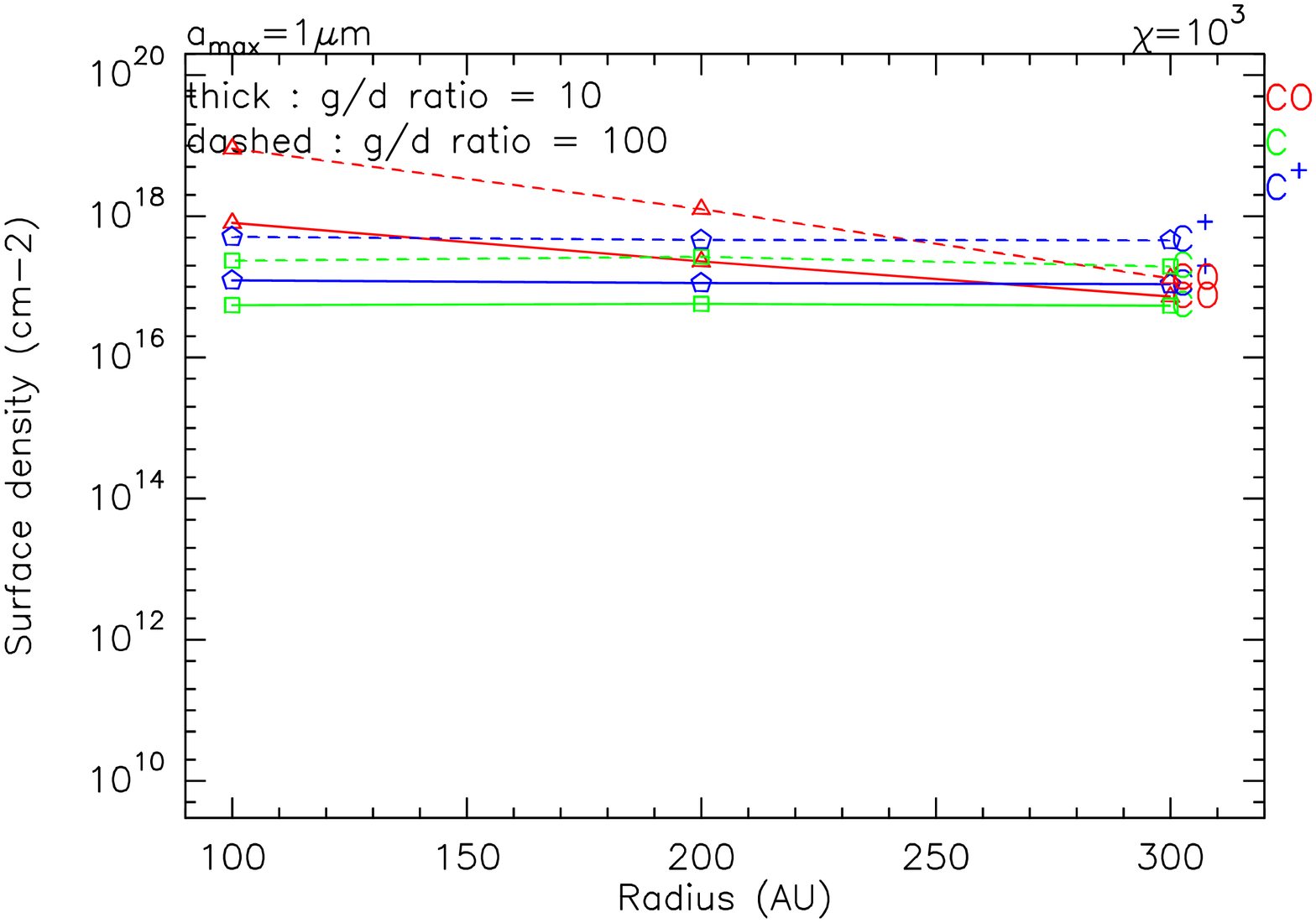}
      \includegraphics[angle=270.0,width=8.0cm]{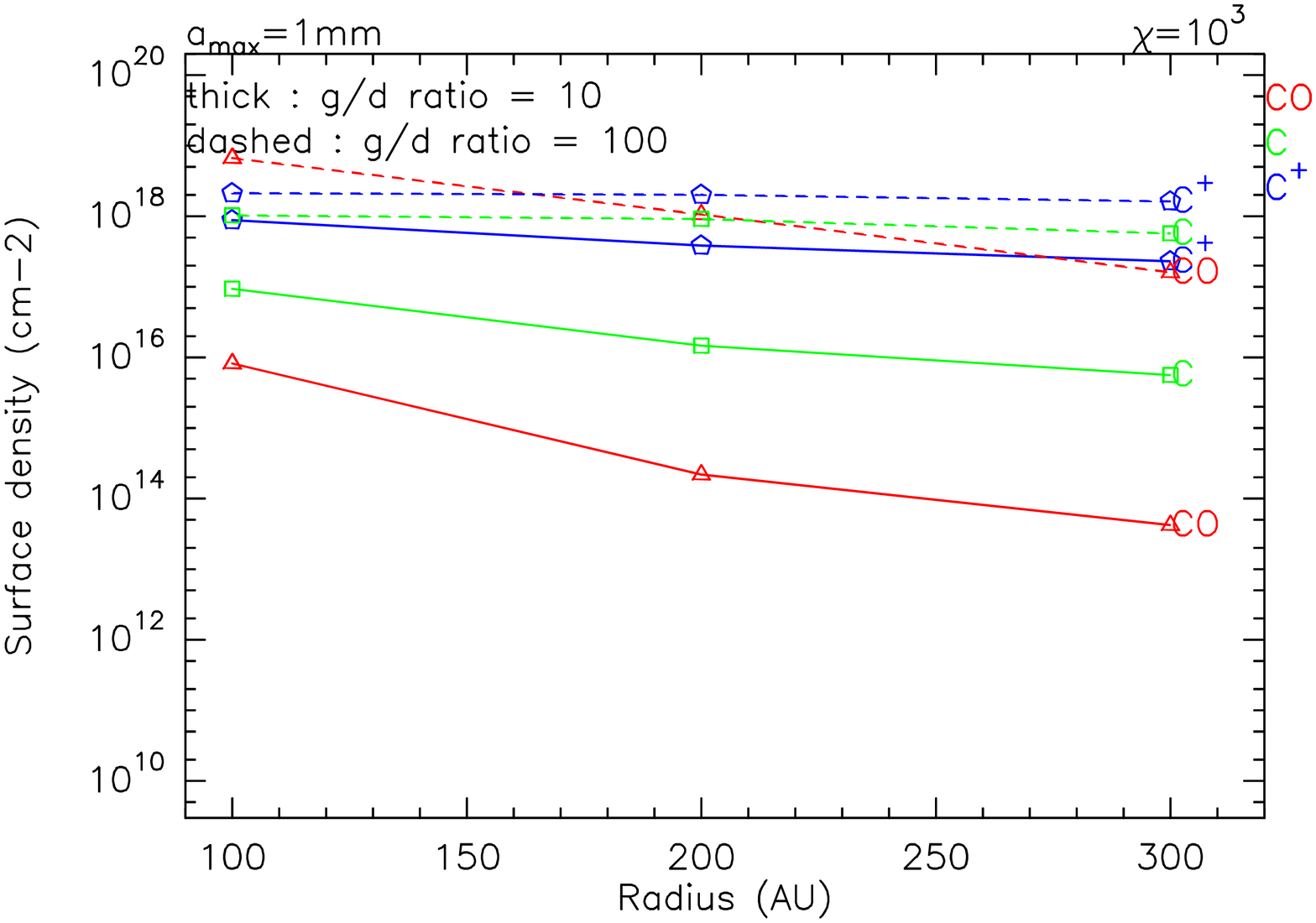}
  \caption{ Grain size effect on column densities for a weak UV field. Radial distribution of the column density of C$^+$, C and CO for the model with weak UV field ($\chi$ =$10^3$ at 100 AU), \app = 1$\mu$m (left) and \app = 1mm (right).}
  \label{fig:coldens-uv1e3}
\end{figure*}

\begin{figure*}[ht]
  \centering
      \includegraphics[angle=270.0,width=8.0cm]{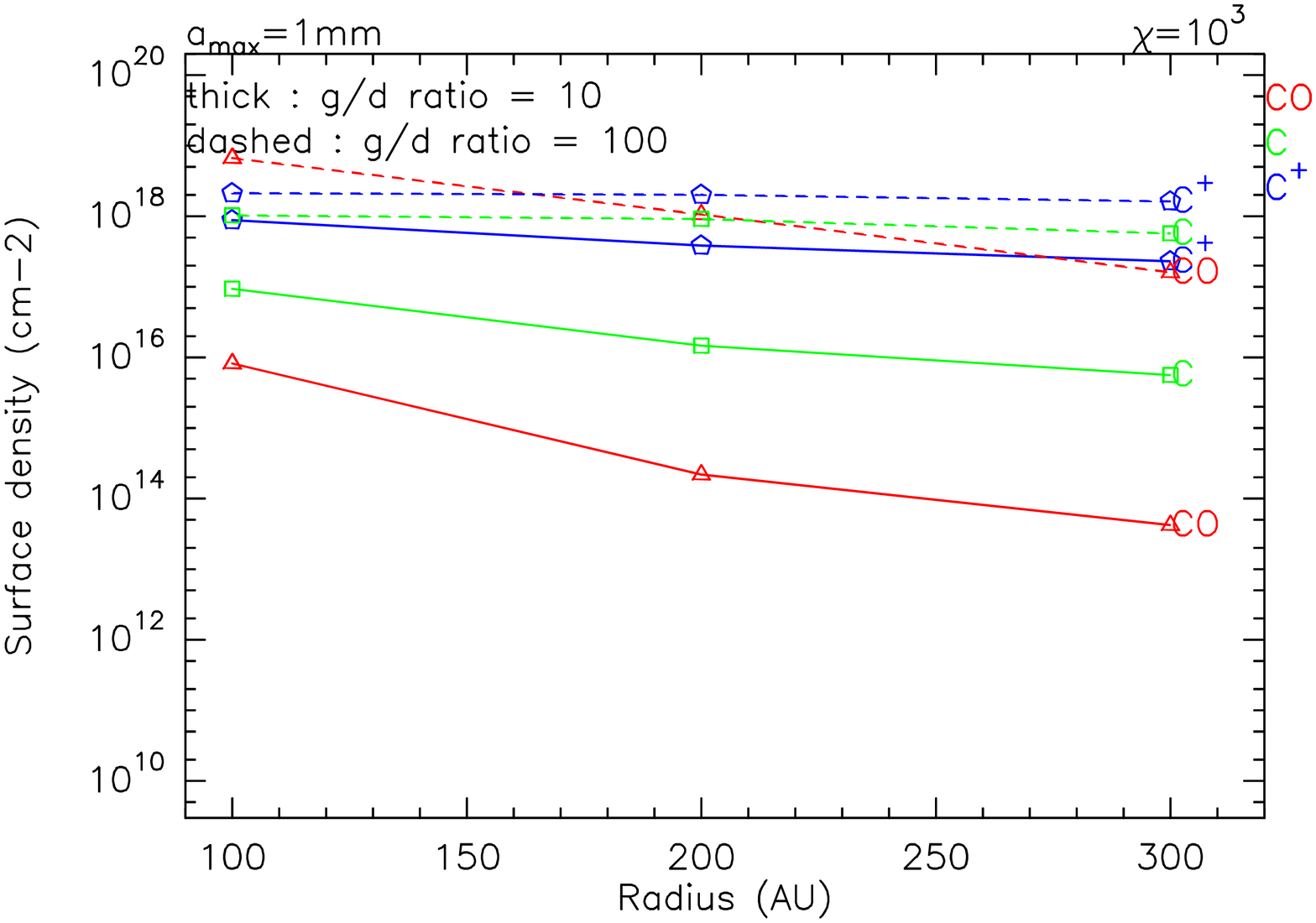}
      \includegraphics[angle=270.0,width=8.0cm]{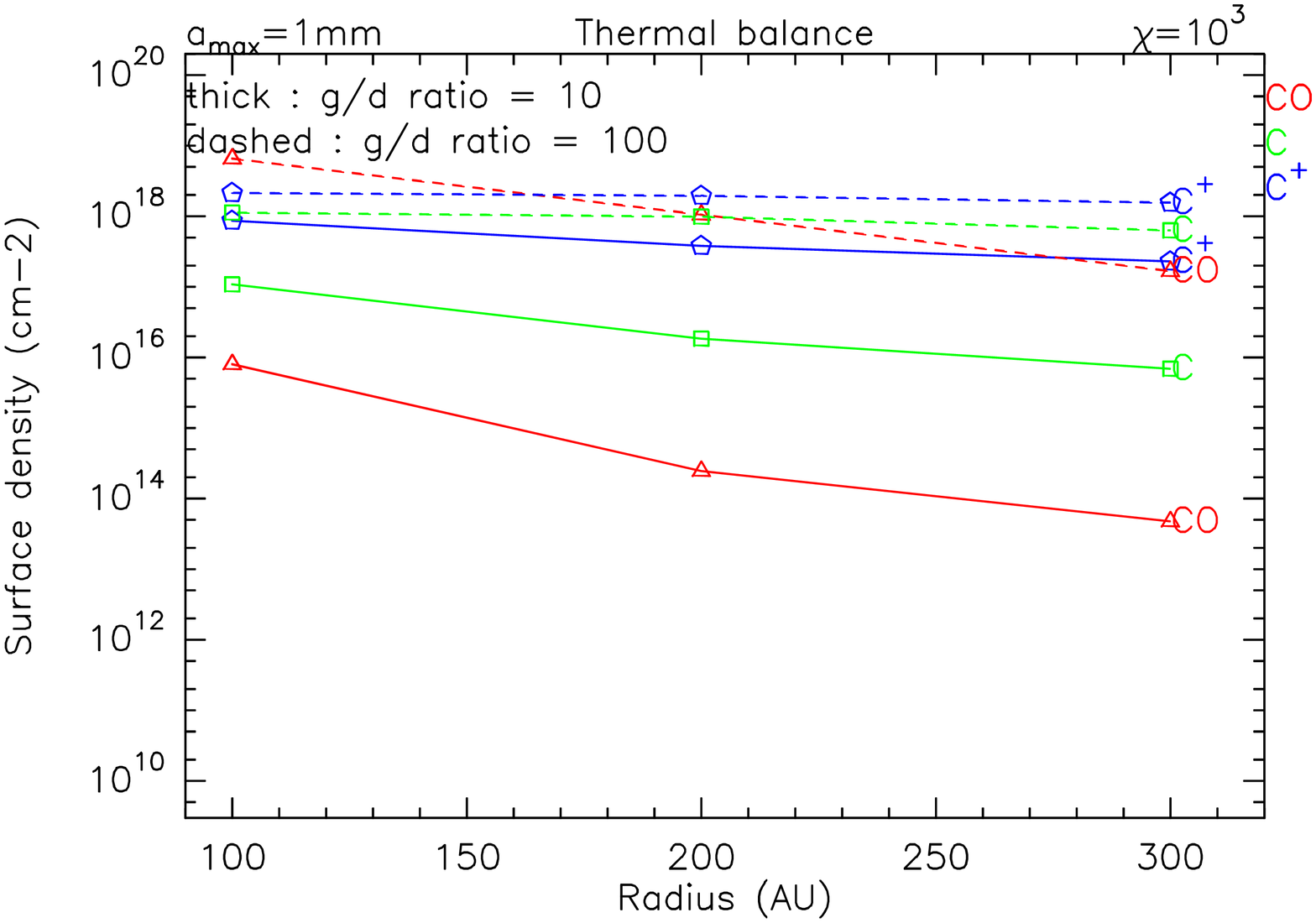}
  \caption{ Thermal balance effect on column densities. Radial distribution of the column density of C$^+$, C and CO  for the model with weak UV field ($\chi =10^3$ at 100 AU), \app = 1mm without (left) and with (right) the thermal balance calculated.}
  \label{fig:1e-1_eqth_coldens_rad}
\end{figure*}

\section{Turbulence}
\label{turbulence}
\begin{figure*}[ht]
  \centering
      \includegraphics[angle=270.0,width=8.0cm]{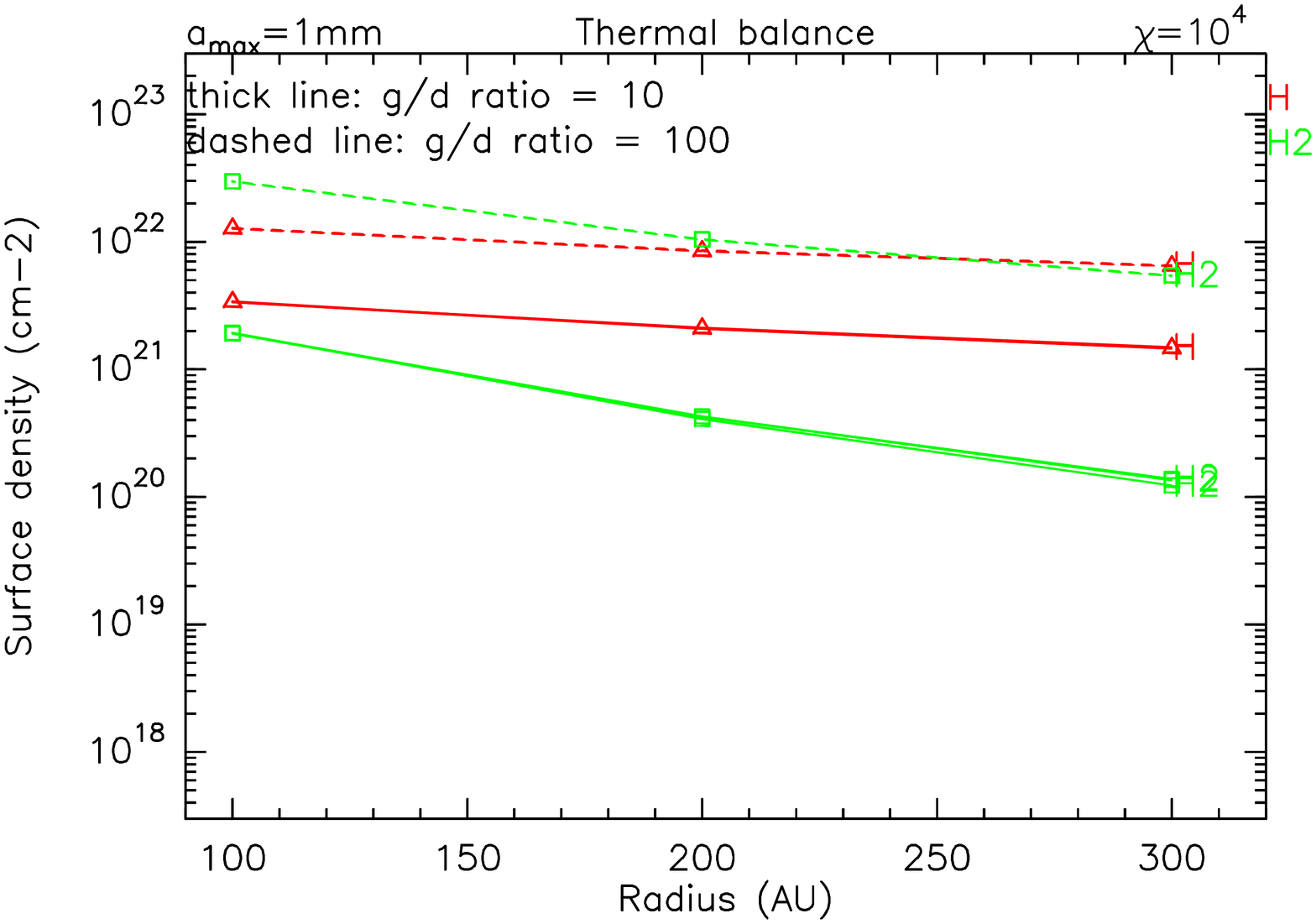}
      \includegraphics[angle=270.0,width=8.0cm]{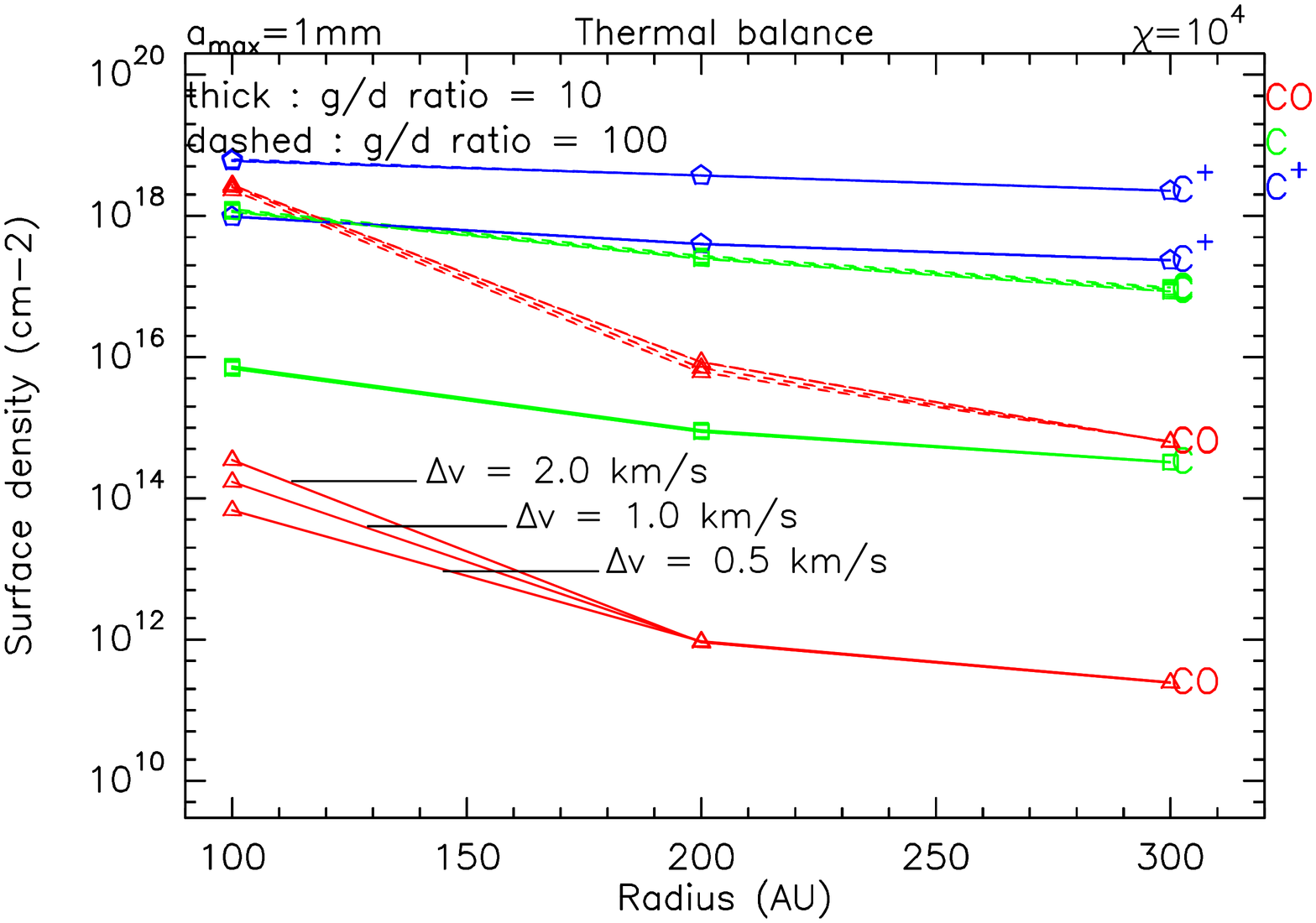}
  \caption{Turbulence effect over column density. Radial distribution of the column density of H, H$_2$ (left), C$^+$, C and CO (right) for three values of the Doppler width with the thermal balance calculated.\app = 1mm (bottom)}
  \label{fig:vturb}
\end{figure*}

We present here the effect of a multiplication and division by a factor 2 of the Doppler width of H$_2$ lines on the surface density. The effect is maximum in the case of big grains with the thermal balance calculated but remain negligible for the C-bearing species (figure \ref{fig:vturb}).

\end{document}